\documentclass[a4paper,11pt]{article}
\usepackage{float} %
\usepackage[section]{placeins} %
\usepackage{subcaption}

\usepackage{tikz}
\usetikzlibrary{arrows.meta}

\usepackage{booktabs}
\usepackage{array}
\usepackage{setspace}                   %
\usepackage{enumitem}
\setlist[itemize]{leftmargin=*} %

\usepackage{amsmath,amssymb,amsthm}
\usepackage{indentfirst}
\usepackage{xcolor}
\usepackage{natbib}

\usepackage[top=2cm,bottom=2cm,left=2.5cm,right=2.5cm,marginparwidth=1.75cm]{geometry}
\usepackage[bookmarks,colorlinks,pagebackref=true]{hyperref} %
\hypersetup{%
           linkcolor=black,%
           filecolor=black,%
           urlcolor=black}
\usepackage[stable]{footmisc} %

\newtheorem{assumption}{Assumption}
\newtheorem{proposition}{Proposition}
\newtheorem{theorem}{Theorem}
\newtheorem{lemma}{Lemma}
\newtheorem{A.assumption}{Assumption}[section]
\newtheorem{A.lemma}{Lemma}[section]
\theoremstyle{definition}
\newtheorem{remark}{Remark}

\newcommand{\cav}{\operatorname{cav}} %

\title{Accelerator and Brake: Dynamic Persuasion with Dead Ends\thanks{Chen: Shandong University, Center of Economic Research, No. 27, South Shanda Road, Jinan, China (email: zhuochen@sdu.edu.cn). Liu: Shandong University, Center of Economic Research, No. 27, South Shanda Road, Jinan, China (email: yliueco@gmail.com). We thank Liang Dai, Jeffrey Ely, Ju Hu, Xu Lang, Jiangtao Li, Chang Liu, Qingmin Liu, Jan Knoepfle, Sivakorn Sanguanmoo, Aidan Smith, Ina Taneva, Wei Zhao, Qiaoxi Zhang, Jie Zheng and Weijie Zhong, as well as participants at the 14th Conference on Economic Design (University of Essex), the 13th World Congress of the Econometric Society (Seoul), and the 27th ACM Conference on Economics and Computation (EC'26), for valuable discussions.}
}

\author{Zhuo Chen \and Yun Liu}
\date{\today} %

\begin{document}
\maketitle

\begin{abstract}
This paper studies dynamic persuasion in a strategic-experimentation relationship in which the principal has a single-peaked preference over the agent's stopping time.
Excessive experimentation may end in a \emph{dead end}.
The principal privately observes project quality, which determines the agent's payoff conditional on success, while both parties learn about feasibility only through the agent's experimentation.
We show that an optimal policy uses at most two one-shot disclosures: an \emph{accelerator} before the principal's ideal stopping time and a \emph{brake} afterward.
A local Arrow--Pratt comparison of induced payoffs over stopping time determines whether the accelerator is concentrated or gradual.
Under common discounting, the comparison yields a one-shot accelerator.
Under heterogeneous discounting, the one-shot result remains robust unless the agent is sufficiently more impatient than the principal, in which case the ranking reverses over an interval and the accelerator can take a one-shot--gradual--one-shot form.


\medskip


\end{abstract}
\large

\section{Introduction}
In many economic relationships, a central conflict between a principal and an agent is deciding when to terminate an ongoing experiment.
Such dilemmas arise naturally when a venture capitalist advises a founder on an R\&D project, or when an academic advisor guides a student's research agenda.
In these settings, the agent continuously weighs the prospective gains from further experimentation against a safe outside option.
Because the two parties evaluate these prospects differently, the agent's preferred stopping time typically diverges from the principal's.

This paper studies how a principal uses dynamic persuasion to influence an agent's experimentation duration.
We focus on informational incentives because standard financial contracting is frequently infeasible in these environments, leaving an informed principal with only informational leverage to align the agent's experimentation with her own interests.
For example, in academic mentorships, intra-firm hierarchies, and policy advisory roles, direct monetary transfers are typically ruled out by institutional rules, social norms, or budget constraints.

This persuasion problem is complicated because the principal's objective is non-monotonic in the agent's stopping time.
Unlike standard persuasion models designed solely to encourage or discourage effort, the principal here faces a two-sided trade-off.
On the one hand, a pessimistic agent may quit too early, abandoning a viable project and destroying surplus.
On the other hand, an optimistic agent may persist too long with an unviable project---wasting resources on a ``dead end'' (cf. \citealp{akcigit2016role,sadler2021dead}).
This tension leaves the principal with conflicting motives over time: she must generate optimism to sustain early experimentation up to her ideal duration, but retain the ability to stop the agent thereafter.

To formally investigate this tension, we study a continuous-time strategic experimentation game based on the exponential bandit framework \citep{keller2005strategic}. 
An agent continuously decides whether to persist with a risky project or irreversibly switch to a safe alternative. 
The project's ultimate \textit{feasibility} is initially unknown to both parties and can only be resolved through a stochastic breakthrough. 
By contrast, the principal is privately informed about the project's \textit{quality}, which determines the agent's payoff upon success, though not her own.
Thus, the principal uses this information to influence the agent's experimentation decisions, aiming to align the agent's stopping time with her own.

We start with a static benchmark in which the principal can disclose information only at the outset.
Standard concavification \citep{kamenica2011bayesian} shows that static disclosure is valuable if and only if the agent's prior belief lies outside an interval containing the aligned belief $\mu^*$ (Proposition~\ref{proposition.static}).
When the agent is too pessimistic (below this interval), the optimal signal reveals the low state with positive probability to boost confidence, and when the agent is too optimistic (above the interval), it reveals the high state with positive probability to discourage excessive continuation.

When information can be disclosed over time, Theorem~\ref{thetheorem} shows that an optimal policy has a surprisingly simple structure: information is disclosed on at most two dates.
The first possible disclosure occurs before the principal's ideal stopping time $t^*$ and acts as an \emph{accelerator}. 
It confirms the low state with positive probability, and thus silence makes the agent more optimistic and willing to continue. 
The second disclosure occurs after $t^*$ and acts as a \emph{brake}. 
It verifies the high state with positive probability, and hence the absence of verification lowers the agent's belief and induces stopping. 
The two disclosures therefore push behavior in opposite directions, aligning the agent's stopping decision with the principal's ideal horizon on both sides of $t^*$.

The central challenge in the dynamic setting is that early motivation and subsequent dissuasion are jointly determined: optimism generated before $t^*$ affects the cost of stopping the agent thereafter, while future disclosure promises feed back into early effort. 
The proof resolves this interaction through a two-step approach centered on the milestone $t^*$.
We first fix the probability of reaching the milestone in each state and the continuation value promised upon reaching it.
These probabilities determine the posterior belief entering the dissuasion subproblem, while the promised continuation value captures how future disclosure affects the agent's willingness to reach the milestone.
Conditional on these milestone variables, we separate the problem into a motivation subproblem leading up to the milestone and a dissuasion subproblem thereafter. 
We solve the two subproblems, each now with a monotone objective, separately; in particular, the dissuasion subproblem reduces to a constrained static persuasion problem \citep{le2019persuasion,doval2024constrained}, whose solution yields the good-news brake. 
Optimizing over the milestone variables then yields the overall optimal policy.

The motivation subproblem yields the one-shot accelerator through a different mechanism.
Because continued experimentation in the high state is mutually beneficial up to $t^*$, the motivation subproblem reduces to designing a stopping-time lottery conditional on the low state.
The key question is whether bad news should be released abruptly at discrete dates or gradually over time.
The two parties differ in both their preferred stopping times and their attitudes toward stopping-time risk, and the optimal design balances the mean stopping time against its risk.
How the bad news is released therefore depends on which party is more sensitive to stopping-time risk.
We show that this choice is governed by a comparison of the players' Arrow--Pratt coefficients computed from their induced payoffs over stopping time. 
Under common discounting, the principal, who prefers the agent to stop later and dislikes risk over stopping time, is uniformly more sensitive to stopping-time risk than the agent, and the optimal accelerator therefore minimizes the stopping-time risk, leading to a one-shot disclosure.

Building on this Arrow--Pratt logic, Appendix \ref{appendix.general} identifies conditions under which the approach extends beyond the exponential-bandit environment to a broader class of dynamic information-design problems.
Appendix~\ref{sec:applications} applies these conditions to existing models of dynamic persuasion, showing how the Arrow--Pratt comparison explains the gradual and one-shot disclosure patterns in their optimal policies.

This same criterion also organizes the case of heterogeneous time preferences.
Unless the agent is sufficiently more impatient than the principal, the principal remains more sensitive to stopping-time risk throughout the motivation region, and thus the accelerator is still one-shot.
If the agent is sufficiently impatient, the ranking reverses on an interval before $t^*$. 
The accelerator can then combine a connected interval of gradual bad-news disclosure with at most one one-shot disclosure on either side, yielding a one-shot--gradual--one-shot structure.
During the gradual phase, the binding continuation constraints pin down the disclosure path in closed form.

Finally, we examine the role of the principal's dynamic commitment.
In contrast to recent dynamic persuasion models \citep[e.g.,][]{koh2024attention,koh2024persuasion} where optimal one-shot disclosure can be implemented without commitment, in our model dynamic commitment is indispensable whenever the accelerator is one-shot.
Because experimentation has instrumental value, confirming the high state does not prompt immediate stopping.
This feature prevents the principal from eliminating the agent's continuation rent during the silence phase leading up to the one-shot accelerator, exposing an uncommitted principal to an inevitable temptation to deviate from the optimal policy.


\subsection{Related Literature}


This paper belongs to the Bayesian persuasion literature initiated by \cite{kamenica2011bayesian}, and is more closely related to dynamic persuasion with commitment and a forward-looking receiver \citep{ely2020moving,orlov2020persuading,ball2023dynamic,liu2023motivating,koh2024attention,koh2024persuasion,saeedi2025getting}.
The closest comparisons clarify three distinct points: what the principal knows, how the optimal disclosure reflects time-risk attitudes, and when commitment is essential.

\citet{ely2020moving} is closely related to our paper because, in both models, the principal first induces optimism and later pessimism.
However, this similarity is only formal.
Ely–Szydlowski study a monotone problem in which both phases of persuasion serve the same end---prolonging effort---and the reversal arises because past effort is a sunk cost.
In our model, the two phases pursue opposing objectives: the early accelerator induces the agent to reach $t^*$, while the brake actively shortens experimentation after $t^*$.

The closest experimentation paper is \citet{liu2023motivating}. 
In both models, information about future rewards is used to motivate effort in a bandit environment, but the theoretical problem is substantively different.
In Liu, quality and feasibility are perfectly correlated, and hence the principal has access from the outset to the information that determines how long the agent should optimally experiment. 
Thus, her only reasonable objective is therefore to prolong effort.
In our model, however, quality and feasibility are decoupled: the principal knows only quality, while feasibility is learned solely through the agent's own experimentation.
The principal therefore also relies on the agent's effort to acquire the feasibility information, which is central to the principal's non-monotone objective.

\citet{koh2024attention} and \citet{saeedi2025getting} are closest to our analysis of attitudes toward stopping-time risk in dynamic disclosure.
Beyond cases in which relative attitudes toward stopping-time risk are globally ordered, both papers study environments in which the relevant curvature ranking varies over time.
In special cases, both obtain a gradual phase followed by an interval of no disclosure and a final one-shot revelation.
We show how the players' relative sensitivity to stopping-time risk determines whether disclosure is gradual or concentrated at a single date within a given interval.
This interval-by-interval perspective provides a unified theoretical framework for understanding how local attitudes toward stopping-time risk shape optimal disclosure across otherwise different dynamic information-design problems.

Our paper is also related to \citet{koh2024attention} and \citet{koh2024persuasion} in its concern with commitment.
They show that, in their setting, the optimal one-shot disclosure can be reconstructed as a sequentially optimal belief martingale, and thus dynamic commitment is dispensable.
Their no-commitment implementation nonetheless relies on a belief-martingale construction that exploits multiplicity in the agent's indifference beliefs. 
In our experimentation environment, effort has instrumental value; that is, even after the high-quality state is known, the agent may want to keep experimenting because a breakthrough can still arrive.
This uniquely pins down the relevant indifference belief by backward induction and prevents the probability-tree surgery used in those papers.

The paper is also related to strategic experimentation with exponential bandits.
\citet{keller2005strategic} formalize the exponential-bandit model, which has been used in contract design \citep{BH11,halac2016optimal}, delegation \citep{guo2016dynamic,escobar2021delegating}, and contest design \citep{HKL17,bimpikis2019designing,ely2023optimal}.
Among these papers, \citet{sadler2021dead} explicitly studies dead ends as the decision to abandon one research idea for another, and shows that stronger incentives need not improve experimentation outcomes because they may induce an agent to persist with an unpromising idea for too long.
Our paper instead studies the role of information disclosure when an agent must decide whether to abandon a project that may be a dead end.

\bigskip
The rest of the paper is organized as follows.
Section \ref{sec:model} introduces the formal model and the static benchmark.
Section \ref{section.main} characterizes the optimal dynamic policy under common discounting and commitment.
Section \ref{section.misaligned} studies heterogeneous discounting. Section \ref{section.commitment} analyzes the value of dynamic commitment. Section \ref{sec:con} concludes.
Appendix \ref{appendix.general} develops the Arrow--Pratt approach and characterizes its scope, while Appendix B contains the proofs.
\section{Model}
\label{sec:model}

\subsection{The Persuasion Problem}

We consider a dynamic interaction between a principal (she) and an agent (he) in continuous time, $t\geq 0$.
There is no monetary transfer.
The agent faces a two-armed bandit problem, following \cite{keller2005strategic}.
The \textit{safe} arm generates known and deterministic flow payoffs for both players whenever it is played.
The present discounted values of these payoff streams are $Z$ for the principal and $z$ for the agent.
The \textit{risky} arm has uncertain \textit{feasibility}; it is either feasible or infeasible.
If infeasible, it yields zero payoff to both players.
If feasible, it yields zero payoff until a stochastic \textit{breakthrough} arrives.
Upon breakthrough, the project yields a lump-sum payment of $Y$ to the principal and $y_{\theta}$ to the agent, where $\theta \in \{H, L\}$ represents the \textit{quality} of the risky arm, with $y_H>y_L$.
We assume quality and feasibility are independent.

Neither player knows whether the risky arm is feasible, and both share the prior probability $p_0\in(0,1)$ that it is feasible.
Whenever the risky arm is feasible and the agent continues to work on it, a breakthrough arrives at a constant Poisson rate $\lambda$.
A breakthrough is publicly observed, and we let $p_t$ denote the common posterior probability at time $t$ that the risky arm is feasible.
If no breakthrough arrives during the interval $[t,t+dt)$, Bayes' rule gives
\[
p_{t+dt}=\frac{p_t(1-\lambda dt)}{p_t(1-\lambda dt)+1-p_t}.
\]
This implies $\dot p_t=-\lambda p_t(1-p_t)$.
Solving the ODE yields the posterior belief conditional on no breakthrough up to time $t$:
\[
p_t=\frac{p_0e^{-\lambda t}}{p_0e^{-\lambda t}+1-p_0}.
\]

The payoff parameters $Y$, $Z$, and $z$ are common knowledge, while the quality state $\theta$, which determines the agent's payoff upon a breakthrough, is privately observed only by the principal.
The agent has a prior belief  $\mu_0\in(0,1)$ that the quality is high.
 
Assume that both players are risk neutral and discount future payoffs at a common rate $r>0$.
Therefore, the net payoff to the agent from continuing from time $t$ to time $s>t$, after which he switches irreversibly to the safe arm, is given by
 \begin{equation*}
 \begin{aligned}
  \bar{v}_\theta(t,s)\equiv \;& p_t\cdot\int_t^s\lambda e^{-(\lambda+r)(s'-t)}ds'\cdot y_\theta+\left(1-p_t+p_t\cdot e^{-\lambda(s-t)}\right)\cdot e^{-r(s-t)}\cdot z-z\\
  =&\left(1-e^{-(\lambda+r)(s-t)}\right)\frac{p_t\lambda y_\theta}{\lambda+r}+\left(1-p_t+p_t\cdot e^{-\lambda (s-t)}\right)\cdot e^{-r(s-t)}\cdot z-z
 \end{aligned}
 \end{equation*}
 where $\theta\in\{H,L\}$.
 Throughout the paper, let $v_\theta(s)\equiv \bar v_\theta(0,s)$ denote the agent's net payoff from experimenting from date $0$ to date $s$ in state $\theta$.
 Accordingly, if the agent works from date $0$ to $s$, the principal's payoff is
 \begin{align*}
  w(s) \equiv \left(1-e^{-(\lambda+r)s}\right)\frac{p_0\lambda Y}{\lambda+r}+\left(1-p_0+p_0\cdot e^{-\lambda s}\right)\cdot e^{-rs}\cdot Z.
 \end{align*}
We focus throughout on the nontrivial case in which the maximizer $t^*$ of $w(s)$ is positive, or equivalently, $p_0\lambda(Y-Z)>rZ$.
In this case, the function $w(s)$ is single-peaked and is maximized at
\[
t^*=\frac{1}{\lambda}\ln\left(\frac{p_0}{1-p_0}\frac{\lambda Y-(\lambda+r)Z}{rZ}\right).
\]

The principal can influence the agent only by revealing information about $\theta$.
To persuade the agent, the principal credibly commits to a dynamic information policy at the beginning of the interaction. 
Following the \textit{effort schedule approach} of \cite{ely2020moving}, we represent the information policy as a joint probability distribution over actions and states.
Formally, an information policy is represented by a pair of cumulative distribution functions $\mathcal{P}=\langle F_H,F_L\rangle$, where $F_\theta(t)$ is the probability that the policy recommends stopping by date $t$ in state $\theta$ along the history with no breakthrough. 
Thus, $dF_\theta(t)$ denotes the probability assigned to stopping at time $t$ given quality $\theta$.\footnote{Throughout the paper, our notation $dF_\theta(t)>0$ covers both atomic and continuous stopping: it denotes either a positive atom at $t$ or a positive density at $t$ when $F_\theta$ is locally absolutely continuous.}

An information policy is implementable if the agent is willing to follow every recommendation prescribed by the policy, and it imposes two classes of incentive constraints.
The first are \emph{continuation constraints}, ensuring that the agent is willing to continue on the risky arm when he is recommended to do so.
Let $\mu_t$ be the belief that the agent assigns to state $H$ when he is not told to stop by time $t$. 
By Bayes' rule,
\begin{equation*}
\mu_t=\frac{\mu_0(1-F_H(t))}{\mu_0(1-F_H(t))+(1-\mu_0)(1-F_L(t))}.
\end{equation*}
Then by the \emph{better-than-no-information} criterion, an information policy $\mathcal{P}=\langle F_H,F_L\rangle$ satisfies the continuation constraints if and only if
 \begin{equation}
 \label{implementable.raw.1}
  \mu_t\int_t^\infty\bar{v}_H(t,s)dF_H(s|s\geq t)+(1-\mu_t)\int_t^\infty\bar{v}_L(t,s)dF_L(s|s\geq t)\geq 0,\quad\forall t\in\mathcal{C}(\mathcal{P}).
 \end{equation}
Here, for any information policy $\mathcal{P}=\langle F_H,F_L\rangle$, $\mathcal{C}(\mathcal{P})$ is the set of times $t$ for which $\mu_0F_H(t)+(1-\mu_0)F_L(t)<1$; in other words, $\mathcal{C}(\mathcal{P})$ is the set of on-path times where a recommendation to continue is still possible.

The second class is the \emph{stopping constraints}, which ensure the agent is willing to quit the risky arm whenever recommended.\footnote{The stopping constraint does not appear in monotone problems, where it is slack.}
Let $\nu_t$ be the agent's posterior belief upon receiving the recommendation to stop at time $t$.
By Bayes' rule,
\begin{equation*}
\nu_t=\frac{\mu_0\,dF_H(t)}{\mu_0\,dF_H(t)+(1-\mu_0)\,dF_L(t)}.
\end{equation*}
The stopping constraint is therefore
 \begin{equation}
 \label{implementable.raw.2}
  V_\text{NI}(t,\nu_t)\equiv\sup_{s\geq t}\bigg(\nu_t\bar{v}_H(t,s)+(1-\nu_t)\bar{v}_L(t,s)\bigg)\leq 0,\quad\forall t\in\mathcal{S}(\mathcal{P}).
 \end{equation}
Here,
\[
\mathcal{S}(\mathcal{P})\equiv
\left\{t:\max\{dF_H(t),dF_L(t)\}>0\right\}
\]
is the set of times at which the policy assigns positive probability to stopping.

Let $\mathcal{I}$ be the set of all policies satisfying the two families of incentive constraints.
The principal's optimization problem can be expressed as 
 \begin{equation}
 \label{optimization}
 \begin{aligned}
  \max_{F_H,F_L}:&\,\int_0^\infty w(t)d\bigg(\mu_0F_H(t)+(1-\mu_0)F_L(t)\bigg)\\
  &\text{subject to: }
   \langle F_H,F_L\rangle\in\mathcal{I}.
 \end{aligned}
 \end{equation}

\subsection{Benchmark: Static Persuasion}


We first consider the case in which the principal can disclose information only at the beginning of the project ($t=0$).
For a belief $\mu$, let $y(\mu)=\mu y_H+(1-\mu)y_L$ denote the agent's expected payoff from a breakthrough.
Given belief $\mu$ and no further disclosure, the agent chooses a stopping time $s\geq0$ to maximize $\mu v_H(s)+(1-\mu)v_L(s)$.
When $p_0\lambda[y(\mu)-z]\leq rz$, his optimal stopping time is $\tau(\mu)=0$.
When $p_0\lambda[y(\mu)-z]>rz$, his optimal stopping time is positive, which is given by
\begin{equation}
  \label{eq.tau}
  \tau(\mu)=\frac{1}{\lambda}\ln\left(\frac{p_0}{1-p_0}\frac{\lambda y(\mu)-(\lambda+r)z}{rz}\right).
\end{equation}
Let
\[
\bar\mu_0\equiv y^{-1}\!\left(\left(1+\frac{r}{p_0\lambda}\right)z\right)
\]
denote the belief threshold below which $\tau(\mu)=0$.
Furthermore, the principal prefers a longer experimentation duration on the risky arm, i.e., $t^*>\tau(\mu)$, if and only if
\[
 \frac{Y}{Z}>\frac{y(\mu)}{z}.
\]

We focus on the case in which the agent stops before $t^*$ in the low-quality state and after $t^*$ in the high-quality state.

 \begin{assumption} \label{asmp:t}
  $\tau(0)<t^*<\tau(1)$.
 \end{assumption}

Clearly, this assumption holds if and only if $y_L/z<Y/Z<y_H/z$.
Also, as a function of the agent's belief $\mu$, the principal's payoff is given by
\begin{align*}
W_{\text{NI}}(\mu)\equiv w(\tau(\mu))=p_0\left(\int_0^{\tau(\mu)}\lambda e^{-(\lambda+r)t}dt\right)\cdot Y+\left(1-p_0+p_0e^{-\lambda \tau(\mu)}\right)e^{-r\tau(\mu)}\cdot Z.
\end{align*}
To characterize the optimal static disclosure, we distinguish two binary signal structures.
A \textit{perfect-bad-news signal} has one realization that fully reveals state $L$ and hence induces posterior $0$, and a \textit{perfect-good-news signal} has one realization that fully reveals state $H$ and hence induces posterior $1$.
The standard concavification argument then yields the following result.

 \begin{proposition}
 \label{proposition.static}
 There exists an interval of beliefs $I=[\mu_L,\mu_H]$, such that:
\begin{enumerate} 
  \item[(i.)] The interval $I$ contains the alignment belief $\mu^*\equiv\tau^{-1}(t^*)$, and the principal benefits from static persuasion if and only if $\mu_0\notin I$.
  \item[(ii.)] If $\mu_0<\mu_L$, the optimal signal is a perfect-bad-news signal, with the other posterior equal to $\mu_L$.
  \item[(iii.)] If $\mu_0>\mu_H$, the optimal signal is a perfect-good-news signal, with the other posterior equal to $\mu_H$.
 \end{enumerate}
 \end{proposition}

When the prior lies in $I=[\mu_L,\mu_H]$, the two parties' preferences are sufficiently aligned, and the static no-information policy is already optimal.
Outside this interval, if the agent is too pessimistic ($\mu_0<\mu_L$), a perfect-bad-news signal raises the posterior conditional on no bad news and prolongs experimentation.
If the agent is too optimistic ($\mu_0>\mu_H$), a perfect-good-news signal lowers the posterior conditional on no good news and induces earlier stopping.
Figure \ref{fig.concavification} illustrates the concavification, where the dashed segments are the relevant parts of the concave envelope of $W_{\text{NI}}$.

 \begin{figure}[ht!]
 \centering
 \begin{tikzpicture}
     \draw[thick,->](0,0)--(9,0) node [below] {$\mu$};
     \draw[thick,->](0,0)--(0,6) node [left] {$W_\text{NI}(\mu)$};
     \draw[thick] (0,1)--(1,1);
     \draw[thick,domain=1:3] plot(\x,{-(\x-3)^2+5});
     \draw[thick,domain=3:5] plot(\x,{1/2*(1+6*\x-\x^2)});
     \draw[thick,domain=5:8] plot(\x,{(\x^2-16*\x+64)/3});
     \draw[thick,dotted] (3,5)--(3,0) node [below] {$\mu^*$};
     \draw[thick,dotted] (1,1)--(1,0) node [below] {$\bar{\mu}_0$};
    \draw[thick,dotted] (4.12702,4.4)--(4.12702,0); 
     \draw[very thick,domain=0:2.23607,dashed] plot(\x,{2*(3-sqrt(5))*\x+1});
     \node[circle,fill=black,inner sep=0pt,minimum size=3.5pt] (a) at (2.23607,4.41641) {}; 
     \node at (2.23607,4.41641) [right] {$A$};
     \draw[thick,dotted] (2.23607,4.5)--(2.23607,0)
        node [below] at (2.23607,-.07) {$\mu_L$}
        node [left] at (0,1) {$Z$}
        node [below] at (4.12702,-.07) {$\mu_H$}
        node [left] at (0,0) {$0$};
    \fill[black, opacity=.3, variable=\x] (0,1) -- (2.23607,4.41641) -- plot[domain=2.23607:3](\x,{-(\x-3)^2+5}) -- plot[domain=3:4.12702](\x,{1/2*(1+6*\x-\x^2)}) --(8,0) -- cycle;
    \draw[very thick,domain=4.12702:8,dashed] plot(\x,{-1.1270166537925839*\x+9.016133230340671});
    \node[circle,fill=black,inner sep=0pt,minimum size=3.5pt] (a) at (4.12702,4.36492) {}; 
    \node at (4.12702,4.36492) [right] {$B$};
    \node [below] at(8,0) {$1$};
    \end{tikzpicture}  
\caption{The optimal static disclosure.
The solid curve represents the principal's payoff function $W_{\text{NI}}(\mu)$, and the shaded area is the convex hull of its graph.
The upper boundary of the shaded region is the concave envelope of $W_{\text{NI}}$.}
    \label{fig.concavification}
 \end{figure}

\section{Optimal Dynamic Persuasion}
\label{section.main}

We now turn to the full dynamic persuasion problem.
We first state the main theorem and its comparative statics, and then we develop a proof sketch that explains why the optimal policy takes this form.

\subsection{The Optimal Policy}


\begin{theorem} \label{thetheorem}
There exists an optimal policy $\mathcal{P}=\langle F_H^*,F_L^*\rangle$ under which information is disclosed at most twice.
If $\mu_0\geq\mu^*$, an optimal policy can be implemented by the optimal static policy characterized in Proposition~\ref{proposition.static}.
If $\mu_0<\mu^*$, there exist probabilities $x_a,x_b\in[0,1]$ and dates $t_b,t_a\geq0$ satisfying $\tau(\mu_0)\leq t_b<t^*\leq t_a\leq\tau(1)$ such that
 \begin{align*}
  F_H^*(t)=\left\{
  \begin{array}{cc}
   0 & t<t_a\\
   x_a & t_a\leq t<\tau(1)\\
   1 & \tau(1)\leq t
  \end{array}
  \right.
  \qquad
  F_L^*(t)=\left\{
  \begin{array}{cc}
   0 & t<t_b\\
   x_b & t_b\leq t<t_a\\
   1 & t_a\leq t
  \end{array}
\right..
\end{align*}
For every $\mu_0<\mu^*$, the participation constraint is binding, and hence
 \begin{equation*}
  \mu_0\bigg(x_av_H(t_a)+(1-x_a)v_H(\tau(1))\bigg)+(1-\mu_0)\bigg(x_bv_L(t_b)+(1-x_b)v_L(t_a)\bigg)=V_{\text{NI}}(\mu_0),
\end{equation*}
where $V_{\text{NI}}(\mu_0)$ is the agent's optimal payoff without disclosure.
Moreover, the agent's stopping constraint at date $t_a$ is binding, which yields
\begin{equation}
\label{eq.constraint}
 x_a=-\frac{1-\mu_0}{\mu_0}\frac{v'_L(t_a)}{v'_H(t_a)}(1-x_b).
\end{equation}
The optimal policy is unique in the sense that every optimal implementation induces the same pair of state-contingent stopping-time distributions $\langle F_H^*,F_L^*\rangle$.
\end{theorem}

As illustrated in Figure \ref{fig-structure}, the policy is implemented as follows:
\begin{itemize}
    \item At time $t_b$, the principal recommends stopping with probability $x_b$ conditional on the low state, and with probability $0$ in the high state; without such a recommendation, the agent's posterior belief updates upward, inducing him to continue experimentation.

\item At time $t_a>t^*$, among projects still under experimentation, the principal recommends stopping with probability $x_a$ in the high state and with probability one in the low state.
If no stopping recommendation is received, the agent's posterior rises to $1$ and he experiments until $\tau(1)$.
\end{itemize}
\begin{figure}[htb!]
     \centering
     \begin{tikzpicture}
     \draw[thick,->](0,0)--(6.6,0) node [below] {$t$};
     \draw[thick,->](0,0)--(0,4.8) node [left] {$F(t)$};
     \draw node [below] at (0,0) {$0$};
     \draw node [left, blue] at (1.4,0.75) {$F_L(t)$};
     \draw node [right, red] at (4.2,0.75) {$F_H(t)$};
     \draw[thick, dashed, -](5.6,4)--(0,4) node [left] {$1$};
     \draw[thick, dashed, -](5.6,1.5)--(0,1.5) node [left] {$x_b$};
     \draw[thick, dashed, -](5.6,2.5)--(0,2.5) node [left] {$x_a$};
     \draw[very thick, -, blue] (0,-0.02)--(1.4,-0.02);
     \draw[very thick, -, blue] (1.4,1.5)--(4.2,1.5);
     \draw[very thick, -, blue] (4.2,4)--(5.6,4);
     \draw[very thick, -, dashed, blue] (1.4,0)--(1.4,1.5);
     \draw[very thick, -, dashed, blue] (4.2,1.5)--(4.2,4);
     \node[circle,fill=black,inner sep=0pt,minimum size=3pt,black] at (5.6,4) {};
     \draw[very thick, -, red] (0,0.02)--(4.2,0.02);
     \draw[very thick, -, red] (4.2,2.5)--(5.6,2.5);
     \draw[thick, dashed, -](4.2,2.5)--(4.2,4);
     \draw[thick, dashed, -](5.6,2.5)--(5.6,0) node [below] {$\tau(1)$};
     \draw[thick, dashed, -, red](5.6,2.5)--(5.6,4);
     \draw[thick, dashed, -](1.4,4)--(1.4,0) node [below] {$t_b$};
     \draw[thick, dashed, -](2.8,4)--(2.8,0) node [below] {$t^*$};
     \draw[thick, dashed, -, red](4.2,2.5)--(4.2,0) node [below, black] {$t_a$};
    \end{tikzpicture}
\caption{\textup{The structure of the optimal dynamic information policy.
When $\mu_0\geq\mu^*$, $x_b=0$ and the accelerator is absent.}}
    \label{fig-structure}
 \end{figure}

Intuitively, these two disclosures act as an \textit{accelerator} and a \textit{brake}. 
The disclosure at $t_b$ acts as an accelerator by filtering out low-quality projects, while the absence of a bad-news stopping recommendation boosts the agent's confidence to reach $t^*$. 
Conversely, the disclosure at $t_a$ acts as a brake by verifying high-quality projects, and the absence of verification depresses the agent's belief by pooling all remaining low states with a fraction of high states, inducing him to stop.

The optimal state-contingent stopping-time distributions are unique, and hence whenever the accelerator or brake is used, its timing and probability are uniquely determined.
Any remaining multiplicity arises only from alternative message implementations that leave stopping behavior unchanged in every state.

The following proposition reveals how the structure of the optimal policy varies with the agent's initial belief $\mu_0$.

\begin{proposition}
\label{prop.comparative}
There exist $\mu_l$ and $\mu_h$ such that $\mu_l<\mu_h<\mu^*$:
\begin{enumerate} 
  \item[(i.)] If $\mu_0<\mu_l$, the optimal policy is full disclosure at time $t_b$ (i.e., $x_b=1$), and $t_b$ increases with $\mu_0$.
\item[(ii.)] If $\mu_0\in[\mu_l,\mu_h)$, we have $x_a,x_b\in(0,1)$, where $x_b$ and $x_a$ are decreasing and increasing with $\mu_0$, respectively.
Also, $t_a$ and $t_b$ in the optimal policy are invariant to changes in $\mu_0$.
  \item[(iii.)] If $\mu_0\in[\mu_h,\mu^*)$, the optimal policy is no disclosure after time $t^*$ (i.e., $x_a=1$).
  Also, $x_b$ and $t_a$ decrease with $\mu_0$, while $t_b$ increases with $\mu_0$.
\end{enumerate}
\end{proposition}

Proposition \ref{prop.comparative} connects the optimal policy to the difficulty of motivation, measured by the agent's initial pessimism (i.e., a lower $\mu_0$). 
Figure \ref{fig-comparative} illustrates these comparative statics by plotting the disclosure probabilities and dates against $\mu_0$.
As the agent becomes more pessimistic (moving from right to left in Figure \ref{fig-comparative}), the principal shifts her strategy through three distinct regions.
As established in Theorem~\ref{thetheorem}, when the agent is sufficiently optimistic ($\mu_0 \ge \mu^*$), motivating the agent to experiment until $t^*$ is effectively costless, and the probability of early stopping is zero ($x_b = 0$).
In this regime, the principal focuses entirely on the brake—dissuading the agent from experimenting beyond $t^*$—which can be implemented via the optimal static disclosure identified in Proposition \ref{proposition.static}.

\begin{figure}[ht!]
     \centering
     \begin{subfigure}[ht]{.48\textwidth}
         \centering
         \begin{tikzpicture}
     \draw[thick,->](0,0)--(5.5,0) node [below] {$\mu_0$};
     \draw[thick,->](0,0)--(0,4.5) node [left] {$x$}
        node [below] at (0,0) {$0$}
        node [left] at (0,4) {$1$}
        node [below] at (5,0) {$1$};
     \draw[thick, dashed, -](0,4)--(5,4);

     \draw[very thick, -, blue] (0,4)--(1,4);
     \draw[domain=1:3,very thick,black] plot(\x,{(\x-3)^2});
     \draw[very thick, -, blue] (3,.01)--(5,.01)
      node [above] at (0.5,4) {$x_b$};
     
     \draw[very thick, -, brown] (2,4)--(4,4) 
      node [above] at (3,4) {$x_a$};
     \draw[domain=1:2,very thick, brown] plot(\x,{4*(\x-1)^2});
     \draw[domain=4:5,very thick, brown] plot(\x,{4*(\x-5)^2});
     \draw[very thick, -, brown] (0,.01)--(1,.01);

     \draw[thick, dashed, -](1,4)--(1,0) node [below] {$\mu_l$};
     \draw[thick, dashed, -](2,4)--(2,0) node [below] {$\mu_h$};
     \draw[thick, dashed, -](3,4)--(3,0) node [below] at (3,0.05) {$\mu^*$};
     \draw[thick, dashed, -](4,4)--(4,0) node [below] {$\mu_H$};
     \draw[thick, dashed, -](5,0)--(5,4);
    \end{tikzpicture}  
     \end{subfigure}
     \begin{subfigure}[ht]{.48\textwidth}
         \centering
      \begin{tikzpicture}
     \draw[thick,->](0,0)--(5.5,0) node [below] {$\mu_0$};
     \draw[thick,->](0,0)--(0,4.5) node [left] {$t$};
     \draw[thick, dashed, -](5,2.88)--(0,2.88)
        node [below] at (0,0) {$0$}
        node [left] at (0,2.88) {$t^*$}
        node [above] at (1.5,.7) {$\tau(\mu_0)$}
        node [left] at (0,4) {$\tau(1)$}
        node [below] at (5,0) {$1$};
     \draw[thick, dashed, -](0,4)--(5,4);
     \draw[domain=0:5,very thick, dashed, -] plot(\x,{-0.08*\x^2+1.2*\x});     
     
     \draw[very thick, -, red] (1,3.6)--(2,3.6);
     \draw[very thick, -, red,dashed] (1,3.6)--(0,3.6);
     \draw[domain=2:3,very thick, red] plot(\x,{0.72*(\x-3)^2+2.88});
     \draw[domain=3:4,very thick, red, -] plot(\x,{-0.08*\x^2+1.2*\x});
     \draw[very thick, -, red] (4,3.52)--(5,3.52);
     \draw[black] node at (3.5,3.55) {$t_a$};

     \draw[domain=0:1,very thick, orange] plot(\x,{-1.8*(\x-1)^2+2.3});
     \draw[very thick, -, orange] (1,2.3)--(2,2.3);
     \draw[very thick, -, orange, dashed] (1,2.3)--(0,2.3);
     \draw[domain=2:3,very thick, orange] plot(\x,{-0.58*(\x-3)^2+2.88});
     \draw[domain=3:5,very thick, orange, dashed] plot(\x,{2.88});
     \draw[black] node [above] at (1.5,2.3) {$t_b$};

     \draw[thick, dashed, -](1,4)--(1,0) node [below] {$\mu_l$};
     \draw[thick, dashed, -](2,4)--(2,0) node [below] {$\mu_h$};
     \draw[thick, dashed, -](3,4)--(3,0) node [below] at (3,0.05) {$\mu^*$};
     \draw[thick, dashed, -](4,4)--(4,0) node [below] {$\mu_H$};
     \draw[thick, dashed, -](5,0)--(5,4);
    \end{tikzpicture}  
     \end{subfigure}
\caption{\textup{Comparative statics.
Panel (a) shows the disclosure probabilities ($x_a, x_b$).
Panel (b) shows the disclosure times ($t_a, t_b$).
The vertical dashed lines indicate the belief thresholds identified in Proposition \ref{prop.comparative}.}}
    \label{fig-comparative}
 \end{figure}

The accelerator appears once $\mu_0$ falls below $\mu^*$.
As is illustrated in Figure \ref{fig-comparative}, once $\mu_0$ falls below $\mu^*$, the principal provides additional motivation through three adjustments.
She raises $x_b$, moves $t_b$ forward, and postpones $t_a$, while $x_a$ remains at one.
A higher $x_b$ makes the absence of a stopping recommendation more favorable news, while an earlier $t_b$ delivers this confidence boost sooner.
Also, postponing $t_a$ extends the experimentation horizon promised to an agent who receives no earlier recommendation.

When $\mu_0$ falls below $\mu_h$, these adjustments no longer provide sufficient motivation while maintaining $x_a=1$, and $x_a$ becomes interior.
Over $[\mu_l,\mu_h)$, $t_a$ and $t_b$ remain fixed, while $x_b$ rises and $x_a$ falls as the agent becomes more pessimistic.
That is, as $\mu_0$ falls, both the accelerator and brake signals become more informative, strengthening the agent's incentive to continue experimenting.

At the lower boundary, the two probability instruments are exhausted.
The state is then fully disclosed at $t_b$, leaving no low-state stopping probability for later disclosure.
As $\mu_0$ falls further, the only remaining adjustment is to move $t_b$ forward.

\subsection{Decomposing the Non-Monotonic Problem}

The primary technical challenge in our analysis is the principal's non-monotonic preference over the agent's stopping time.
Unlike standard persuasion problems that globally maximize or minimize effort, our principal desires experimentation up to $t^*$ but discourages it thereafter.
This suggests decomposing the global optimization problem at the principal's ideal stopping time $t^*$, because her payoff is monotonic on either side of this date.

However, the principal cannot design either the motivation policy or the dissuasion policy in isolation, because each policy also affects the other subproblem.
Designing the motivation policy solely to encourage experimentation may excessively boost the agent's posterior belief at $t^*$, making him more reluctant to stop and subsequent dissuasion more costly.
Conversely, a policy designed solely to induce stopping may disclose too little information, and the resulting continuation value at $t^*$ may be insufficient to motivate the agent before $t^*$.

Our strategy is therefore to solve the two subproblems separately for fixed values of the variables that connect them, and then choose these values to maximize the principal's total payoff.
Specifically, for any information policy $\mathcal{P}=\langle F_H,F_L\rangle$, let $\pi_\theta=1-F_\theta(t^*)$ be the probability that the agent continues to experiment with the risky arm at $t^*$ in state $\theta$.
These probabilities determine the interim belief $\mu_{t^*}$.
We also fix $u$, the continuation value promised to the agent if he reaches $t^*$, which helps motivate experimentation before $t^*$.
Taking $t^*$ as the milestone, we refer to $\langle\pi_H,\pi_L,u\rangle$ as the three \emph{milestone} variables.

To formulate the two subproblems, we express each policy in terms of its conditional stopping-time distributions on either side of $t^*$.
Given $\pi_\theta$, any distribution of stopping time $F_\theta$ can be decomposed into
\begin{align*}
    F_\theta^b(t)=\left\{
  \begin{array}{cc}
  \frac{F_\theta(t)}{1-\pi_\theta} & t\leq t^*\\
  1 & t>t^*
  \end{array}
  \right.
  \qquad  \qquad
  F_\theta^a(t)=\left\{
  \begin{array}{cc}
  0 & t<t^*\\
  \frac{F_\theta(t)-(1-\pi_\theta)}{\pi_\theta} & t\geq t^*
  \end{array}
  \right.
\end{align*}
where $F_\theta^b$ and $F_\theta^a$ are the respective conditional persuasion policies \emph{before} and \emph{after} $t^*$.
For fixed milestone variables, we show that we can then decompose the entire persuasion problem into the following two subproblems, each with a monotonic objective:
\begin{itemize}
    \item The \emph{motivation subproblem} chooses $\langle F_H^b,F_L^b\rangle$ to maximize the principal's payoff before her ideal stopping time $t^*$, subject to the specified continuation probabilities $\pi_H$ and $\pi_L$ and the agent's incentive constraints given the promised continuation payoff $u$.
    \item The \emph{dissuasion subproblem} chooses $\langle F_H^a,F_L^a\rangle$ to maximize the principal's payoff after her ideal stopping time $t^*$, subject to the promised continuation value $u$ at $t^*$.
\end{itemize}
That is, if policy $\langle F_H,F_L\rangle$ is optimal, then $\langle F_H^b,F_L^b\rangle$ and $\langle F_H^a,F_L^a\rangle$ must be the solutions to the two preceding subproblems.

For given milestone variables, the two subproblems determine the optimal motivation and dissuasion policies.
Optimizing over these variables then balances the cost of sustaining experimentation before $t^*$ against the payoff from inducing stopping thereafter.
The next two subsections characterize the accelerator and the brake, respectively, the combination of which forms the optimal policy in Theorem~\ref{thetheorem}.


\subsection{The Accelerator: Persuasion as Risk Sharing}


In the motivation subproblem, conditional on the high state ($\theta=H$), the optimal policy never recommends stopping before $t^*$ since continued experimentation is mutually beneficial.\footnote{In a problem without a deadline, this conclusion is immediate because continued experimentation in the high state is optimal for both parties.
Here it applies only within the motivation subproblem, which ends at $t^*$.
The reduction must therefore also ensure that the agent is willing to stop if he receives a stopping recommendation at $t^*$, and the proof shows that the same reduction remains feasible despite this additional stopping constraint.}
The remaining choice is the conditional low-state stopping-time distribution $F\equiv F_L^b$, and the reduced problem is
\begin{equation}
\label{eq.reduction}
 \begin{aligned}
 \max_{F}:\,&\int_0^{\infty}w(t)dF(t)\\
  &\text{subject to: }\int_t^{\infty}\bar{v}_L(t,s)dF(s)\geq C(t)\quad\forall t\in[\tau(\mu_0),t^*),
 \end{aligned}
\end{equation}
where $C(t)$ is the continuation requirement induced by the milestone variables fixed in the decomposition, with its explicit expression derived in the proof of Theorem \ref{thetheorem}.

Problem (\ref{eq.reduction}) admits a contract-theoretic interpretation.
The principal makes a take-it-or-leave-it offer of a stopping-time lottery subject to a continuum of the agent's sequential participation constraints.
The optimal policy therefore lies on the Pareto frontier of feasible stopping-time lotteries for the two parties.
Observe that the two parties exhibit different preferences over two distinct dimensions---the \emph{mean} of the stopping time and their \textit{attitudes toward stopping-time risk}. 
On the one hand, the principal's payoff function $w(\cdot)$ increases for $t\in[0,t^*)$, while the agent's payoff function $v_L(\cdot)$ decreases over the interval $t\in[\tau(\mu_0),t^*)$.\footnote{By the decomposition, it is optimal to disclose no information before time $\tau(\mu_0)$.}
On the other hand, the principal is risk-averse regarding the stopping time since $w(\cdot)$ is concave for $t\in[0,t^*)$; meanwhile, the agent's payoff function $v_L(\cdot)$ is initially concave but subsequently convex, implying his risk-seeking behavior when the average stopping time is relatively later.\footnote{When $\tau(\mu_0)$ is sufficiently small, there is also a concave portion of $v_L$ in the relevant interval. However, both parties prefer contraction on this portion, and thus it does not affect the conclusion.}

The optimal contract balances these two dimensions.
If the principal is more sensitive to stopping-time risk than the agent, she prefers to remove risk and compensate the agent with an earlier average stopping time; this gives a one-shot disclosure.
If the agent is more sensitive, the principal can use a riskier stopping-time lottery to buy a later average stopping time; this gives gradual disclosure when the relevant constraints bind.
Lemma \ref{lemma.simplified} shows that this sensitivity is captured by the Arrow--Pratt coefficient $R(f,t)\equiv -f''(t)/f'(t)$, computed from the players' induced payoff functions over stopping time.

\begin{lemma}
\label{lemma.simplified}
 Let $R(w,t)$ and $R(v_L,t)$ denote the Arrow--Pratt coefficients of (absolute) risk aversion for the two players, respectively.
Let function $F^*$ be the solution to problem (\ref{eq.reduction}).
Then:
  \begin{enumerate}
    \item[(i)] Suppose $R(w,t)>R(v_L,t)$ for all $t\in[\underline{t},\bar{t}]\subset[\tau(\mu_0),t^*]$. 
    Then $F^*$ must be one-shot in $[\underline{t},\bar{t}]$; that is, there exists $\hat{t}\in[\underline{t},\bar{t}]$ such that $dF^*(t)=0$ for all $t\in[\underline{t},\bar{t}]\setminus\{\hat{t}\}$.
   \item[(ii)] Suppose $R(w,t)<R(v_L,t)$ for all $t\in[\underline{t},\bar{t}]\subset[\tau(\mu_0),t^*]$. 
   Then if there exists $t\in(\underline{t},\bar{t})$ such that $dF^*(t)>0$, the incentive constraint at time $t$ must be binding, i.e., $\int_{t}^\infty\bar{v}_L(t,s)dF^*(s)=C(t)$.
  \end{enumerate}
\end{lemma}

Evaluating these Arrow--Pratt coefficients confirms that the principal is uniformly more sensitive to stopping-time risk than the agent prior to $t^*$, which, by Lemma \ref{lemma.simplified}(i), guarantees that the optimal disclosure before $t^*$ is one-shot.

Lemma \ref{lemma.simplified} provides a local criterion for whether stopping-time risk should be concentrated or spread out.
When $R(w,t)>R(v_L,t)$, the principal can replace a nondegenerate stopping-time lottery with its certainty-equivalent date under the agent's payoff function, thereby leaving the agent's continuation value unchanged while reducing stopping-time risk.
Because the agent dislikes this reduction in risk, the principal compensates him with an earlier average stopping time.
Being more sensitive to stopping-time risk than the agent, the principal gains more from removing this risk than she loses from the earlier stopping date.
The contraction therefore raises her payoff, and hence optimal disclosure is one-shot on the relevant interval.

When $R(w,t)<R(v_L,t)$, a profitable deviation instead spreads stopping-time risk.
Suppose the continuation constraint is slack at a disclosure date $t$.
The principal can move a small amount of stopping probability at $t$ to two nearby dates, one earlier and one later, such that the agent's continuation value from this stopping probability remains unchanged.
Because the original constraint is slack, the perturbation can be made small enough that all nearby continuation constraints remain satisfied.
The agent is indifferent between the original stopping probability and its spread, while the principal benefits because she is less sensitive to stopping-time risk.
An optimal policy therefore cannot leave positive continuation rent at a disclosure date, and thus the continuation constraint must bind whenever disclosure occurs.
If one can further show that these binding disclosure dates fill an interval without gaps, disclosure is gradual on that interval.

In the closest related models, the global Arrow--Pratt rankings are determined entirely by differences in time preferences \citep{ely2020moving,liu2023motivating,koh2024attention,saeedi2025getting}.
Observe
\[
R(v_L,t)-R(w,t)
=
\frac{d}{dt}\log\left(\frac{w'(t)}{-v_L'(t)}\right).
\]
The ratio $w'(t)/[-v_L'(t)]$ measures the principal's marginal gain from delaying stopping relative to the agent's marginal loss.
In the cases where the Arrow--Pratt coefficients can be globally ranked, the instantaneous payoff from continued engagement is constant.\footnote{For example, in Liu's model, conditional on state $L$, the task is completed at a constant rate $\lambda_L$.
Until completion, the principal earns an expected reward $\lambda_L b_L$ per unit of time, while the agent incurs an expected net cost $z-\lambda_L y_L>0$.
Writing the players' discount rates as $r_P$ and $r_A$, their marginal payoffs from delaying stopping are
\[
w'(t)=\lambda_L b_L e^{-(\lambda_L+r_P)t},
\qquad
-v_L'(t)=(z-\lambda_L y_L)e^{-(\lambda_L+r_A)t}.
\]
The common probability $e^{-\lambda_Lt}$ that the task remains unfinished cancels in the ratio, giving
\[
\frac{w'(t)}{-v_L'(t)}
=
\frac{\lambda_L b_L}{z-\lambda_L y_L}e^{(r_A-r_P)t}
\quad\Longrightarrow\quad
R(v_L,t)-R(w,t)=r_A-r_P.
\]
}
Therefore, differences in discounting determine how the principal's marginal gain from delaying stopping changes relative to the agent's marginal loss over time.

In our model, the instantaneous payoffs from continued experimentation vary over time.
Even conditional on project quality, feasibility remains uncertain and is learned through experimentation.
As experimentation continues without a breakthrough, the posterior probability of feasibility $p_t$ declines, and the principal's marginal gain from delaying stopping relative to the agent's marginal loss in the low state is
\[
\frac{w'(t)}{-v_L'(t)}
=
\frac{\lambda p_t(Y-Z)-rZ}{rz-\lambda p_t(y_L-z)}.
\]
Here, the declining probability of feasibility changes the expected gains from a breakthrough relative to the fixed opportunity costs of forgoing the safe alternative.
The resulting change in the marginal-payoff ratio depends on how each player values these two outcomes:
\[
\frac{d}{dt}\log\left(\frac{w'(t)}{-v_L'(t)}\right)
=
\frac{\lambda r(zY-Zy_L)\dot p_t}
{\bigl[\lambda p_t(Y-Z)-rZ\bigr]\bigl[rz-\lambda p_t(y_L-z)\bigr]}.
\]
Because $\dot p_t<0$, the direction of this change is determined by the comparison between $Y/Z$ and $y_L/z$.
Under our assumption $Y/Z>y_L/z$, the ratio declines over time, yielding $R(w,t)>R(v_L,t)$.
Thus, differences in the players' valuations generate different attitudes toward stopping-time risk even under common discounting.

The Arrow--Pratt argument extends beyond the present model.
Appendix \ref{appendix.general} develops the approach for dynamic information-design problems in which disclosure determines stopping times across several states.
The approach is especially useful when the relevant Arrow--Pratt ranking is local or changes over time, because spreading and contraction can be studied interval by interval.
However, applying the approach requires specific structural prerequisites.
For this comparison to apply consistently, we first need the agent's Arrow--Pratt coefficient over stopping times to be invariant across dates.
Even when this prerequisite holds, however, it is more demanding to implement contraction because  collapsing stopping probability from an entire interval to a single date creates two no-disclosure gaps that can disrupt dynamic incentives elsewhere.
Appendix~\ref{appendix.general} identifies the exact conditions under which both spreading and contraction remain valid in general multi-state environments.

\begin{remark}
Appendix \ref{sec:applications} applies this framework to \citet{liu2023motivating}, \citet{saeedi2025getting}, and \citet{koh2024attention}.
These applications illustrate both the reach and the boundary of the approach.
The Arrow--Pratt comparison sharply simplifies several results, including settings in which the Arrow--Pratt ranking varies across dates. 
The only exception we know of is the heterogeneous-prior setting in \citet{saeedi2025getting}, where payoffs cannot be scaled uniformly across states. 
A joint certainty-equivalent date across states is therefore unavailable, and the setting falls outside this method.
\end{remark}

\subsection{The Brake: Dissuasion via Constrained Static Persuasion}
\label{ssec:brake}

In the dissuasion problem, the following lemma shows that the optimal policy can always be implemented by a static policy, under which all information scheduled after $t^*$ is disclosed through a single signal at $t^*$.

\begin{lemma}
  \label{lem.static.reduction}
For every globally optimal policy, there exists an equally optimal policy under which all information scheduled after $t^*$ is disclosed through a single static signal at $t^*$.
Conditional on each posterior generated by this signal, no further information about quality is disclosed.
\end{lemma}

The intuition behind Lemma \ref{lem.static.reduction} is that any disclosure scheduled strictly after $t^*$ can be folded back into a static disclosure at $t^*$ by reducing compound lotteries. 
The agent weakly benefits because the new disclosure policy provides more timely information.
The principal also weakly benefits because any disclosure scheduled after $t^*$ would give the agent an additional incentive to postpone stopping, a delay from which the principal derives no benefit.

Consequently, we can model the optimal information policy after $t^*$ as a static signal sent at $t^*$, which induces a distribution of posterior beliefs over the state, and the dissuasion subproblem can therefore be written as the following constrained static persuasion problem:
\begin{equation}
\label{eq.kg}
 \begin{aligned}
  \max_{P\in\Delta([0,1])}:\,\int_0^1W_{\text{NI}}(t^*,&\,\mu)P(d\mu)\\
  \text{subject to: }&\int_0^1\mu P(d\mu)=\mu_{t^*}\quad\quad\quad\quad\quad\quad\quad\quad\quad\text{(BP)}\\
  &\int_0^1V_\text{NI}(t^*,\mu) P(d\mu)=u.\,\quad\quad\quad\quad\quad\quad\text{(U)}
 \end{aligned}    
\end{equation}
Here, conditional on reaching $t^*$ with posterior belief $\mu$, $V_\text{NI}(t^*,\mu)$ and $W_\text{NI}(t^*,\mu)$ are the agent's and principal's respective indirect payoffs. 
Constraint (BP) is the standard Bayes plausibility condition, and constraint (U), inherited from our decomposition, ensures that the policy delivers the promised continuation value $u$ to the agent.

Problem (\ref{eq.kg}) is a static Bayesian persuasion problem with an additional linear constraint, and this type of problem has been extensively studied in the recent literature \citep{le2019persuasion,boleslavsky2020bayesian,doval2024constrained}.
The problem can be solved by concavifying the Lagrangian \cite[Theorem 3.2]{doval2024constrained}. 
However, as is widely observed in the literature, a generic constrained persuasion problem with binary states could have a ternary signal because the multiplier on constraint (U) is endogenous.
Fortunately, for the payoff functions defined above, there are only two possible patterns of the Lagrangian, and their concavification is supported by at most two posterior beliefs.
In fact, the only candidates for the optimal disclosure are full disclosure or a perfect-good-news signal, and Step~8 of the proof of Theorem~\ref{thetheorem} further rules out full disclosure once the dissuasion and motivation policies are combined.
An active brake therefore takes the form of a nondegenerate perfect-good-news signal.

\bigskip
Taken together, the two subproblems explain the structure of the optimal policy in Theorem~\ref{thetheorem}: the comparison of attitudes toward stopping-time risk yields a one-shot accelerator, while constrained static persuasion yields a one-shot brake.
Optimizing over the milestone variables coordinates these two disclosures and determines their timing and probabilities.

\section{Misaligned Time Preferences}
\label{section.misaligned}

The baseline model keeps discount rates common, and thus the ranking of sensitivity to stopping-time risk comes only from payoff differences across the risky and safe arms. 
We now allow the agent and the principal to discount time differently. 
Denote by $r_P$ and $r_A$ the discount rates of the principal and the agent, respectively.
In this setting, the two parties' conflict of interest is jointly determined by the relative valuations of the project and their divergent discount rates.
Without information, the agent prefers to experiment longer than the principal, if and only if
\begin{align*}
 \frac{y(\mu)}{z}\geq\frac{r_A}{r_P}\frac{Y}{Z}+\left(1-\frac{r_A}{r_P}\right).
\end{align*}
That is, a more patient player (i.e., one with a lower discount rate) naturally prefers a later stopping time, ceteris paribus.

\begin{theorem}
\label{proposition.gradual}
 Suppose $\mathcal{P}=\langle F_H,F_L\rangle$ is the optimal information policy.
 When $r_A$ is sufficiently large, there exists $[\underline{t}_g,\bar{t}_g]\subset[\tau(\mu_0),t^*)$ such that:
 \begin{enumerate}
\item[(i)] For all $t\in[\tau(\mu_0),t^*)$, $R(w,t)\leq R(v_L,t)$ if and only if $t\in[\underline{t}_g,\bar{t}_g]$.
 \item[(ii)] If nonempty, $\mathcal{S}(\mathcal{P})\cap[\underline{t}_g,\bar{t}_g]$ is an interval.
 \item[(iii)] Whenever $dF_L(t)>0$ for $t \in [\underline{t}_g,\bar{t}_g]$, the continuation constraint at time $t$ is binding, which indicates that 
      \begin{equation}
      \label{eq.closeform}
      F_L(t)=F_L^*(t)\equiv 1-\frac{\mu_0}{1-\mu_0}\left|\frac{v'_H(t)}{v'_L(t)}\right|.
      \end{equation}
 \item[(iv)] Each of the sets $\mathcal{S}(\mathcal{P})\cap[\tau(\mu_0),\underline{t}_g)$ and $\mathcal{S}(\mathcal{P})\cap(\bar{t}_g,t^*)$ contains at most one disclosure date.
 Hence the optimal accelerator has a one-shot--gradual--one-shot structure.
\end{enumerate}
\noindent Also, there exist $x_a\in[0,1]$ and $t_a\in[t^*,\tau(1)]$ such that, conditional on reaching $t^*$, the agent stops at $t_a$ with probability $x_a$ in state $H$ and with probability one in state $L$, while the remaining high-state probability stops at $\tau(1)$.
\end{theorem}

To our knowledge, this is the only model in the literature where the optimal policy can exhibit gradual disclosure on a strictly internal interval, with stopping probability both before and after that interval. 
This possibility rests on the conjunction of two features of the reduced motivation subproblem.
First, the ranking of the Arrow--Pratt coefficients can vary over time, and hence the principal may be more sensitive to stopping-time risk at early and late dates but less sensitive in between.
Second, $\bar v_L(t,t)$, the agent's date-$t$ net payoff from stopping immediately under state $L$, is $0$, and therefore reallocating stopping probability within a bounded interval does not make stopping more attractive at later dates.
The latter property is abstracted by Assumption~\ref{ass:continuation-order} in Appendix \ref{appendix.general}, but it fails in settings where the stopping payoff varies with the posterior belief upon stopping, such as in \citet{koh2024attention} and \citet*{saeedi2025getting}.

When $\tau(0)>0$, the Arrow--Pratt comparison under heterogeneous discounting separates into two components:
\[
R(v_L,t)-R(w,t)=r_A-r_P+\lambda\left(\frac{1}{1-e^{\lambda(t-\tau(0))}}-\frac{1}{1-e^{\lambda(t-t^*)}}\right).
\]
The term $r_A-r_P$ captures the direct effect of relative discounting, while the term in parentheses captures the remaining time variation and reflects the distance from the two payoff peaks.
This difference diverges to $-\infty$ as $t$ approaches either $\tau(0)$ or $t^*$, and attains an interior maximum in between.
Because greater impatience on the agent's side raises the additive term $r_A-r_P$, the ranking may reverse over an interior interval.

Where $R(v_L,t)<R(w,t)$, the principal remains more sensitive to stopping-time risk, and contraction implies that optimal disclosure is concentrated at no more than one date.
Where the ranking reverses to $R(v_L,t)>R(w,t)$, the principal's payoff is strictly convex with respect to the agent's continuation utility, and she strictly benefits from introducing dispersion into the stopping-time lottery.
Whenever a continuation constraint is slack at $t\in\mathcal{C}(\mathcal P)$, the principal can replace the deterministic recommendation with a mean-preserving spread of stopping dates, preserving the agent's expected payoff while strictly increasing her own.
To maximize the value of this risk dispersion, the principal pushes the spread until all continuation rents are exhausted, so that the continuation constraint must bind at every disclosure date.

We further show that these binding disclosure dates cannot be separated by a gap.
On a no-disclosure interval bounded by two binding dates, the agent's continuation surplus is zero at both endpoints.
This surplus is strictly convex over the interval and therefore negative at every interior date, violating the continuation constraint.
Hence the binding disclosure dates fill an interval, converting the no-rent conclusion into gradual disclosure.

Crucially, the trajectory $F_L^*(t)$ in \eqref{eq.closeform} satisfies $F_L^*(\tau(\mu_0))=0$ and $F_L^*(\tau(1))=1$, independent of the boundaries of the gradual phase.
Because the reversal interval $[\underline{s},\bar{s}]\subseteq[\underline{t}_g,\bar{t}_g]\subsetneq(\tau(\mu_0),t^*)$ is identified entirely by the local Arrow--Pratt ranking, this gradual schedule starts strictly after $\tau(\mu_0)$ and ends well before $t^*$.
Consequently, its cumulative stopping probability starts strictly above zero ($F_L^*(\underline{s})>0$) and typically falls short of the milestone stopping requirement $1-\pi_L$ at $\bar{s}$, necessitating a one-shot disclosure on both sides of the gradual phase.
The optimal policy is therefore characterized by the quadruple $\langle s_1, \underline{s}, \bar{s}, s_2 \rangle$ as illustrated in Figure \ref{fig-convex}:
it starts with a one-shot disclosure at $s_1$ to absorb the initial stopping probability $F_L^*(\underline{s})$; it is followed by the gradual phase over $[\underline{s}, \bar{s}]$ where information is disclosed continuously according to $F_L^*(t)$; and if the cumulative stopping probability at $\bar{s}$ falls short of $1-\pi_L$, a final discrete disclosure occurs at $s_2$.
The quadruple $\langle s_1,\underline{s},\bar{s},s_2\rangle$ serves a role parallel to that of the accelerator in Theorem \ref{thetheorem}.\footnote{As $r_A$ increases, the interval $[\underline{t}_g,\bar{t}_g]$ expands. In the limit as $r_A\to\infty$, it covers the whole motivation region $[\tau(\mu_0),t^*)$, the boundary mass points disappear, and the policy converges to the continuous disclosure rule in the \textit{beep problem} of \cite{ely2017beeps}.}

\begin{figure}[H]
     \centering
     \begin{tikzpicture}
     \draw[thick,->](0,0)--(11,0) node [below] {$t$};
     \draw[thick,->](0,0)--(0,6) node [left] {$v_L(t)$/$w(t)$};
     \draw[thick,->](10,0)--(10,6) node [right] {$F_L(t)$};
     \draw[domain=2:5,very thick, dashed] plot(\x,{5*(1-e^((\x-2)/4))/(1-e^2)});
     \draw[domain=5:7,very thick,black] plot(\x,{5*(1-e^((\x-2)/4))/(1-e^2)});
     \draw[domain=7:10,very thick,dashed] plot(\x,{5*(1-e^((\x-2)/4))/(1-e^2)});
     \draw[domain=0:2,thick] plot(\x,{-3*\x^2/40+9/2});
     \draw[domain=2:10,thick] plot(\x,{3*\x^2/160-3*\x/8+39/8})
        node [below] at (0,0) {$0$}
        node [left] at (0,4.5) {$v_L(0)$}
        node [above] at (6,3.3) {$v_L(t)$}
        node [below] at (10,0) {$t^*$}
        node [right] at (10,5) {$\pi_L$}
        node at (9.5,3.5) {$F_L^*(t)$};
     \draw[thick, dashed, -](0,5)--(10,5);
     \draw[thick, dashed, -](10,3)--(0,3) node [left] {$v_L(\tau(1))$};
     \draw[thick, dashed, -](0,1.94891)--(10,1.94891) node [right] {$F_L^*(\bar{s})$};
     \draw[thick, dashed, -](0,0.874151)--(10,0.874151) node [right] {$F_L^*(\underline{s})$};
     \draw[domain=0:10,very thick,densely dotted] plot(\x,{1.64229*ln(2*\x+1)}) node at (6,4.5) {$w(t)$};
     \draw[very thick, -, red] (0,0)--(3,0);
     \draw[very thick, -, red] (3,0.874151)--(5,0.874151);
     \draw[very thick, -, red] (7,1.94891)--(9,1.94891);
     \draw[very thick, -, red] (9,5)--(10,5);
     \draw[thick, dashed, -](1,5)--(1,0) node [below] {$\tau(0)$};
     \draw[thick, dashed, -](2,5)--(2,0) node [below] {$\tau(\mu_0)$};
     \draw[thick, dashed, -](3,5)--(3,0.874151) node [below] at (3,-0.1) {$s_1$};
     \draw[thick, dashed, -, red](3,0)--(3,0.874151);
     \draw[thick, dashed, -](4,5)--(4,0) node [below] {$\underline{t}_g$};
     \draw[thick, dashed, -](5,5)--(5,0) node [below] at (5,-.1) {$\underline{s}$};
     \draw[thick, dashed, -](7,5)--(7,0) node [below] at (7,-.1) {$\bar{s}$};
     \draw[thick, dashed, -] (8,5)--(8,0) node [below] {$\bar{t}_g$};
     \draw[thick, dashed, -] (9,0)--(9,1.94891) node [below] at (9,-0.1) {$s_2$};
     \draw[thick, dashed, -, red] (9,5)--(9,1.94891);
     \node[circle,fill=black,inner sep=0pt,minimum size=3.2pt] (a) at (10,5) {};
    \end{tikzpicture}  
    \caption{\textup{A representative CDF of the optimal information policy  with gradual disclosure before $t^*$ ($\underline{t}_g>\tau(\mu_0)$).}} %
    \label{fig-convex}
 \end{figure}

Finally, the optimal brake after $t^*$ remains binary under heterogeneous discounting, but for a fundamentally different reason than in Section~\ref{ssec:brake}.
When the agent is sufficiently impatient ($r_A>2(r_P+\lambda)$), the Lagrangian becomes convex--concave--convex, so that the continuation problem for fixed milestone variables genuinely admits a ternary signal supported on $\{0,\mu_a,1\}$.
What eliminates this ternary signal here is joint optimization across the milestone.
Under such a ternary signal, posterior $0$ stops low-quality projects immediately at $t^*$.
The principal can then strictly improve her payoff through a cross-milestone perturbation: she slightly postpones an earlier low-state disclosure date $s<t^*$ toward $t^*$, while compensating the agent by moving a fraction of the stopping probability at $t^*$ slightly earlier to $t^*-\delta$.
Because $w'(s)>0$ while $w'(t^*)=0$, the principal reaps a first-order gain from prolonging early experimentation while suffering only a vanishing second-order loss at the milestone.
Consequently, the brake robustly preserves its binary message structure.

\section{The Value of Dynamic Commitment}
\label{section.commitment}
We return to the common-discounting baseline and study the role of dynamic commitment.
We ask whether the optimal policy in Theorem~\ref{thetheorem} can be implemented by a sequentially optimal information policy.
Here, an information policy is \textit{sequentially optimal} if, after every history reached with positive probability, the principal cannot improve her payoff by replacing the policy's continuation with another feasible information policy that the agent is willing to follow.

This question is motivated by \cite{koh2024attention} and \cite{koh2024persuasion}, who show that in their settings, the optimal policy can be implemented without dynamic commitment.
That is, even though the optimal outcome distribution may not be achievable through a sequentially optimal direct-recommendation implementation, it can still be implemented by other sequentially optimal disclosure policies.\footnote{See \citet[Theorem 3]{koh2024attention} and \citet[Theorem 2]{koh2024persuasion}.}

The following proposition shows when such an implementation is possible in our setting and characterizes the sequentially optimal policy when it is not.
\begin{proposition}
\label{proposition.nocommitment}
Suppose the principal has no dynamic commitment power.
\begin{itemize}
  \item The optimal policy in Theorem \ref{thetheorem} (i.e. the optimal static policy in Proposition \ref{proposition.static}) is implementable when $\mu_{0}\geq\mu^*$, but cannot be implemented when $\mu_0<\mu^*$.

  \item When $\mu_0<\mu^*$, there exists a sequentially optimal information policy $\langle F_H,F_L\rangle$, such that
\begin{align*}
 F_H(t)=\left\{
 \begin{array}{cc}
  0 & t<t^*\\
  1 & t\geq t^*
 \end{array}
 \right.
  \qquad  \qquad
 F_L(t)=\left\{
 \begin{array}{cc}
 0 & t< \tau(\mu_0)\\
 1-\frac{\mu_0}{1-\mu_0}\left|\frac{v'_H(t)}{v'_L(t)}\right| & t\in[\tau(\mu_0),t^*)\\
 1 & t\geq t^*
 \end{array}
 \right..
\end{align*}
\end{itemize}
\end{proposition}

Thus, when the agent is initially optimistic ($\mu_0\geq\mu^*$), the optimal policy identified in Theorem \ref{thetheorem} (and Proposition \ref{proposition.static}) is implementable. 
This is because a sequentially optimal policy can be derived by backward induction, solving a local persuasion problem at each instant by concavifying the continuation value taking future disclosure as given. 
Crucially, the concavification results are time-invariant when the belief is weakly greater than $\mu^*$, ensuring that the static optimal disclosure remains dynamically consistent.

When the agent is initially pessimistic, the optimal policy with dynamic commitment power in Theorem \ref{thetheorem} involves a one-shot disclosure before $t^*$, and hence its direct implementation is not sequentially optimal.
This is because a one-shot disclosure necessarily creates an interval of silence with no information arrival. 
On such an interval, the agent's continuation constraint cannot remain continuously binding. 
Wherever the constraint is strictly slack, an uncommitted principal inevitably has an incentive to expropriate by revising her disclosure policy.

To understand why alternative implementations are also impossible, observe that in the settings of \citet{koh2024attention} and \citet{koh2024persuasion}, the agent's payoff is decreasing over time in every state, and the only motive for the agent to wait is the prospect of subsequent information.
Thus, the agent will quit regardless of the state once his belief becomes degenerate.
This assumption is crucial for the implementation of a one-shot optimal policy without dynamic commitment power.
Accordingly, they reconstruct the belief martingale through a recursive \emph{probability-tree surgery}, which partitions the information flow such that the agent remains indifferent between continuing and stopping at each decision node reached with positive probability.

However, in our setting, the agent's decision concerns effort exertion rather than waiting for information, and therefore the agent is willing to continue when $\theta=H$ is confirmed.
This instrumental value of effort ensures that the belief at each instant that makes the agent indifferent between continuing and stopping, by backward induction, is unique. 
It therefore precludes the use of probability-tree surgery to reconstruct a belief martingale in our persuasion environment.

Conversely, by the policy described in Proposition \ref{proposition.nocommitment}, when $\mu_0<\mu^*$, the agent stops no later than $t^*$ with probability $1$, which, as identified by Theorem \ref{thetheorem}, is suboptimal when the principal has dynamic commitment power.
This is because the principal's inability to commit to future disclosure forces her to provide the optimal incentives at present, which essentially extracts all of the agent's rent at each instant before termination.
According to this strategy, in each infinitesimal time interval for $t\in[\tau(\mu_0),t^*)$, a small probability mass $dF_L(t)$ corresponds to bad news, fully revealing that the state is low; if this piece of information is not received, the agent's belief grows continuously, keeping his incentive constraint binding.
Equation~(\ref{eq.closeform}) thus arises in both settings---no commitment and impatient agent---because both impose continuously binding incentive constraints.

\section{Conclusion}
\label{sec:con}

This paper studies dynamic persuasion in a bandit experimentation problem with a dead end. 
The principal wants the agent to continue until her ideal stopping time and then stop. 
The optimal policy has a simple structure: a one-shot bad-news accelerator before $t^*$ and a one-shot good-news brake after $t^*$.

Our main contributions are twofold.
The first contribution is to characterize the optimal policy in this non-monotonic persuasion problem.
The key difficulty is that the accelerator and the brake are jointly determined. 
Optimism generated before $t^*$ affects the cost of dissuasion after $t^*$, while the continuation value promised after $t^*$ feeds back into the agent's incentives before $t^*$. 
We resolve this interaction through a milestone decomposition: fixing the interim belief and promised continuation value at $t^*$ separates the problem into a motivation subproblem and a dissuasion subproblem. 
The dissuasion problem admits a static solution, and therefore can be written as a constrained static persuasion problem.

The second contribution is to use a local comparison of the players' Arrow--Pratt coefficients to characterize how stopping-time risk shapes disclosure.
An information policy determines both the average stopping time and the risk around it.
The relative Arrow--Pratt coefficients identify which player is more sensitive to that risk and therefore whether disclosure should be concentrated or spread out locally.
In the common-discounting baseline, this ranking is uniform throughout the motivation region, and hence the accelerator is one-shot.
When heterogeneous discounting reverses the ranking over an interior interval, the optimal accelerator can include a connected gradual-disclosure phase within that interval, with at most one discrete disclosure on either side.
The same pointwise comparison therefore explains both the baseline policy and the one-shot--gradual--one-shot structure, and it provides a unified analytical tool for organizing related results in dynamic persuasion.

\bibliographystyle{ecta} %
\bibliography{ref_bandit}

@article{keller2005strategic,
  title={Strategic experimentation with exponential bandits},
  author={Keller, Godfrey and Rady, Sven and Cripps, Martin},
  journal={Econometrica},
  volume={73},
  number={1},
  pages={39--68},
  year={2005},
  publisher={Wiley Online Library}
}

@article{kamenica2011bayesian,
  title={Bayesian persuasion},
  author={Kamenica, Emir and Gentzkow, Matthew},
  journal={American Economic Review},
  volume={101},
  number={6},
  pages={2590--2615},
  year={2011}
}

@article{ely2017beeps,
  title={Beeps},
  author={Ely, Jeffrey C},
  journal={American Economic Review},
  volume={107},
  number={1},
  pages={31--53},
  year={2017},
  publisher={American Economic Association 2014 Broadway, Suite 305, Nashville, TN 37203}
}

@article{le2019persuasion,
  title={Persuasion with limited communication capacity},
  author={Le Treust, Ma{\"e}l and Tomala, Tristan},
  journal={Journal of Economic Theory},
  volume={184},
  pages={104940},
  year={2019},
  publisher={Elsevier}
}

@article{bimpikis2019designing,
  title={Designing dynamic contests},
  author={Bimpikis, Kostas and Ehsani, Shayan and Mostagir, Mohamed},
  journal={Operations Research},
  volume={67},
  number={2},
  pages={339--356},
  year={2019},
  publisher={INFORMS}
}

@article{boleslavsky2020bayesian,
  title={Bayesian persuasion and moral hazard},
  author={Boleslavsky, Raphael and Kim, Kyungmin},
  journal={Working paper},
  year={2020}
}

@article{ely2020moving,
  title={Moving the goalposts},
  author={Ely, Jeffrey C and Szydlowski, Martin},
  journal={Journal of Political Economy},
  volume={128},
  number={2},
  pages={468--506},
  year={2020},
  publisher={The University of Chicago Press Chicago, IL}
}

@article{orlov2020persuading,
  title={Persuading the principal to wait},
  author={Orlov, Dmitry and Skrzypacz, Andrzej and Zryumov, Pavel},
  journal={Journal of Political Economy},
  volume={128},
  number={7},
  pages={2542--2578},
  year={2020},
  publisher={The University of Chicago Press Chicago, IL}
}

@article{escobar2021delegating,
  title={Delegating learning},
  author={Escobar, Juan F and Zhang, Qiaoxi},
  journal={Theoretical Economics},
  volume={16},
  number={2},
  pages={571--603},
  year={2021},
  publisher={Wiley Online Library}
}

@article{sadler2021dead,
  title={Dead ends},
  author={Sadler, Evan},
  journal={Journal of Economic Theory},
  volume={191},
  pages={105167},
  year={2021},
  publisher={Elsevier}
}

@article{ball2023dynamic,
  title={Dynamic information provision: Rewarding the past and guiding the future},
  author={Ball, Ian},
  journal={Econometrica},
  volume={91},
  number={4},
  pages={1363--1391},
  year={2023},
  publisher={Wiley Online Library}
}

@article{ely2023optimal,
  title={Optimal feedback in contests},
  author={Ely, Jeffrey C and Georgiadis, George and Khorasani, Sina and Rayo, Luis},
  journal={Review of Economic Studies},
  volume={90},
  number={5},
  pages={2370--2394},
  year={2023},
  publisher={Oxford University Press}
}

@article{liu2023motivating,
  title={Motivating effort with information about future rewards},
  author={Liu, Chang},
  journal={Working paper},
  year={2025},
  note={Version dated December 28, 2025, arXiv:2110.05643v4},
  url={https://arxiv.org/abs/2110.05643v4}
}

@article{doval2024constrained,
  title={Constrained information design},
  author={Doval, Laura and Skreta, Vasiliki},
  journal={Mathematics of Operations Research},
  volume={49},
  number={1},
  pages={78--106},
  year={2024},
  publisher={INFORMS}
}

@article{koh2024attention,
  title={Attention capture},
  author={Koh, Andrew and Sanguanmoo, Sivakorn},
  journal={Working paper},
  year={2024}
}

@article{koh2024persuasion,
  title={Persuasion and optimal stopping},
  author={Koh, Andrew and Sanguanmoo, Sivakorn and Zhong, Weijie},
  journal={Working paper},
  year={2024}
}

@article{saeedi2025getting,
  title={Getting the agent to wait},
  author={Saeedi, Maryam and Shen, Yikang and Shourideh, Ali},
  journal={Working paper},
  year={2025}
}

@article{HKL17,
  title={Contests for experimentation},
  author={Halac, Marina and Kartik, Navin and Liu, Qingmin},
  journal={Journal of Political Economy},
  volume={125},
  number={5},
  pages={1523--1569},
  year={2017},
  publisher={University of Chicago Press Chicago, IL}
}

@article{BH11,
  title={Collaborating},
  author={Bonatti, Alessandro and H{\"o}rner, Johannes},
  journal={American Economic Review},
  volume={101},
  number={2},
  pages={632--63},
  year={2011}
}

@article{guo2016dynamic,
  title={Dynamic delegation of experimentation},
  author={Guo, Yingni},
  journal={American Economic Review},
  volume={106},
  number={8},
  pages={1969--2008},
  year={2016},
  publisher={American Economic Association 2014 Broadway, Suite 305, Nashville, TN 37203}
}

@article{halac2016optimal,
  title={Optimal contracts for experimentation},
  author={Halac, Marina and Kartik, Navin and Liu, Qingmin},
  journal={The Review of Economic Studies},
  volume={83},
  number={3},
  pages={1040--1091},
  year={2016},
  publisher={Wiley-Blackwell}
}

@article{akcigit2016role,
  title={The role of information in innovation and competition},
  author={Akcigit, Ufuk and Liu, Qingmin},
  journal={Journal of the European Economic Association},
  volume={14},
  number={4},
  pages={828--870},
  year={2016},
  publisher={Oxford University Press}
}

\appendix
\part*{Appendix}
\section{The Arrow--Pratt Approach}
\label{appendix.general}

This appendix establishes conditions under which the Arrow--Pratt approach developed in the main text applies to a broader class of dynamic information-design problems.

We call the party controlling the information policy the principal and the party choosing when to stop the agent.
In this problem, the state is persistent.
Also, the principal's payoff is state-irrelevant and increasing in the stopping time.

Compared to the main text, the number of states is not necessarily binary.
Let $\Theta$ be a finite state space with prior $\mu_0\in\Delta(\Theta)$. 
A policy is a collection $F=\{F_\theta\}_{\theta\in\Theta}$, where $F_\theta(t)$ is the probability of a stopping recommendation by date $t$ in state $\theta$.  
Given that no stopping recommendation has arrived by time $t$, the agent's posterior belief is
\begin{equation*}
 \mu_t(\theta)
 =\frac{\mu_0(\theta)\bigl(1-F_\theta(t)\bigr)}
 {\sum_{\theta'\in\Theta}\mu_0(\theta')\bigl(1-F_{\theta'}(t)\bigr)}.
\end{equation*}

The problem is formally expressed as
\begin{equation}
\label{eq:multistate-problem}
\begin{aligned}
 \max_{\{F_\theta\}_{\theta\in\Theta}}\quad
 \sum_{\theta\in\Theta}\mu_0(\theta)&
   \int_0^\infty w(t)\,dF_\theta(t)\\
 \text{subject to}\quad
 &\sum_{\theta\in\Theta}\mu_0(\theta)
   \int_{[t,\infty)}
   \bigl(\hat v_\theta(t,s)-\psi(\mu_t)\bigr)\,dF_\theta(s)
   \geq C(t),\\
 &\mu_t=\Pi\left(\mu_0,
   \{F_\theta(s)\}_{\theta\in\Theta,\,s\in[0,t)}\right).
\end{aligned}
\end{equation}
Here, 
 \begin{itemize}
  \item $w(t)$ is the payoff of the principal if the agent stops at time $t$, which is increasing, continuous, and twice differentiable;
  \item $\hat{v}_\theta(t,s)$ is the payoff of the agent at date $t$ when he stops at time $s\geq t$, which is continuous and twice differentiable in both arguments, and $\hat{v}_\theta(t,\cdot)$ is decreasing;
  \item $\psi(\mu_t)$ is the payoff when the agent stops directly at time $t$ with belief $\mu_t$, which is continuous in beliefs;
  \item $C(t)$ is the requirement for the agent to continue at time $t$, which is continuous and twice differentiable;
 \item $\Pi$ specifies how belief $\mu_t$ evolves according to policy $F$, which is locally continuous in the information policy under weak convergence of the induced stopping-time distributions.
 \end{itemize}

Problem \eqref{eq:multistate-problem} generalizes the model in the main text and a series of problems in the literature, including \cite{liu2023motivating}, \cite{koh2024attention} and \cite{saeedi2025getting}.

\subsection{The Agent is More Sensitive to Stopping-Time Risk}

We begin with the case in which the agent is more sensitive to stopping-time risk, the curvature ranking that underlies gradual disclosure in the main text and in the related models discussed in Appendix \ref{sec:applications}.

\begin{A.assumption}
\label{ass:statewise-separable}
For every $\theta\in\Theta$, there exist a positive function $A_\theta$, a strictly decreasing, continuous, and smooth function $v$, and a function $B_\theta$ such that for all $t$ and $s$ with $t\leq s$,
\begin{equation*}
 \hat v_\theta(t,s)=A_\theta(t)v(s)+B_\theta(t).
\end{equation*}
\end{A.assumption}

Assumption \ref{ass:statewise-separable} imposes the defining feature of the Arrow--Pratt approach.
Under this assumption, the payoff functions entering the family of continuation constraints share a common Arrow--Pratt coefficient over the stopping date.
That is, the Arrow--Pratt coefficient of $\hat v_\theta(t,\cdot)$ equals $R(v,s)$ for every current date $t$ and state $\theta$.

\begin{A.lemma}
\label{lem:no-rent}
Suppose Assumption \ref{ass:statewise-separable} holds and $R(w,t)<R(v,t)$ for any $t\in[\underline{t},\bar{t}]$.
Under any optimal policy, if $t\in(\underline{t},\bar t)$ belongs to $\operatorname{supp}(dF_\theta)$ for some state $\theta$, the continuation constraint binds at $t$.
\end{A.lemma}

That is, if the agent is more sensitive to stopping-time risk, the principal cannot leave positive rent at any disclosure date. 
If the continuation constraint were slack at a disclosure date, the principal could spread a small amount of stopping probability to nearby dates while preserving the agent's continuation value and maintaining feasibility.
Since $R(w,t)<R(v,t)$, this perturbation would strictly raise the principal's payoff.

\begin{proof}
\medskip
\noindent\textit{Step 1. Construct a local spread.}

Take a disclosure date $t\in(\underline t,\bar t)\cap\operatorname{supp}(dF_\theta)$ for some state $\theta$, and suppose its continuation constraint is strictly slack.
Fix $\varepsilon>0$ small enough that $[t-\varepsilon,t+\varepsilon]\subset[\underline t,\bar t]$ and the continuation constraints have a common positive slack
there. 
Suppose first that $F_\theta(t)-\lim_{s\uparrow t}F_\theta(s)>0$, and construct the alternative policy $\tilde F_\theta$ by
\begin{align*}
 \tilde F_\theta(s)=\left\{
 \begin{array}{cl}
  F_\theta(s)+\alpha\beta & t-\varepsilon\leq s<t,\\
  F_\theta(s)-(1-\alpha)\beta & t\leq s<t+\varepsilon,\\
  F_\theta(s) &\text{otherwise},
 \end{array}\right.
\end{align*}
leaving $F_{\theta'}$ unchanged for every $\theta'\neq\theta$, where $\beta\in(0,\bar\beta(\varepsilon)]$ is no larger than the jump at $t$, with $\bar\beta(\varepsilon)$ specified in Step 2, and $\alpha$ solves
\begin{align*}
 \alpha v(t-\varepsilon)+(1-\alpha)v(t+\varepsilon)=v(t)\;\Longrightarrow\;\alpha=\frac{v(t)-v(t+\varepsilon)}{v(t-\varepsilon)-v(t+\varepsilon)}.
\end{align*}
Since $v$ is strictly decreasing, $v(t+\varepsilon)<v(t)<v(t-\varepsilon)$ and hence $\alpha\in(0,1)$. 
This upper bound on $\beta$ makes $\tilde F_\theta$ nondecreasing at $t$, and thus it is a well-defined distribution. 
The perturbation moves mass $\beta$ from $t$ to the two dates $t\pm\varepsilon$, preserving both state $\theta$'s total stopping probability and its expectation under $v$.

If instead $F_\theta(t)=\lim_{s\uparrow t}F_\theta(s)$, the same construction applies after replacing the
date $t$ by a neighborhood: fix $\delta\in(0,\varepsilon)$, let
\begin{equation*}
 m\equiv\int_{[t-\delta,t+\delta]}dF_\theta(s)>0,
 \qquad
 \bar v\equiv\frac1m\int_{[t-\delta,t+\delta]}v(s)\,dF_\theta(s),
\end{equation*}
where $m>0$ because $t\in\operatorname{supp}(dF_\theta)$, remove the fraction $\beta/m$ of $dF_\theta$ on $[t-\delta,t+\delta]$ in place of the atom, and let $\alpha$ solve $\alpha v(t-\varepsilon)+(1-\alpha)v(t+\varepsilon)=\bar v$. 
Because $v$ is strictly decreasing and $\delta<\varepsilon$ we have $v(t+\varepsilon)<\bar v<v(t-\varepsilon)$, and hence $\alpha\in(0,1)$; and since the probability removed is a fraction of $dF_\theta$ itself, $d\tilde F_\theta\geq0$ for every
$\beta\leq m$. 
The two cases differ only in whether the probability is removed from an atom at $t$ or from a small neighborhood around it.
In either case, the perturbation preserves both state $\theta$'s total stopping probability and its average value of $v(s)$, while reallocating the removed probability to $t-\varepsilon$ and $t+\varepsilon$, with the former earlier and the latter later than every date from which probability is removed.

\medskip
\noindent\textit{Step 2. Verify feasibility.}

Assumption \ref{ass:statewise-separable} makes the change in every continuation payoff at dates $t'\leq t-\varepsilon$ equal to zero: the perturbation lies entirely in $[t',\infty)$, leaves $F_\theta$ unchanged on $[0,t')$ and hence $\mu_{t'}$ unaffected, and preserves both state $\theta$'s stopping probability and the average of $v(s)$ (equal to $v(t)$ in the atomic case and to $\bar v$ in the non-atomic case).
Dates $t'>t+\varepsilon$ face the original policy, since the reallocated probability lies at dates no later than $t+\varepsilon$ and the probability removed equals the probability added. 
For $t'$ inside $[t-\varepsilon,t+\varepsilon]$, the induced change in the continuation surplus---including the effect on beliefs through
$\Pi$---is bounded by a quantity that depends only on $\beta$ and vanishes uniformly in $t'$ as $\beta\downarrow0$, by continuity of $\hat v_\theta$ and $\psi$ on the relevant compacta and local continuity of $\Pi$. 
Let $\bar\beta(\varepsilon)>0$ be small enough that this bound falls below the common slack; then every continuation constraint remains satisfied, and thus the perturbed policy is feasible.\footnote{Let $t'\in\{t-\varepsilon,t+\varepsilon\}$ be either new stopping date and let $\nu$ be the posterior induced by its recommendation. For every $s>t'$, the definition of $\psi$ and Assumption \ref{ass:statewise-separable} give
\[
 \psi(\nu)-\sum_{\theta\in\Theta}\nu(\theta)\hat v_\theta(t',s)
 =\sum_{\theta\in\Theta}\nu(\theta)
 \bigl[\hat v_\theta(t',t')-\hat v_\theta(t',s)\bigr]>0.
\]
Hence both new stopping recommendations are obedient.}

\medskip
\noindent\textit{Step 3. Compare the principal's payoff.}

The perturbation holds the agent's $v$-certainty equivalent fixed. 
Since $R(w,s)<R(v,s)$ on $[t-\varepsilon,t+\varepsilon]$, the function $g:=w\circ v^{-1}$
is strictly convex there, and the two-point distribution on $\{v(t-\varepsilon),v(t+\varepsilon)\}$ with weights $(\alpha,1-\alpha)$ is a non-degenerate mean-preserving spread after applying $v$ to the stopping dates.
The principal's payoff therefore changes by
\begin{equation*}
 \mu_0(\theta)\,\beta\Bigl[\alpha\,g\bigl(v(t-\varepsilon)\bigr)
 +(1-\alpha)\,g\bigl(v(t+\varepsilon)\bigr)-g\bigl(v(t)\bigr)\Bigr]>0,
\end{equation*}
which contradicts the optimality of the original policy. 
(In the non-atomic case $g(v(t))$ is replaced by $\int_{[t-\delta,t+\delta]}g(v(s))\,dF_\theta(s)/m$, and the same mean-preserving-spread argument gives the same strict inequality.)
The continuation constraint must therefore bind at every $t\in(\underline t,\bar t)\cap\operatorname{supp}(dF_\theta)$.
\end{proof}

The contract-theoretic interpretation of Lemma \ref{lem:no-rent} has been discussed in the main text. 
Although the lemma shows that the continuation constraint binds at each disclosure date, the policy may still contain gaps in its support. 
A gradual disclosure result therefore requires an additional argument showing that the binding constraints cannot be separated by an interval with no disclosure. 

\subsection{The Principal is More Sensitive to Stopping-Time Risk}

Now we consider the inverse case where the principal is more sensitive to stopping-time risk than the agent.
One might expect an analogous result: disclosure can always be contracted into a single date, improving the principal's payoff while preserving the agent's incentive constraints.
In a multi-state environment, however, several difficulties arise when attempting such a contraction, and we therefore require stronger assumptions.
First, we strengthen Assumption~\ref{ass:statewise-separable} as follows:

\begingroup
\addtocounter{A.assumption}{-1}
\expandafter\def\csname theA.assumption\endcsname{\thesection.\arabic{A.assumption}$'$}
\expandafter\def\csname theHA.assumption\endcsname{\thesection.\arabic{A.assumption}.prime}
\begin{A.assumption}
\label{ass:multi-separable}
There exist a positive function $A$, a strictly decreasing, continuous, and smooth function $v$, and state-dependent functions $B_\theta$ such that
\begin{equation}
\label{eq:multi-a1}
 \hat v_\theta(t,s)=A(t)v(s)+B_\theta(t)
\end{equation}
for every $\theta\in\Theta$ and every $s\geq t$.
\end{A.assumption}
\endgroup

Contracting stopping probability from all states to the same date requires a joint certainty-equivalent date for the state-contingent stopping distributions as a whole.  
Although Assumption \ref{ass:statewise-separable} makes the Arrow--Pratt coefficient common across current dates and states, it does not provide the global comparability needed for moving the stopping probability from all states to the same date.
Assumption \ref{ass:multi-separable} guarantees that the contraction leaves every continuation constraint at or before the beginning of the contracted interval unchanged, and thus all these constraints remain satisfied.
State dependence may still enter through $B_\theta(t)$ because the contraction preserves each state's total stopping probability.

Even with a common contraction date available under Assumption~\ref{ass:multi-separable}, this contraction creates a no-disclosure segment between the beginning of the interval and the contracted date, on which the continuation constraints need not remain satisfied.
The next assumption addresses this difficulty by requiring the constraint at the beginning of the interval to be the tightest along the no-disclosure segment.

\begin{A.assumption}
\label{ass:no-disclosure}
Consider any policy $F=\{F_\theta\}_{\theta\in\Theta}$ and any interval $I=[\underline{t},\bar{t})$ on which $F_\theta(t)=F_\theta(\underline{t})$ for every $\theta$ and every $t\in I$.
Let $\{\mu_t\}_t$ be the no-information belief path beginning at $\mu_{\underline{t}}$.
If $\underline{t}$ is no earlier than the agent's voluntary stopping time with no information, then, for every $t\in I$,
\begin{align}
 &\frac{
 \sum_{\theta\in\Theta}\mu_0(\theta)
 \bigl(1-F_\theta(\underline{t})\bigr)
 \bigl[\psi(\mu_{\underline{t}})-B_\theta(\underline{t})\bigr]+C(\underline{t})}{A(\underline{t})}
 \notag\\
 &\qquad\geq
 \frac{
 \sum_{\theta\in\Theta}\mu_0(\theta)
 \bigl(1-F_\theta(\underline{t})\bigr)
 \bigl[\psi(\mu_t)-B_\theta(t)\bigr]+C(t)}{A(t)}.
\label{eq:no-disclosure-ordering}
\end{align}
\end{A.assumption}

Assumption \ref{ass:no-disclosure} requires that waiting for a fixed future disclosure become no harder as that disclosure approaches.
This is a natural regularity condition for a well-behaved specification of the agent's intertemporal incentives.
Formally, after substituting \eqref{eq:multi-a1} into the continuation constraint and dividing by $A(t)$, the two sides of \eqref{eq:no-disclosure-ordering} compare the required payoff of continuation at the interval's start with that at date $t$.
Assumption \ref{ass:no-disclosure} then makes the continuation constraint at $\underline{t}$ the tightest, and hence feasibility at $\underline{t}$ implies feasibility throughout the no-disclosure segment.

\begin{A.lemma}
\label{lem:final-interval}
Suppose Assumptions \ref{ass:multi-separable} and \ref{ass:no-disclosure} hold and $R(w,t)>R(v,t)$ for every $t\in I=[\underline t,\bar t]$, where $\bar t$ may be infinite.
Suppose $\underline t$ is no earlier than the agent's voluntary stopping time with no information.
If an optimal policy $F^*$ satisfies $F_\theta^*(\bar t)=1$ for every $\theta\in\Theta$, then $F^*$ places positive stopping probability on at most one date in $I$.
\end{A.lemma}

\begin{proof}
Suppose $F^*$ places positive stopping probability on at least two distinct dates in $I$.
Define the joint certainty-equivalent date $\hat t$ by
\begin{equation}
\label{eq:joint-ce}
 v(\hat{t})
 =\frac{\sum_{\theta\in\Theta}\mu_0(\theta)
   \int_I v(s)\,dF^*_\theta(s)}
 {\sum_{\theta\in\Theta}\mu_0(\theta)\int_I dF^*_\theta(s)}.
\end{equation}
Continuity and strict monotonicity of $v$ make $\hat t$ well-defined and unique, and it is straightforward that $\hat t\in I$.
Move every state's stopping probability in $I$ to $\hat t$, and denote the resulting policy by $\widetilde F$.

At every date weakly before $\underline t$, the two policies have the same history.
The contraction changes only the stopping dates within $I$, and thus $\int_I d\widetilde F_\theta=\int_I dF_\theta^*$ for every state $\theta$.
Assumption \ref{ass:multi-separable} therefore makes the $B_\theta(t)$ terms cancel, and the difference between the two continuation payoffs is
\begin{align*}
 A(t)\biggl[
 &\sum_{\theta\in\Theta}\mu_0(\theta)
 \int_I v(\hat t)\,dF^*_\theta(s)
 -\sum_{\theta\in\Theta}\mu_0(\theta)
 \int_I v(s)\,dF^*_\theta(s)
 \biggr]=0.
\end{align*}
Equation \eqref{eq:joint-ce} gives the equality.
Thus $\widetilde F$ preserves every constraint at or before $\underline t$.

The contracted policy sends no message on $[\underline t,\hat t)$.
Date $\underline t$ is both the last date covered by the preceding calculation and the first date of this no-disclosure interval.
The preceding calculation preserves its constraint, and Assumption \ref{ass:no-disclosure} makes that constraint the tightest one on $[\underline t,\hat t)$.
Hence, feasibility at $\underline t$ implies feasibility of every constraint before $\hat t$.

The stopping recommendation at $\hat t$ may pool several states.
Under its induced posterior, the payoff difference between stopping at $\hat t$ and stopping at any $s>\hat t$ is
\[
 A(\hat t)\bigl[v(\hat t)-v(s)\bigr]>0.
\]
The posterior expectation of $B_\theta(\hat t)$ cancels from this comparison.
The pooled recommendation is therefore obedient.

Because the interval exhausts all remaining stopping probability, $\widetilde F$ assigns no stopping probability after the common date.
No later continuation constraint remains.

Finally, the identity
\[
 (w\circ v^{-1})''(v(s))
 =\frac{w'(s)}{(v'(s))^2}\bigl(R(v,s)-R(w,s)\bigr)
\]
and the maintained Arrow--Pratt ranking imply that $w\circ v^{-1}$ is strictly concave.
Jensen's inequality shows that reallocating the stopping probability to $\hat t$ strictly raises the principal's payoff, contradicting optimality.
\end{proof}

Lemma \ref{lem:final-interval} establishes the one-shot contraction when disclosure occurs on a final interval.
Assumption \ref{ass:multi-separable} makes a common contraction date available, while Assumption \ref{ass:no-disclosure} preserves all continuation constraints up to that date, including those on the no-disclosure segment created by the contraction.  
The result requires the contracted interval to exhaust all remaining stopping probability in every state, a condition under which there will be no continuation constraint after the common date.

When stopping probability remains beyond the contracted interval $[\underline{t},\bar{t}]$, moving the stopping probability within the interval to an earlier date $\hat{t}$ alters the distribution of surviving states and the agent's posterior beliefs on $(\hat{t},\bar{t}]$, which may tighten continuation constraints on this segment.
We therefore need a condition that preserves these constraints while leaving the stopping probability beyond the interval unchanged.
To formulate this condition, given a policy $F$, let $q_\theta(t):=\mu_0(\theta)(1-F_\theta(t))$ denote the prior-weighted survival probability in state $\theta$, and let $q(t):=(q_\theta(t))_{\theta\in\Theta}$.
For every nonzero $q\geq0$, define
\begin{equation}
\label{eq:psi-perspective}
 \Psi(q)
 :=
 \left(\sum_{\theta\in\Theta}q_\theta\right)
 \psi\left(
 \frac{q}{\sum_{\theta\in\Theta}q_\theta}
 \right),
 \qquad \Psi(0):=0.
\end{equation}
Continuity of $\psi$ on the simplex implies that $\Psi$ is continuous on $\mathbb R_+^{|\Theta|}$.

\begin{A.assumption}
\label{ass:continuation-order}
The function $\Psi$ is differentiable, and
\begin{equation}
\label{eq:continuation-pointwise}
 \hat v_\theta(t,t)
 \leq
 \partial_{q_\theta}\Psi(q)
\end{equation}
for every date $t$, every nonzero $q\geq0$, and every state $\theta$ with $q_\theta>0$.
\end{A.assumption}

The derivative $\partial_{q_\theta}\Psi(q)$ is the marginal contribution of state-$\theta$ surviving probability to the payoff from stopping under the pooled belief.
At a full-support belief $\mu=q/\sum_{\theta'}q_{\theta'}$,
\[
 \partial_{q_\theta}\Psi(q)
 =\psi(\mu)+\nabla\psi(\mu)\cdot(\delta_\theta-\mu).
\]
When immediate stopping after learning the state gives $\hat v_\theta(t,t)=\psi(\delta_\theta)$, Assumption \ref{ass:continuation-order} requires
\begin{equation}
 \psi(\mu)+\nabla\psi(\mu)\cdot(\delta_\theta-\mu)
 \geq \psi(\delta_\theta)
\end{equation}
for every state $\theta$, which implies that $\psi$ is concave.
However, if $\psi$ is an indirect decision payoff obtained by maximizing expected utility over available actions, it is convex, and Assumption \ref{ass:continuation-order} can hold throughout the belief simplex only if $\psi$ is affine, ruling out the case that information has a strictly positive instrumental value.

\begin{A.lemma}
\label{lem:bounded-interval}
Suppose Assumptions \ref{ass:multi-separable}, \ref{ass:no-disclosure}, and \ref{ass:continuation-order} hold and $R(w,t)>R(v,t)$ for every $t\in I=[\underline t,\bar t]$, where $\bar t<\infty$.
If $\underline t$ is no earlier than the agent's no-information stopping date at that history, an optimal policy places positive stopping probability on at most one date in $I$.
\end{A.lemma}

\begin{proof}
Suppose an optimal policy has a nondegenerate stopping distribution in $I$, and apply the contraction from Lemma \ref{lem:final-interval} by moving every state's stopping probability in $I$ to the same certainty-equivalent date $\hat t$.
The proof of Lemma \ref{lem:final-interval} applies unchanged through $\hat t$ and preserves every incentive constraint up to and including that date.
In particular, Assumption \ref{ass:multi-separable} preserves all constraints at or before $\underline t$, Assumption \ref{ass:no-disclosure} preserves those on $[\underline t,\hat t)$, and the same payoff comparison makes the pooled stopping recommendation at $\hat t$ obedient.
It remains only to check the continuation constraints after $\hat t$.

Fix $t\in(\hat t,\bar t)$ that the agent reaches with positive probability under the contracted policy.
Under the original policy $F^*$, write $q:=q(t)$.
Let
\[
 m_\theta:=\mu_0(\theta)\int_{[t,\bar t]}dF_\theta^*(s),
 \qquad
 m:=(m_\theta)_{\theta\in\Theta}.
\]
Here $m_\theta$ is the probability of stopping in $[t,\bar t]$ in state $\theta$ under $F^*$, multiplied by the prior probability $\mu_0(\theta)$.
In every state $\theta$, the contraction moves this stopping probability from $[t,\bar t]$ to $\hat t<t$.
At date $t$, the contraction replaces $q$ with $q-m$ and removes the stopping probability in $[t,\bar t]$ from the integral in the continuation constraint.
Because the original policy is feasible, it is enough to show that
\begin{equation}
 \Psi(q)-\Psi(q-m)
 -
 \sum_{\theta\in\Theta}
 \mu_0(\theta)
 \int_{[t,\bar t]}
 \hat v_\theta(t,s)\,dF_\theta^*(s)
 \geq0.
\end{equation}
Assumption \ref{ass:multi-separable} and the decrease of $v$ imply $\hat v_\theta(t,s)\leq\hat v_\theta(t,t)$ for every $s\geq t$.
For every $\delta\in(0,1)$, Assumption \ref{ass:continuation-order} gives
\[
 \Psi(q)-\Psi(q-m+\delta m)
 =
 \int_\delta^1
 \left\langle
 \nabla\Psi(q-m+\sigma m),m
 \right\rangle d\sigma
 \geq
 (1-\delta)
 \sum_{\theta\in\Theta}
 m_\theta\hat v_\theta(t,t).
\]
Letting $\delta\downarrow0$ and using continuity of $\Psi$ gives
\[
 \Psi(q)-\Psi(q-m)
 \geq
 \sum_{\theta\in\Theta}m_\theta\hat v_\theta(t,t)
 \geq
 \sum_{\theta\in\Theta}
 \mu_0(\theta)
 \int_{[t,\bar t]}
 \hat v_\theta(t,s)\,dF_\theta^*(s).
\]
From $\bar t$ onward, the continuation problem is unchanged.
The Arrow--Pratt comparison and Jensen's inequality then yield a strict payoff improvement, contradicting optimality.
\end{proof}

The preceding proof establishes sufficiency.
Assumption \ref{ass:continuation-order} need not be necessary for a particular contraction with slack constraints, but it is tight among conditions imposed separately at each $(\theta,t,q)$.
Any such condition that guarantees that moving stopping probability from after $t$ to before $t$ preserves later constraints must imply Assumption \ref{ass:continuation-order}.
To see this, fix $\theta,t,q$ and move a mass $\varepsilon$ in state $\theta$ from $t+\eta$ to a date before $t$.
The required ordering is
\[
 \Psi(q)-\Psi(q-\varepsilon e_\theta)
 -\varepsilon\hat v_\theta(t,t+\eta)
 \geq0.
\]
Dividing by $\varepsilon$ and letting $\varepsilon,\eta\downarrow0$ gives \eqref{eq:continuation-pointwise}.

The misaligned-time-preference model in Section 4 shows why Lemma \ref{lem:bounded-interval} is needed.
Its one-shot--gradual--one-shot proof contracts a bounded interval while stopping probability remains after that interval.
Lemma \ref{lem:final-interval} does not cover this case, whereas Lemma \ref{lem:bounded-interval} uses Assumption \ref{ass:continuation-order} to preserve the later continuation constraints.

\section{Proofs}

\subsection{Proof of Proposition \ref{proposition.static}}

 By (\ref{eq.tau}), $\tau(\mu)=0$ if and only if
 \begin{align*}
  \mu\leq\bar{\mu}_0 \equiv y^{-1}\left[\left(1+\frac{r}{p_0\lambda}\right)z\right].
 \end{align*}
If so, $W_{\text{NI}}(\mu)\equiv Z$.
If not,
 \begin{align*}
  W_{\text{NI}}(\mu)=&w(\tau(\mu))=p_0\left(\int_0^{\tau(\mu)}\lambda e^{-(\lambda+r)t} dt \right)\cdot Y+\left(1-p_0+p_0e^{-\lambda \tau(\mu)}\right)e^{-r\tau(\mu)}\cdot Z\\
   =&p_0\left(1-e^{-(\lambda+r)\tau(\mu)}\right)\cdot\frac{\lambda Y}{\lambda+r}+\left(1-p_0+p_0e^{-\lambda \tau(\mu)}\right)e^{-r\tau(\mu)}\cdot Z\\
   =&(1-\zeta(\mu))\cdot\frac{p_0\lambda Y}{\lambda+r}+\zeta(\mu)\cdot\frac{p_0\lambda(y(\mu)-z)Z}{rz},
 \end{align*}
 where
 \begin{equation*}
  \zeta(\mu)=e^{-(\lambda+r)\tau(\mu)}=\left(\frac{p_0}{1-p_0}\frac{\lambda y(\mu)-(\lambda+r)z}{r z}\right)^{-\frac{\lambda+r}{\lambda}}.
 \end{equation*}
Since
\[
\tau'(\mu)=\frac{y_H-y_L}{\lambda y(\mu)-(\lambda+r)z},
\qquad
\frac{\zeta'(\mu)}{\zeta(\mu)}
=-\frac{(\lambda+r)(y_H-y_L)}
{\lambda y(\mu)-(\lambda+r)z},
\]
direct differentiation gives
 \begin{align*}
  W_{\text{NI}}'(\mu)&=-\zeta'(\mu)\left(\frac{p_0\lambda Y}{\lambda+r}-\frac{p_0\lambda(y(\mu)-z)Z}{rz}\right)+\zeta(\mu)\frac{p_0\lambda(y_H-y_L)Z}{rz}\\
  &=\zeta(\mu)\cdot\frac{p_0\lambda(y_H-y_L)}{z}\frac{Yz-y(\mu)Z}{\lambda y(\mu)-(\lambda+r)z}.
 \end{align*}
 Hence, $W_{\text{NI}}(\mu)$ is increasing if and only if $\mu\leq y^{-1}\left(\frac{Yz}{Z}\right)=\tau^{-1}(t^*).$
 Also, we have
 \begin{align*}
  W_{\text{NI}}''(\mu)=&\frac{p_0\lambda(y_H-y_L)}{z}\left[\zeta'(\mu)\frac{Yz-y(\mu)Z}{\lambda y(\mu)-(\lambda+r)z}+\zeta(\mu)\left(\frac{Yz-y(\mu)Z}{\lambda y(\mu)-(\lambda+r)z}\right)'\right]\\
  =&\zeta(\mu)\cdot\frac{p_0\lambda(y_H-y_L)^2}{z}\frac{(\lambda+r)(y(\mu)+z)Z-(2\lambda+r)Yz}{(\lambda y(\mu)-(\lambda+r)z)^2},
 \end{align*}
 which is positive if and only if
 \begin{align*}
  \mu\geq\mu_2^{\text{NI}}\equiv y^{-1}\left[\frac{z}{Z}\frac{(2\lambda+r)Y-(\lambda+r)Z}{\lambda+r}\right].
 \end{align*}
 The threshold is greater than $\tau^{-1}(t^*)$ if $t^*>0$.
 Thus, $W_{\text{NI}}(\mu)$ is equal to $Z$ when $\mu\leq\bar{\mu}_0$, increasing and concave when $\bar{\mu}_0\leq \mu\leq\tau^{-1}(t^*)$, decreasing and concave when $\tau^{-1}(t^*)\leq\mu\leq\mu_2^{\text{NI}}$ and decreasing and convex when $\mu\geq\mu_2^{\text{NI}}$.
 Let $\cav\circ W_{\text{NI}}(\cdot)$ be the concavification of $W_{\text{NI}}$,\footnote{The concavification of $W_{\text{NI}}$ is the smallest concave function that weakly dominates $W_{\text{NI}}$.} and define
 \begin{align*}
  \mu_L\equiv\inf_{0<\mu < 1}\,\left\{\cav\circ W_{\text{NI}}(\mu)=W_{\text{NI}}(\mu)\right\} \text{ and }\mu_H\equiv\sup_{0 <\mu<1}\,\left\{\cav\circ W_{\text{NI}}(\mu)=W_{\text{NI}}(\mu)\right\}.
 \end{align*}
The result follows directly from \cite{kamenica2011bayesian}.

\subsection{A Useful Lemma}

We start with an intermediate result that is helpful throughout this paper. 

\begin{A.lemma}
 \label{lemma.useful}
    For any $t_1<s<t_2$, we have $\bar{v}_\theta(t_1,t_2)=\bar{v}_\theta(t_1,s)+\gamma(t_1,s)\cdot \bar{v}_\theta(s,t_2)$, where $\gamma(t,s)=\frac{p_t}{p_{s}}\cdot e^{-(\lambda+r)(s-t)}$.
\end{A.lemma}
\begin{proof}
  We know
  \begin{align*}
   \bar{v}_\theta(t_1,t_2)=&p_{t_1}\cdot\frac{\lambda}{\lambda+r}\left(1-e^{-(\lambda+r)(t_2-t_1)}\right)\cdot y_\theta\\ 
   &+\bigg(\left(1-p_{t_1}+p_{t_1}\cdot e^{-\lambda(t_2-t_1)}\right)\cdot e^{-r(t_2-t_1)}-1\bigg)\cdot z.
  \end{align*}
  Thus, 
  \begin{align*}
   \bar{v}_\theta&(t_1,t_2)-\bar{v}_\theta(t_1,s)-\gamma(t_1,s)\cdot\bar{v}_\theta(s,t_2)\\
   =&\left(p_{t_1}\left(1-e^{-(\lambda+r)(t_2-t_1)}\right)-p_{t_1}\left(1-e^{-(\lambda+r)(s-t_1)}\right)-\gamma(t_1,s)p_{s}\left(1-e^{-(\lambda+r)(t_2-s)}\right)\right)\frac{\lambda y_\theta}{\lambda+r}\\
   &+\bigg(\left(1-p_{t_1}+p_{t_1}\cdot e^{-\lambda(t_2-t_1)}\right)\cdot e^{-r(t_2-t_1)}-\left(1-p_{t_1}+p_{t_1}\cdot e^{-\lambda(s-t_1)}\right)\cdot e^{-r(s-t_1)}\\
   &-\gamma(t_1,s)\left(\left(1-p_{s}+p_{s}\cdot e^{-\lambda(t_2-s)}\right)\cdot e^{-r(t_2-s)}-1\right)\bigg)z.
  \end{align*}
  By the expression of $\gamma(t_1,s)$, the first term is reduced to
  \begin{align*}
   \left(\left(1-e^{-(\lambda+r)(t_2-t_1)}\right)-\left(1-e^{-(\lambda+r)(s-t_1)}\right)-e^{-(\lambda+r)(s-t_1)}\left(1-e^{-(\lambda+r)(t_2-s)}\right)\right)\frac{p_{t_1}\lambda y_\theta}{\lambda+r}=0.
  \end{align*}
  Also, the second term becomes
  \begin{align*}
   &\bigg((1-p_{t_1})\left(e^{-r(t_2-t_1)}-e^{-r(s-t_1)}\right)+p_{t_1}\left(e^{-(\lambda+r)(t_2-t_1)}-e^{-(\lambda+r)(s-t_1)}\right)\\
   &+\frac{p_{t_1}}{p_s}e^{-(\lambda+r)(s-t_1)}(1-e^{-r(t_2-s)})+p_{t_1}e^{-(\lambda+r)(s-t_1)}e^{-r(t_2-s)}\left(1-e^{-\lambda(t_2-s)}\right)\bigg)z\\
   =&\left(e^{-r(t_2-t_1)}-e^{-r(s-t_1)}\right)\bigg(1-p_{t_1}+p_{t_1}e^{-\lambda(s-t_1)}-\frac{p_{t_1}}{p_s}e^{-\lambda(s-t_1)}\bigg)z.
  \end{align*}
  Using the expressions of $p_{t_1}$ and $p_{s}$, we have
  \begin{align*}
   & 1-p_{t_1}+p_{t_1}e^{-\lambda(s-t_1)}-\frac{p_{t_1}}{p_s}e^{-\lambda(s-t_1)}=0,%
  \end{align*}
  which completes the proof.
\end{proof}

\subsection{Proof of Theorem \ref{thetheorem}}


We complete the proof in eleven steps.

\subsubsection*{Step 1: Simplifying the incentive constraints}
We simplify incentive constraints (\ref{implementable.raw.1}) and (\ref{implementable.raw.2}).
For the continuation constraints, consider any policy $\langle F_H,F_L\rangle$.
Using
 \begin{align*}
  \mu_t=\frac{\mu_0(1-F_H(t))}{\mu_0(1-F_H(t))+(1-\mu_0)(1-F_L(t))}
 \end{align*}
 and
 \begin{align*}
  F_\theta(s|s\geq t)=\frac{F_\theta(s)-F_\theta(t)}{1-F_\theta(t)},\quad\forall\theta\in\{H,L\},
 \end{align*}
we can reduce the continuation constraints (\ref{implementable.raw.1}) to 
 \begin{equation}
 \label{implementable.1}
  \mu_0\int_{t}^\infty\bar{v}_H(t,s)dF_H(s)+(1-\mu_0)\int_{t}^\infty \bar{v}_L(t,s)dF_L(s)\geq 0,\quad\forall t\in\mathcal{C}(\mathcal{P}).
 \end{equation}

For the stopping constraints, Lemma \ref{lemma.useful} allows us to write (\ref{implementable.raw.2}) as
\begin{align*}
 \nu_tv'_H(t)+(1-\nu_t)v'_L(t)\leq0\Rightarrow\nu_t\leq-\frac{v'_L(t)}{v'_H(t)-v'_L(t)}\equiv \bar{\nu}(t).
\end{align*}
Using Bayes' rule, 
\begin{align*}
 \nu_t=\frac{\mu_{t-dt}dF_H(t|t-dt)}{\mu_{t-dt}dF_H(t|t-dt)+(1-\mu_{t-dt})dF_L(t|t-dt)}=\frac{\mu_0dF_H(t)}{\mu_0dF_H(t)+(1-\mu_0)dF_L(t)},
\end{align*}
which permits us to formalize the second class of incentive constraints as
\begin{equation}
\label{implementable.2}
 \mu_0dF_H(t)(1-\bar{\nu}(t))\leq(1-\mu_0)dF_L(t)\cdot\bar{\nu}(t),\quad\forall t\in\mathcal{S}(\mathcal{P}).
\end{equation}

\subsubsection*{Step 2: $F_H(t)\equiv 0$ for all $t\in[\tau(\mu_0),t^*)$}

Suppose that $\langle F_H,F_L\rangle$ is the optimal information policy, but there exists $t<t^*$ such that $dF_H(t)$ and $dF_L(t)$ are both strictly positive. 
Define
 \begin{align*}
  \phi(t)&\equiv\frac{\mu_0}{1-\mu_0}\frac{1-\bar{\nu}(t)}{\bar{\nu}(t)}\\&=-\frac{\mu_0}{1-\mu_0}\frac{v'_H(t)}{v'_L(t)}=-\frac{\mu_0}{1-\mu_0}\frac{p_0e^{-\lambda t}(\lambda y_H-(\lambda+r)z)-(1-p_0)rz}{p_0e^{-\lambda t}(\lambda y_L-(\lambda+r)z)-(1-p_0)rz},
 \end{align*}
and then we have 
\begin{align*}
  \phi'(t)=-\frac{\mu_0}{1-\mu_0}\frac{p_0(1-p_0)\lambda^2 r e^{\lambda t}(y_H-y_L)z}{(p_0(\lambda y_L-(\lambda+r)z)-(1-p_0)re^{\lambda t}z)^2}<0.
 \end{align*}

Given function $\phi$, fix a small distance $\varepsilon>0$ and consider the alternative information policy $\langle G_H,G_L\rangle$, such that
 \begin{align*}
  G_H(s)=\left\{
   \begin{array}{cc}
       F_H(s)-dF_H(t) & s\in[t,t+\varepsilon) \\
       F_H(s) & \text{otherwise}
   \end{array}
  \right.,
 \end{align*}
 and
 \begin{align*}
  G_L(s)=\left\{
   \begin{array}{cc}
       F_L(s)-\phi(t+\varepsilon)dF_H(t) & s\in[t,t+\varepsilon) \\
       F_L(s) & \text{otherwise}
   \end{array}
  \right..
 \end{align*}
 That is, we extract all the probability mass at time $t$ in state $H$ and some probability mass $\phi(t+\varepsilon)dF_H(t)$ at time $t$ in state $L$, and add them to time $t+\varepsilon$.\footnote{When $dF_H(t)>0$ denotes a density rather than an atom, use the neighborhood construction in the proof of Lemma~\ref{lem:no-rent}; the same inequalities follow by continuity.}
 Here, we guarantee that the stopping constraint at time $t+\varepsilon$ is binding.
Because $\phi(\cdot)$ is decreasing, $dF_H(t)\phi(t+\varepsilon)<dF_H(t)\phi(t)$.
At $t\in\mathcal{S}(\mathcal{P})$, the stopping constraint (\ref{implementable.2}) gives $dF_H(t)\phi(t)\leq dF_L(t)$, and hence the extraction is feasible.

 Here, the new policy is more desirable to the principal, and the stopping constraints are always satisfied.
 Now, it suffices to show that the continuation constraints at all $t\geq 0$ are also satisfied.
 First, the incentive constraints after $t+\varepsilon$ are completely identical; second, when $s<t$, it suffices to show that
 \begin{align*}
  \mu_0dF_H(t)&\bigg(\bar{v}_H(s,t+\varepsilon)-\bar{v}_H(s,t)\bigg)\\&+(1-\mu_0)\phi(t+\varepsilon)dF_H(t)\bigg(\bar{v}_L(s,t+\varepsilon)-\bar{v}_L(s,t)\bigg)\geq 0.
 \end{align*}
 By Lemma \ref{lemma.useful}, it is equivalent to
 \begin{align*}
  \mu_0\bigg(v_H(t+\varepsilon)-v_H(t)\bigg)+(1-\mu_0)\phi(t+\varepsilon)\bigg(v_L(t+\varepsilon)-v_L(t)\bigg)\geq 0.
 \end{align*}
 Substituting the expression of function $\phi$, it is equivalent to
 \begin{align*}
  v_H'(t+\varepsilon)\bigg(v_L(t+\varepsilon)-v_L(t)\bigg)-v_L'(t+\varepsilon)\bigg(v_H(t+\varepsilon)-v_H(t)\bigg)\geq 0.
 \end{align*}
 We can directly calculate the left-hand side as
 \begin{align*}
  \underbrace{\frac{p_0(1-p_0)\lambda e^{-(2\lambda+r)(t+\varepsilon)}(y_H-y_L)z}{\lambda+r}}_{\geq 0}\cdot\bigg(\lambda(1-e^{r\varepsilon})+r(e^{\lambda\varepsilon}-1)e^{r\varepsilon}\bigg).
 \end{align*}
 Note that 
 \begin{align*}
  \frac{\partial}{\partial \varepsilon}\bigg(\lambda(1-e^{r\varepsilon})+r(e^{\lambda\varepsilon}-1)e^{r\varepsilon}\bigg)=r(\lambda+r)(e^{\lambda\varepsilon}-1)e^{r\varepsilon}>0,
 \end{align*}
 and thus 
 \begin{equation}
 \label{eq.oh?}
    \lambda(1-e^{r\varepsilon})+r(e^{\lambda\varepsilon}-1)e^{r\varepsilon}>\lambda(1-e^{r\varepsilon})+r(e^{\lambda\varepsilon}-1)e^{r\varepsilon}\bigg|_{\varepsilon=0}=0.
 \end{equation}
 This shows that all the continuation constraints at time $s<t$ are satisfied.
 When $t\leq s<t+\varepsilon$, it suffices to show that
 \begin{align*}
  \mu_0dF_H(t)\bar{v}_H(s,t+\varepsilon)+(1-\mu_0)\phi(t+\varepsilon)dF_H(t)\bar{v}_L(s,t+\varepsilon)\geq 0.
 \end{align*}
 Again, by Lemma \ref{lemma.useful} and the expression for $\phi$, it suffices to verify that, for every $t'\geq 0$,
 \begin{align*}
  v_H'(t')v_L(t')-v_L'(t')v_H(t')\geq 0.
 \end{align*}
 The left-hand side is
 \begin{align*}
  \underbrace{\frac{p_0(1-p_0)\lambda z (y_H-y_L) e^{-(\lambda+2r)t'} }{\lambda+r}}_{\geq 0}\cdot \underbrace{\bigg(\lambda(1-e^{rt'})+r(e^{\lambda t'}-1)e^{rt'}\bigg)}_{\geq 0\text{ by (\ref{eq.oh?})}}\geq 0.
 \end{align*}
 Thus, the continuation constraint at time $s$ is satisfied when $t\leq s<t+\varepsilon$.
 Since $\langle G_H,G_L\rangle$, an information policy that is more desirable to the principal, is implementable, $\langle F_H,F_L\rangle$ cannot be optimal, which completes the proof.

\subsubsection*{Step 3: $F_H(t)=F_L(t)\equiv 0$ for all $t<\min\{\tau(\mu_0),t^*\}$}
Suppose instead that an optimal information policy $\langle G_H,G_L\rangle$ has $dG_L(t)>0$ for some $t<\min\{t^*,\tau(\mu_0)\}$.
Note that the agent chooses to experiment even though there is no subsequent information when $t<\min\{t^*,\tau(\mu_0)\}$.
Consider an alternative policy $\langle F_H,F_L\rangle$, such that for any $\theta\in\{H,L\}$,
\begin{align*}
 F_\theta(t)=\left\{
  \begin{array}{cc}
   0 & t<\min\{t^*,\tau(\mu_0)\}\\
   G_\theta(t) & t\geq\min\{t^*,\tau(\mu_0)\}
\end{array} 
 \right..
\end{align*}
This policy obviously makes the principal better off since it postpones the stopping time when $t<t^*$.
We show that $\langle F_H,F_L\rangle$ is also implementable. 
Before $\min\{t^*,\tau(\mu_0)\}$, recommending the agent to continue with probability $1$ is obviously implementable.
At $\min\{t^*,\tau(\mu_0)\}$, stopping remains optimal because the pooled posterior combines only stopping posteriors $\nu_s$ satisfying $\tau(\nu_s)\leq s\leq\min\{t^*,\tau(\mu_0)\}$.
After $\min\{t^*,\tau(\mu_0)\}$, policy $\langle F_H,F_L\rangle$ is identical to policy $\langle G_H,G_L\rangle$, which was assumed to be implementable, and thus the constraints are satisfied.
Given that $\langle G_H,G_L\rangle$ cannot be optimal, there is no information before $\min\{t^*,\tau(\mu_0)\}$.

\subsubsection*{Step 4: The decomposition}

We use the following lemma to further simplify the optimization problem (\ref{optimization}).

\begin{A.lemma}

If the information policy $\langle F_H,F_L\rangle$ is the optimal information policy, set $\pi_\theta=1-F_\theta(t^*)$ and define $u$ by
\begin{align*}
\mu_0\int_{t^*}^\infty\bar{v}_H(t^*,s)dF_H(s)+(1-\mu_0)\int_{t^*}^\infty\bar{v}_L(t^*,s)dF_L(s)=\big(\mu_0\pi_H+(1-\mu_0)\pi_L\big)\cdot u.
\end{align*}\footnote{For notational simplicity, we first consider the nondegenerate case in which $\pi_\theta\in(0,1)$ for each $\theta\in\{H,L\}$.
Boundary cases are handled by the following convention.
If $\pi_\theta=0$, the conditional distribution $F_\theta^a$ is payoff- and constraint-irrelevant and can be chosen arbitrarily.
If $\pi_\theta=1$, the conditional distribution $F_\theta^b$ is payoff- and constraint-irrelevant and can be chosen arbitrarily.
Equivalently, all formulas below are interpreted through the weighted measures
$(1-\pi_\theta)dF_\theta^b$ and $\pi_\theta dF_\theta^a$, and thus the boundary cases follow by continuity.}

Also, define
\begin{align*}
F_\theta^b(t)=\left\{
\begin{array}{cc}
\frac{F_\theta(t)}{F_\theta(t^*)} & t\leq t^*\\
1 & t>t^*
\end{array}
\right.
\text{ and  }\,
F_\theta^a(t)=\left\{
\begin{array}{cc}
0 & t<t^*\\
\frac{F_\theta(t)-F_\theta(t^*)}{1-F_\theta(t^*)} & t\geq t^*
\end{array}
\right..
\end{align*}
Then the pair $\langle F_H^a,F_L^a\rangle$ must be the solution to
\begin{equation}
\label{eq.after}
\begin{aligned}
\max_{F_H^a,F_L^a}:\,&\mu_{t^*}\int_{t^*}^\infty w(t)dF^a_H(t)+(1-\mu_{t^*})\int_{t^*}^\infty w(t)dF^a_L(t)\\
\text{s.t: }&\mu_{0}\pi_H\int_{t^*}^\infty\bar{v}_H(t^*,s)dF_H^a(s)+(1-\mu_{0})\pi_L\int_{t^*}^\infty\bar{v}_L(t^*,s)dF_L^a(s)\\
&\quad=\big(\mu_0\pi_H+(1-\mu_0)\pi_L\big)u,\\
&\mu_{0}\pi_H\int_{t}^\infty\bar{v}_H(t,s)dF_H^a(s)+(1-\mu_{0})\pi_L\int_{t}^\infty\bar{v}_L(t,s)dF_L^a(s)\geq0,\,\forall t\in\mathcal{C}(\mathcal{P}^a),\\
&\mu_{0}\pi_HdF^a_H(t)(1-\bar{\nu}(t))\leq(1-\mu_{0})\pi_LdF^a_L(t)\cdot\bar{\nu}(t),\quad\forall t\in\mathcal{S}(\mathcal{P}^a),
\end{aligned}
\end{equation}
and the pair $\langle F_H^b,F_L^b\rangle$ must be the solution to
\begin{equation}
\label{eq.before}
\begin{aligned}
\max_{F_H^b,F_L^b}:\,&\mu_{0}(1-\pi_H)\int_0^{t^*} w(t)dF^b_H(t)+(1-\mu_{0})(1-\pi_L)\int_0^{t^*}w(t)dF^b_L(t)\\
\text{s.t.: }  &\mu_0(1-\pi_H)\int_t^{t^*}\bar{v}_H(t,s)dF_H^b(s)+(1-\mu_0)(1-\pi_L)\int_t^{t^*}\bar{v}_L(t,s)dF_L^b(s)+\mu_0\pi_H\bar{v}_H(t,t^*)\\
&\quad+(1-\mu_0)\pi_L\bar{v}_L(t,t^*)+\gamma(t,t^*)\big(\mu_0\pi_H+(1-\mu_0)\pi_L\big)\cdot u\geq 0\quad\forall t\in\mathcal{C}(\mathcal{P}^b),\\
&\mu_0(1-\pi_H)dF^b_H(t)(1-\bar{\nu}(t))\leq(1-\mu_0)(1-\pi_L)dF^b_L(t)\cdot\bar{\nu}(t),\quad\forall t\in\mathcal{S}(\mathcal{P}^b).
\end{aligned}
\end{equation}
\end{A.lemma}

\begin{proof}
In fact, we can write the objective function of the principal as
\begin{align*}
 &\mu_0\int_0^\infty w(t)\bigg((1-\pi_H)dF_H^b(t)+\pi_HdF_H^a(t)\bigg)\\
 &\quad+(1-\mu_0)\int_0^\infty w(t)\bigg((1-\pi_L)dF_L^b(t)+\pi_LdF_L^a(t)\bigg)\\
 =&\underbrace{\mu_0(1-\pi_H)\int_0^\infty w(t)dF_H^b(t)+(1-\mu_0)(1-\pi_L)\int_0^\infty w(t)dF_L^b(t)}_{\text{independent of }\langle F_H^a,F_L^a\rangle}\\
 &\quad+\underbrace{\mu_0\pi_H\int_0^\infty w(t)dF_H^a(t)+(1-\mu_0)\pi_L\int_0^\infty w(t)dF_L^a(t)}_{\text{independent of }\langle F_H^b,F_L^b\rangle}.
\end{align*}

Now we show that all the constraints in (\ref{optimization}) are either independent of $\langle F_H^a,F_L^a\rangle$ or independent of $\langle F_H^b,F_L^b\rangle$.
First, for any $t<t^*$, the continuation constraint at time $t$ given by (\ref{implementable.1}) can be written as
 \begin{align*}
  &\mu_0\left(\int_t^{t^*}\bar{v}_H(t,s)dF_H(s)+\int_{t^*}^\infty\bar{v}_H(t,s)dF_H(s)\right)\\
  &+(1-\mu_0)\left(\int_t^{t^*}\bar{v}_L(t,s)dF_L(s)+\int_{t^*}^\infty\bar{v}_L(t,s)dF_L(s)\right)\geq 0.
 \end{align*}
 By Lemma \ref{lemma.useful}, 
 \begin{align*}
  &\mu_0\int_{t^*}^{\infty}\bar{v}_H(t,s)dF_H(s)+(1-\mu_0)\int_{t^*}^\infty\bar{v}_L(t,s)dF_L(s)\\
  =&\left(\mu_0\int_{t^*}^\infty\bar{v}_H(t,t^*)dF_H(s)+(1-\mu_0)\int_{t^*}^\infty\bar{v}_L(t,t^*)dF_L(s)\right)\\&+\gamma(t,t^*)\left(\mu_0\int_{t^*}^\infty\bar{v}_H(t^*,s)dF_H(s)+(1-\mu_0)\int_{t^*}^\infty\bar{v}_L(t^*,s)dF_L(s)\right)\\
  =&\mu_0\pi_H\bar{v}_H(t,t^*)+(1-\mu_0)\pi_L\bar{v}_L(t,t^*)\\
  &+\gamma(t,t^*)\left(\mu_0\pi_H\int_{t^*}^\infty\bar{v}_H(t^*,s)dF^a_H(s)+(1-\mu_0)\pi_L\int_{t^*}^\infty\bar{v}_L(t^*,s)dF^a_L(s)\right),\\
  =&\mu_0\pi_H\bar{v}_H(t,t^*)+(1-\mu_0)\pi_L\bar{v}_L(t,t^*)+\gamma(t,t^*)\big(\mu_0\pi_H+(1-\mu_0)\pi_L\big)\cdot u.
 \end{align*}
That is, the continuation constraint at time $t<t^*$ in (\ref{implementable.1}) is given by
\begin{equation}
\label{eq:decomposed-early-constraint}
\begin{aligned}
 &\mu_0\int_t^{t^*}\bar{v}_H(t,s)dF_H(s)+(1-\mu_0)\int_t^{t^*}\bar{v}_L(t,s)dF_L(s)+\mu_0\pi_H\bar{v}_H(t,t^*)\\&+(1-\mu_0)\pi_L\bar{v}_L(t,t^*)+\gamma(t,t^*)\big(\mu_0\pi_H+(1-\mu_0)\pi_L\big)\cdot u\geq 0,
\end{aligned}
\end{equation}
which is irrelevant to $\langle F_H^a,F_L^a\rangle$ when $u$ is fixed.
For $t\geq t^*$, the left-hand side of the continuation constraint at time $t$ in (\ref{implementable.1}) is given by
 \begin{align*}
  &\mu_0\int_{t}^\infty\bar{v}_H(t,s)d(F^a_H(s)\pi_H +(1-\pi_H))+(1-\mu_0)\int_{t}^\infty \bar{v}_L(t,s)d(F^a_L(s)\pi_L+(1-\pi_L))\\
  &=\mu_0\pi_H\int_{t}^\infty\bar{v}_H(t,s)dF_H^a(s)+(1-\mu_0)\pi_L\int_{t}^\infty\bar{v}_L(t,s)dF_L^a(s),
 \end{align*}
which is also irrelevant to $\langle F_H^b,F_L^b\rangle$.
 
For the stopping part, it is straightforward that the constraint (\ref{implementable.2}) can be written as
\begin{align*}
 \mu_0\pi_HdF_H^a(t)\cdot(1-\bar{\nu}(t))\leq (1-\mu_0)\pi_LdF_L^a(t)\cdot\bar{\nu}(t)
\end{align*}
if $t\geq t^*$, and
\begin{align*}
 \mu_0(1-\pi_H)dF_H^b(t)\cdot(1-\bar{\nu}(t))\leq (1-\mu_0)(1-\pi_L)dF_L^b(t)\cdot\bar{\nu}(t)
\end{align*}
if $t<t^*$, both of which are independent.
\end{proof}

\subsubsection*{Step 5: The optimal disclosure before $t^*$ is one-shot when $\mu_0<\mu^*$}

By Step 4, if $\langle F_H,F_L\rangle$ solves the optimization problem (\ref{optimization}), then $\langle F_H^b,F_L^b\rangle$ solves the subproblem (\ref{eq.before}).
Also, by Steps~2 and 3, we know that $F_H^b(t)\equiv0$ for every $t<t^*$.

It suffices to characterize $F_L$ before $t^*$, and since the belief is always updated to $0$ if the agent is recommended to stop, the incentive constraint of stopping can be ignored.
By the decomposition, $F_L$ is the solution to problem (\ref{eq.reduction}), where
\begin{align*}
 C(t)=-\frac{\gamma(t,t^*)\big(\mu_0\pi_H+(1-\mu_0)\pi_L\big)\cdot u+\mu_0\bar{v}_H(t,t^*)+(1-\mu_0)\pi_L\bar{v}_L(t,t^*)}{(1-\mu_0)(1-\pi_L)}.
\end{align*}
Here, 
\begin{align*}
 R(w,t)=\frac{p_0(\lambda+r)(\lambda Y-(\lambda+r)Z)-(1-p_0)r^2e^{\lambda t}Z}{p_0(\lambda Y-(\lambda+r)Z)-(1-p_0)re^{\lambda t}Z}.
\end{align*}
and
\begin{align*}
 R(v_L,t)=\frac{p_0(\lambda+r)(\lambda y_L-(\lambda+r)z)-(1-p_0)r^2e^{\lambda t}z}{p_0(\lambda y_L-(\lambda+r)z)-(1-p_0)re^{\lambda t}z}.
\end{align*}
Since $Y/Z>y_L/z$, we must have $R(v_L,t)<R(w,t)$ for all $t\in[\tau(\mu_0),t^*)$. 

By Lemma \ref{lemma.useful}
\begin{align*}
 \bar{v}_L(t,s)=\frac{1}{\gamma(0,t)}\cdot v(s)-\frac{v(t)}{\gamma(0,t)},
\end{align*}
which verifies Assumption \ref{ass:multi-separable} for this reduced single-state problem.
To verify Assumption \ref{ass:no-disclosure}, fix an interval on which no disclosure occurs.
Lemma \ref{lemma.useful} rewrites the continuation constraint at every date $t$ in this interval as
\begin{align*}
 &\mu_0\int_t^\infty v_H(s)dF_H(s)
 +(1-\mu_0)\int_t^\infty v_L(s)dF_L(s)\\
 &\qquad\geq
 \mu_0\bigl(1-F_H(t)\bigr)v_H(t)
 +(1-\mu_0)\bigl(1-F_L(t)\bigr)v_L(t).
\end{align*}
Because no stopping occurs within the interval, both integrals on the left-hand side and the survival probabilities $1-F_H(t)$ and $1-F_L(t)$ are constant.
Dividing the right-hand side by the total survival probability, $\mu_0\bigl(1-F_H(t)\bigr)+(1-\mu_0)\bigl(1-F_L(t)\bigr)$, gives $\mu v_H(t)+(1-\mu)v_L(t)$, where $\mu$ is the posterior conditional on reaching the interval.
This payoff is maximized at $\tau(\mu)$ and decreases thereafter.
Hence, on any no-disclosure interval $[\underline t,\bar t)$ with $\underline t\geq\tau(\mu)$, the continuation constraint is tightest at $\underline t$, which verifies Assumption \ref{ass:no-disclosure}.
By Step~3 and the definition of $F_L^b$, the distribution $F$ in \eqref{eq.reduction} satisfies $F(t)=0$ for all $t<\tau(\mu_0)$ and $F(t^*)=1$.
Hence $[\tau(\mu_0),t^*]$ is its final disclosure interval, and Lemma \ref{lem:final-interval} implies that $F_L$ assigns positive stopping probability to at most one date in $[\tau(\mu_0),t^*)$.

\subsubsection*{Step 6: Proof of Lemma \ref{lem.static.reduction}}

Let $F=\langle F_H,F_L\rangle$ be an optimal implementable policy, and let $F^a=\langle F_H^a,F_L^a\rangle$ be the conditional policy after $t^*$ obtained from the decomposition in Step 4.
Since $F^a$ is implementable, there exists a strategy of information disclosure after $t^*$, together with an incentive-compatible stopping rule, that induces $F_H^a$ and $F_L^a$.

By Bayesian updating, the belief process $\{\mu_t\}_{t\geq t^*}$ induced by any implementable continuation policy is a martingale.
Fix any such belief martingale that implements $F^a$.
Since the agent stops no later than $\tau(1)$ with probability $1$, define the terminal belief as $\mu_\infty\equiv\mu_{\tau(1)}$.

Because $\mu_\infty$ is the terminal value of the belief martingale, revealing it at $t^*$ and disclosing no further information Blackwell-dominates the original disclosure policy and therefore weakly increases the agent's continuation value at $t^*$.
Holding the policy on $[0,t^*)$ and the continuation probabilities $\pi_H$ and $\pi_L$ fixed, equation~\eqref{eq:decomposed-early-constraint} then implies that every continuation constraint at $t<t^*$ remains satisfied, while every stopping constraint at those dates is unchanged.

It remains to compare the principal's payoffs under the original and reduced policies.
The original continuation policy induces a distribution of terminal beliefs.
Conditional on any terminal belief $\mu_\infty=\mu$ in its support, let $\widetilde t$ denote the random stopping time under the original policy.
Under the reduced policy, the agent learns $\mu$ at $t^*$ and receives no further information, and hence he stops at $\max\{t^*,\tau(\mu)\}$.

We show that $\widetilde t\geq\max\{t^*,\tau(\mu)\}$.
Suppose instead that some $\widetilde t$ in the support of the conditional distribution satisfies $\widetilde t<\tau(\mu)$.
If the agent rejects the stopping recommendation at $\widetilde t$ and continues until $\tau(\mu)$, his payoff is
\begin{align*}
&\mu\bar v_H\bigl(\widetilde t,\tau(\mu)\bigr)+(1-\mu)\bar v_L\bigl(\widetilde t,\tau(\mu)\bigr)\\
&\quad=\frac{\mu\bigl[v_H(\tau(\mu))-v_H(\widetilde t)\bigr]+(1-\mu)\bigl[v_L(\tau(\mu))-v_L(\widetilde t)\bigr]}{\gamma(0,\widetilde t)}>0.
\end{align*}
The equality follows from Lemma~\ref{lemma.useful}, and the strict inequality follows because $\tau(\mu)$ uniquely maximizes $\mu v_H(t)+(1-\mu)v_L(t)$.
This profitable deviation contradicts obedience under the original policy.
Together with $\widetilde t\geq t^*$, this establishes that $\widetilde t\geq\max\{t^*,\tau(\mu)\}$ almost surely.
Because $w$ is strictly decreasing on $[t^*,\infty)$, if $\widetilde t>\max\{t^*,\tau(\mu)\}$ on a set of positive probability, the reduction would give the principal a strictly higher payoff, contradicting optimality.

\subsubsection*{Step 7: Solving Bayesian persuasion problem (\ref{eq.kg})}

We now solve the subproblem \eqref{eq.after}.
We write $V_{\text{NI}}(t^*,\mu)$ and $W_{\text{NI}}(t^*,\mu)$ as the agent's and the principal's indirect payoffs when (i) the agent's belief is $\mu$ at $t^*$ and (ii) no information is disclosed thereafter.
By Lemma \ref{lem.static.reduction}, the subproblem has an optimal implementation that reveals a static signal at $t^*$ and induces a distribution $P$ over posterior beliefs $\mu\in[0,1]$, subject to delivering continuation payoff $u$.
Thus, \eqref{eq.after} is equivalent to \eqref{eq.kg}, a conventional Bayesian persuasion problem with an additional participation constraint.

The solution to (\ref{eq.kg}) is determined by the curvature of $V_{\text{NI}}(t^*,\mu)$ and $W_{\text{NI}}(t^*,\mu)$.
Given posterior belief $\mu$, the agent is willing to continue on the risky arm if and only if $\mu\geq\mu^*$. 
Thus, by dynamic consistency, the optimal stopping time is $\tau(\mu)$ if $\mu>\mu^*$, and is $t^*$ if $\mu\leq\mu^*$.
Therefore, $V_{\text{NI}}(t^*,\mu)=0$ if $\mu\leq\mu^*$, and otherwise, it is equal to
 \begin{align*}
  &\mu \bar v_H\left(t^*,\tau(\mu)\right)+(1-\mu)\bar v_L\left(t^*,\tau(\mu)\right)\\
  &=\left(1-e^{-(\lambda+r)(\tau(\mu)-{t^*})}\right)\frac{p_{t^*}\lambda y(\mu)}{\lambda+r}+\left(1-p_{t^*}+p_{t^*}\cdot e^{-\lambda(\tau(\mu)-{t^*})}\right)\cdot e^{-r(\tau(\mu)-{t^*})}\cdot z-z.
 \end{align*}
Also, the principal's indirect payoff $W_{\text{NI}}(t^*,\mu)=Z$ if $\mu\leq\mu^*$, and otherwise, it is equal to
  \begin{align*}
  \left(1-e^{-(\lambda+r)(\tau(\mu)-{t^*})}\right)\frac{p_{t^*}\lambda Y}{\lambda+r}+\left(1-p_{t^*}+p_{t^*}\cdot e^{-\lambda(\tau(\mu)-{t^*})}\right)\cdot e^{-r(\tau(\mu)-{t^*})}\cdot Z.
 \end{align*}

By \citet[Theorems 3.2 and 3.3]{doval2024constrained}, the optimal signal $P$ maximizes the Lagrangian
 \begin{align*}
     \int_0^1\bigg(W_{\text{NI}}(t^*,\mu)+\kappa\big(V_{\text{NI}}(t^*,\mu)-u\big)\bigg)P(d\mu),
 \end{align*}
for some Lagrangian multiplier $\kappa$. 
When $\kappa=0$, the Lagrangian objective coincides with the unconstrained static-persuasion objective, while constraint (U) in \eqref{eq.kg} still requires the signal to deliver continuation payoff $u$.
Define $\mathcal{U}(\mu)\equiv W_{\text{NI}}(t^*,\mu)+\kappa V_{\text{NI}}(t^*,\mu)$, which equals $Z$ when $\mu\leq\mu^*$, and otherwise,
 \begin{align*}
  \mathcal{U}(\mu)=&\left(1-e^{-(\lambda+r)(\tau(\mu)-{t^*})}\right)\frac{p_{t^*}\lambda(Y+\kappa y(\mu))}{\lambda+r}\\
  &\quad+\left(1-p_{t^*}+p_{t^*}\cdot e^{-\lambda(\tau(\mu)-{t^*})}\right)\cdot e^{-r(\tau(\mu)-{t^*})}\cdot(Z+\kappa z)-\kappa z.
 \end{align*}

Using $\tau'(\mu)=y'(\mu)/(\lambda y(\mu)-(\lambda+r)z)$ and $\tau''(\mu)=-\lambda y'(\mu)^2/(\lambda y(\mu)-(\lambda+r)z)^2$, it is straightforward to calculate that when $\mu>\mu^*$, $\mathcal{U}''(\mu)>0$ if and only if 
\begin{equation}
\label{eq:lagrangian-curvature-threshold}
  \mu\geq y^{-1}\left[\frac{z((2\lambda+r)Y-(\lambda+r)(Z-\kappa z))}{(r+\lambda)Z+\kappa\lambda z}\right].
\end{equation}
 
Note that $\mathcal{U}'(\mu^*)=0$; therefore, there are two possible shapes of function $\mathcal{U}(\mu)$. 
Figure \ref{Figure.concavification} illustrates the two cases.
First, if $\kappa\geq Z/z$, the function $\mathcal{U}(\mu)$ is increasing and convex for all $\mu\in[0,1]$.
In this case, the concavification of function $\mathcal{U}(\mu)$ is a straight line connecting points $(0,\mathcal{U}(0))$ and $(1,\mathcal{U}(1))$; that is, the optimal signal is full disclosure at $t^*$. 
Second, if $\kappa<Z/z$, the right-hand side of \eqref{eq:lagrangian-curvature-threshold} exceeds $\mu^*$.
Thus, $\mathcal{U}(\mu)$ is concave and decreasing near $\mu^*$, and becomes convex once $\mu$ exceeds this threshold.
In this case, full disclosure can also be optimal,\footnote{This occurs when $\mathcal{U}(1)\geq Z$, in which case the line connecting the two endpoints lies weakly above $\mathcal{U}$.} and if not, the optimal signal is non-disclosure when $\mu_{t^*}$ is small and provides perfect good news when $\mu_{t^*}$ is large.
In either case, the optimal signal contains at most two posterior beliefs.

\begin{figure}[ht!]
  \centering 
  \begin{subfigure}[ht]{.45\textwidth}
    \centering
    \begin{tikzpicture}
      \draw[thick,->](0,0)--(5.5,0) node [below] {$\mu$};
      \draw[thick,->](0,0)--(0,5) node [right] {$\mathcal{U}(\mu)$};
      \draw[domain=0:2] plot(\x,{1});
      \draw[domain=2:5] plot(\x, {\x^2/3-4*\x/3+7/3});
      \draw[dashed] (0,1) -- (5,4);
      \draw[dashed] (5,4.5) -- (5,0) node [below] {$1$};
      \draw[dashed] (2,4.5) -- (2,0) node [below] {$\mu^*$};
      \draw[dashed] (0,4.5) -- (5,4.5);
    \end{tikzpicture}
    \caption{$\kappa\ge Z/z$: Full disclosure is optimal}
  \end{subfigure}
  \hfill
  \begin{subfigure}[ht]{.45\textwidth}
    \centering
    \begin{tikzpicture}
      \draw[thick,->](0,0)--(5.5,0) node [below] {$\mu$};
      \draw[thick,->](0,0)--(0,5) node [right] {$\mathcal{U}(\mu)$};
      \draw (0,4)--(1,4);
      \draw[domain=1:3] plot(\x,-\x^2/2+\x+7/2);
      \draw[domain=3:5] plot(\x, {\x^2/2-5*\x+25/2} );
      \draw[domain=0:3,dashed] (5,0) -- (1.15901,4.5);
      \node[circle,fill=black,inner sep=0pt,minimum size=3pt] at (2.1715728752538097,3.31371) {};
      \draw[dashed] (5,4.5) -- (5,0) node [below] {$1$};
      \draw[dashed] (2.1715728752538097,4.5) -- (2.1715728752538097,0) node [below] {$\mu_a$};
      \draw[dashed] (1,4.5) -- (1,0) node [below] {$\mu^*$};
      \draw[dashed] (0,4.5) -- (5,4.5);
    \end{tikzpicture}
    \caption{$\kappa < Z/z$: Perfect good news is optimal}
  \end{subfigure}
  
  \caption{The potential optimal structures.}
  \label{Figure.concavification}
\end{figure}

\subsubsection*{Step 8: Integrating the solutions to the two subproblems}

Steps 5 and 7 leave two possible forms for the optimal policy.
Under full disclosure in the dissuasion subproblem, the agent stops at $t^*$ whenever $\theta=L$ and he has not stopped earlier.
Under the signal inducing posterior beliefs $\mu_a$ and $1$, he instead stops at $t_a>t^*$ when $\theta=L$ and he has not stopped earlier.

First, suppose that the static signal in the dissuasion subproblem is full disclosure.
Combining this signal with the solution to the motivation subproblem gives
\begin{align*}
F_H(t)
&=
\begin{cases}
0, & t<t^*,\\
1-\pi_H, & t^*\leq t<\tau(1),\\
1, & \tau(1)\leq t,
\end{cases}
&
F_L(t)
&=
\begin{cases}
0, & t<t_b,\\
x_b, & t_b\leq t<t^*,\\
1, & t^*\leq t.
\end{cases}
\end{align*}

Second, suppose that the static signal induces posterior beliefs $\mu_a$ and $1$.
The lower contact point satisfies $\mu_a>\mu^*$, and hence $t_a=\tau(\mu_a)>t^*$.
In this case, the combined policy is
\begin{align*}
F_H(t)
&=
\begin{cases}
0, & t<t^*,\\
1-\pi_H, & t^*\leq t<t_a,\\
x_a, & t_a\leq t<\tau(1),\\
1, & \tau(1)\leq t,
\end{cases}
&
F_L(t)
&=
\begin{cases}
0, & t<t_b,\\
x_b, & t_b\leq t<t^*,\\
1-\pi_L, & t^*\leq t<t_a,\\
1, & t_a\leq t.
\end{cases}
\end{align*}

We now show that the agent does not stop at $t^*$ in either state.
We first consider the high state.
Suppose that the principal recommends stopping at $t^*$ with positive probability when $\theta=H$.
Obedience requires her to send the same recommendation with positive probability when $\theta=L$, and thus the recommendation induces a posterior belief $\nu_{t^*}$ satisfying $0<\nu_{t^*}\leq\mu^*$.
In either case, this recommendation and the disclosure in the dissuasion subproblem form a static disclosure with three posterior beliefs ($\nu_{t^*}$, $0$, and $1$ under full disclosure, and $\nu_{t^*}$, $\mu_a$, and $1$ under the signal inducing posterior beliefs $\mu_a$ and $1$), whereas Step 7 shows that an optimal static signal contains at most two posterior beliefs.
Hence the agent does not stop at $t^*$ in the high state, and $\pi_H=1$.

Given $\pi_H=1$, a recommendation to stop at $t^*$ when $\theta=L$ induces posterior belief $0$.
If the signal in the dissuasion subproblem induces posterior beliefs $\mu_a$ and $1$, combining that signal with this recommendation produces three posterior beliefs, $0$, $\mu_a$, and $1$, contradicting Step 7.
Thus the agent can stop at $t^*$ when $\theta=L$ only if the signal in the dissuasion subproblem is full disclosure.
If $x_b>0$, the agent stops with positive probability at both $t_b$ and $t^*$ when $\theta=L$, and the contraction used in Step 5 combines these probabilities and strictly increases the principal's payoff.
It remains to rule out full disclosure at $t^*$ when $x_b=0$.

Under this policy, the agent stops at $t^*$ when $\theta=L$ and at $\tau(1)$ when $\theta=H$.
For a sufficiently small $\epsilon>0$, move all the stopping probability conditional on $\theta=L$ from $t^*$ to $t^*-\epsilon$.
Conditional on $\theta=H$, move
\begin{align*}
\frac{1-\mu_0}{\mu_0}
\frac{v_L(t^*-\epsilon)-v_L(t^*)}{v_H(\tau(1))-v_H(t^*-\epsilon)}
\end{align*}
of the stopping probability from $\tau(1)$ to $t^*-\epsilon$.
The agent's expected payoff under the new policy is
\begin{align*}
&(1-\mu_0)v_L(t^*-\epsilon)
+\mu_0\frac{1-\mu_0}{\mu_0}
\frac{v_L(t^*-\epsilon)-v_L(t^*)}{v_H(\tau(1))-v_H(t^*-\epsilon)}v_H(t^*-\epsilon)\\
&\quad+\mu_0\left[1-\frac{1-\mu_0}{\mu_0}
\frac{v_L(t^*-\epsilon)-v_L(t^*)}{v_H(\tau(1))-v_H(t^*-\epsilon)}\right]v_H(\tau(1))\\
&=(1-\mu_0)v_L(t^*-\epsilon)+\mu_0v_H(\tau(1))\\
&\quad-(1-\mu_0)
\frac{v_L(t^*-\epsilon)-v_L(t^*)}{v_H(\tau(1))-v_H(t^*-\epsilon)}
\bigl[v_H(\tau(1))-v_H(t^*-\epsilon)\bigr]\\
&=(1-\mu_0)v_L(t^*)+\mu_0v_H(\tau(1)),
\end{align*}
which is exactly his expected payoff under the original policy.
The recommendation at $t^*-\epsilon$ pools the low-state stopping probability moved from $t^*$ with the high-state stopping probability moved from $\tau(1)$, while no recommendation reveals that $\theta=H$ and leads the agent to continue until $\tau(1)$; since this deviation preserves his expected payoff, Lemma~\ref{lemma.useful} ensures that every continuation constraint for $t<t^*-\epsilon$ remains satisfied. 

For the stopping constraint at $t^*-\epsilon$, Bayes' rule and the state-contingent stopping probabilities moved to this date give
\begin{align*}
\nu_{t^*-\epsilon}
=
\frac{v_L(t^*-\epsilon)-v_L(t^*)}
{v_H(\tau(1))-v_H(t^*-\epsilon)
 +v_L(t^*-\epsilon)-v_L(t^*)}.
\end{align*}
As $\epsilon\downarrow0$, $\nu_{t^*-\epsilon}\to0$.
Since $\tau(0)<t^*$, continuity implies $\tau(\nu_{t^*-\epsilon})<t^*-\epsilon$ for sufficiently small $\epsilon>0$.
By Lemma~\ref{lemma.useful}, immediate stopping is therefore optimal, and thus $V_{\mathrm{NI}}(t^*-\epsilon,\nu_{t^*-\epsilon})=0$, satisfying the stopping constraint.

Having established that the perturbed policy is implementable, we now compare the principal's payoff.
The change in her payoff, divided by the positive factor $1-\mu_0$, is
\begin{align*}
w(t^*-\epsilon)-w(t^*)
+\frac{v_L(t^*-\epsilon)-v_L(t^*)}{v_H(\tau(1))-v_H(t^*-\epsilon)}
\bigl[w(t^*-\epsilon)-w(\tau(1))\bigr].
\end{align*}
This expression equals zero at $\epsilon=0$, and its derivative with respect to $\epsilon$ at zero is
\begin{align*}
\frac{-v_L'(t^*)\bigl[w(t^*)-w(\tau(1))\bigr]}
{v_H(\tau(1))-v_H(t^*)}>0.
\end{align*}
Choose $\epsilon>0$ sufficiently small that the required high-state stopping probability is available and the derivative of the displayed payoff change remains strictly positive between zero and $\epsilon$.
The mean value theorem then implies that the displayed payoff change is strictly positive.
Thus the perturbed policy strictly improves the principal's payoff, contradicting optimality.

Thus the agent does not stop at $t^*$ in either state.
Full disclosure in the dissuasion subproblem is ruled out, while the signal inducing posterior beliefs $\mu_a$ and $1$ satisfies $\pi_H=1$, $\pi_L=1-x_b$, and $t_a>t^*$.

\subsubsection*{Step 9: Optimal disclosure when $\mu_0<\mu^*$}

So far, we have already identified that the information policy discloses information at most twice.
\begin{itemize}
 \item Before $t^*$, the principal selects a date $t_b$ at which she recommends that the agent stop with probability $x_b$ when $\theta=L$.
 No recommendation is made when $\theta=H$, and if the agent does not receive this recommendation, he updates his belief to
 \begin{align*}
  \mu_b=\frac{\mu_0}{\mu_0+(1-\mu_0)(1-x_b)}\geq\mu_0.
 \end{align*}
 \item After $t^*$, the principal selects a date $t_a$ at which she recommends that the agent stop with probability $x_a$ when $\theta=H$ and with probability $1$ when $\theta=L$.
 Therefore, if the agent is not recommended to stop, he updates his belief to $1$.
\end{itemize}
By this policy, the principal's payoff is given by
\begin{align*}
 \mu_0\bigg(x_aw(t_a)+(1-x_a)w(\tau(1))\bigg)+(1-\mu_0)\bigg(x_bw(t_b)+(1-x_b)w(t_a)\bigg).
\end{align*}

We now identify which continuation and stopping constraints bind. 
First, $\mathcal{S}(\mathcal{P})\subseteq\{t_b,t_a,\tau(1)\}$, and any remaining stopping probability in state $H$ is assigned to $\tau(1)$. 
When the posterior is one, this is the agent's optimal stopping date without further information, and hence the stopping constraint is satisfied.
Since $t_b\geq\tau(\mu_0)\geq\tau(0)$ and $\nu_{t_b}=0$, the stopping constraint at time $t_b$ is satisfied automatically.
Meanwhile, since information policy $\langle F_H^a,F_L^a\rangle$ can be implemented by a static information structure, there is no information between $t^*$ and $t_a$. 
Thus, the stopping constraint at time $t_a$ must be binding at $t_a$, since otherwise the agent would have stopped earlier when he received the bad message from the static information structure.

The following lemma locates the tightest continuation constraint on an interval containing a single disclosure date.

\begin{A.lemma}
\label{lemma.iambinding}
Suppose that under information policy $\mathcal{P}=\langle F_H,F_L\rangle$, $t_2$ is the only disclosure date in $[t_1,t_3)$, where $t_1<t_2<t_3$.
The belief therefore remains constant at $\mu$ on $[t_1,t_2)$.
 Then the continuation constraints at time $t\in[t_1,t_2)$ are all satisfied if and only if they are satisfied at time
 \begin{align*}
  \tau^*=\operatorname*{arg\,max}_{t\in[t_1,t_2]}\left\{\mu v_H(t)+(1-\mu)v_L(t)\right\}.
 \end{align*}
 \footnote{If $\tau^*=t_2$, the constraint at $\tau^*$ is understood as the limiting constraint as $t\uparrow t_2$.}
\end{A.lemma}

That is, in any time interval $I$, if there is a one-shot information disclosure ahead of $I$, it is sufficient to guarantee that the incentive constraint is satisfied at the voluntary stopping time in the absence of future information, and all the other continuation constraints in $I$ are automatically satisfied.
In particular, the tightest continuation constraint in $I$ is the one at $\tau(\mu)$ if $\tau(\mu)\in I$, and the one at the nearest endpoint of $I$ otherwise.

\begin{proof}
 For any $t\in[t_1,t_2)$, by Lemma \ref{lemma.useful}, the continuation constraint at time $t$ can be written as
 \begin{align*}
  \mu_0\int_{t}^\infty\bar{v}_H(0,s)dF_H(s)&+(1-\mu_0)\int_{t}^\infty\bar{v}_L(0,s)dF_L(s)\\&\geq\mu_0(1-F_H(t))v_H(t)+(1-\mu_0)(1-F_L(t))v_L(t).
 \end{align*}
Since $t_2$ is the only disclosure date in $[t_1,t_3)$, the left-hand side can be rewritten as
 \begin{align*}
 \mu_0&dF_H(t_2)v_H(t_2)+(1-\mu_0)dF_L(t_2)v_L(t_2)\\
 &+\mu_0\int_{t_3}^\infty\bar{v}_H(0,s)dF_H(s)+(1-\mu_0)\int_{t_3}^\infty\bar{v}_L(0,s)dF_L(s),
 \end{align*}
 which is irrelevant to $t$.
Thus, the continuation constraints in $I$ are all satisfied if and only if they are satisfied where the right-hand side is maximized.
Finally, in the interval $[t_1,t_2)$, we have
 \begin{align*}
  \mu_t=\frac{\mu_0(1-F_H(t))}{\mu_0(1-F_H(t))+(1-\mu_0)(1-F_L(t))}\equiv\mu,
 \end{align*}
 and thus maximizing the right-hand side is equivalent to maximizing $\mu v_H(t)+(1-\mu)v_L(t)$, which completes the proof.
\end{proof}

Therefore, when $t<t_b$, $\tau^*=\tau(\mu_0)$, and the continuation constraints at all $t<t_b$ are satisfied if and only if the continuation constraint at time $\tau(\mu_0)$ is satisfied.
Also, there is no further disclosure after $t_a$, and it remains to specify the continuation conditions in $[t_b,t_a)$. 

When $t\in[t_b,t_a)$, the continuation constraints are given by
\begin{align*}
\mu_0\bigg(x_a\bar{v}_H(t,t_a)+(1-x_a)\bar{v}_H(t,\tau(1))\bigg)+(1-\mu_0)(1-x_b)\bar{v}_L(t,t_a)\geq 0,\quad\forall t\in[t_b,t_a).
\end{align*}
By the proof of Lemma \ref{lemma.iambinding}, there exists a time point $t'\in[t_b,t_a]$, such that all these constraints hold if and only if
\begin{equation}
\label{eq.t'}
\mu_0\bigg(x_a\bar{v}_H(t',t_a)+(1-x_a)\bar{v}_H(t',\tau(1))\bigg)+(1-\mu_0)(1-x_b)\bar v_L(t',t_a)\geq 0.
\end{equation}
Since the policy is implementable, (\ref{eq.t'}) holds.

We now prove by contradiction that the continuation constraint at $\tau(\mu_0)$ must bind.
Suppose instead that it is slack, and consider the alternative policy $\langle x_b,x_a,t_b+\varepsilon,t_a\rangle$ for sufficiently small $\varepsilon>0$.
Since all the constraints in $[t_b+\varepsilon,t_a)$ remain unchanged when $t_b$ is postponed, it is straightforward that all the constraints in $[t_b+\varepsilon,t_a)$ still hold. 
Also, the constraints in $[\tau(\mu_0),t_b+\varepsilon)$ hold when $\varepsilon$ is sufficiently small.
Obviously, this makes the principal better off but is still implementable, which violates its optimality.
Thus, the continuation constraint at $\tau(\mu_0)$ must be binding, which completes the proof.

\subsubsection*{Step 10: Optimal disclosure when $\mu_0\geq\mu^*$}

When $\mu_0\geq\mu^*$, Step~3 rules out disclosure at every $t<t^*$.
Meanwhile, Lemma~\ref{lem.static.reduction} shows that all disclosure scheduled at $t>t^*$ can be implemented by a static signal at $t^*$. 
Combining this signal with any signal already sent at $t^*$, the entire information policy can be represented by a single static signal at $t^*$. 
Since no information has previously been disclosed, its prior belief is $\mu_0$, and the dynamic problem therefore reduces to a static Bayesian persuasion problem at $t^*$.

Because $\mu_0\geq\mu^*$, the agent voluntarily experiments until $t^*$, and thus constraint~(U) does not affect the principal's choice of the static signal.
Thus, it remains to show that the unconstrained optimum at time $t^*$ can be implemented by the optimal static signal at the outset.
For $\mu\leq\mu^*$, $W_{\mathrm{NI}}(t^*,\mu)=Z$, and hence posteriors below $\mu^*$ cannot improve the concavification at any prior $\mu_0\geq\mu^*$. 
On $[\mu^*,1]$, Lemma~\ref{lemma.useful} gives
\[
W_{\mathrm{NI}}(t^*,\mu)-W_{\mathrm{NI}}(t^*,\widehat\mu)
=
\frac{W_{\mathrm{NI}}(\mu)-W_{\mathrm{NI}}(\widehat\mu)}{\gamma(0,t^*)}.
\]
Thus the concavification of $W_{\mathrm{NI}}(t^*,\cdot)$ has the same contact point $\mu_H$ and the same optimal signal as the concavification of $W_{\mathrm{NI}}$ in Proposition~\ref{proposition.static}. 
Sending this signal at $t^*$ induces the same stopping dates and principal payoff as sending it at the outset. 

\subsubsection*{Step 11: Uniqueness}

If $\mu_0\geq\mu^*$, then $\tau(\mu_0)\geq t^*$, and thus Steps~2 and 3 give $x_b=0$, and Proposition~\ref{proposition.static} uniquely determines $F_H$ and $F_L$.
We therefore consider $\mu_0<\mu^*$.
We first establish uniqueness among optimal policies identified by a four-tuple $\langle x_a,x_b,t_a,t_b\rangle$, where $x_a,x_b\in[0,1]$, $t_a\in(t^*,\tau(1)]$, and $t_b\in[\tau(\mu_0),t^*)$.
A date assigned zero stopping probability does not affect either $F_H$ or $F_L$.

Because the participation constraint binds, every candidate optimum delivers the same payoff $V_{\mathrm{NI}}(\mu_0)$ to the agent.
Suppose that two multipliers support two candidate policies.
Adding the two inequalities that define their respective Lagrangian maximizers shows that each policy also maximizes the Lagrangian under the other multiplier.
It therefore suffices to prove that, for any fixed multiplier, the dates and probabilities are unique.

We first show that the two dates are unique.
Let $\kappa$ be the multiplier on the binding participation constraint.
At an interior solution, $t_b$ satisfies $w'(t_b)+\kappa v_L'(t_b)=0$.
Since
\begin{align*}
\frac{d}{dt}\log\!\left(-\frac{w'(t)}{v_L'(t)}\right)=R(v_L,t)-R(w,t)<0,
\end{align*}
$t_b$ is unique for any $\kappa$.
Next let $\mathcal U_\kappa(\mu)=W_{\mathrm{NI}}(\mu)+\kappa V_{\mathrm{NI}}(\mu)$.
The curvature calculation in Step~7 implies that the tangent to $\mathcal U_\kappa$ passing through $(1,\mathcal U_\kappa(1))$ has at most one interior contact point $\mu_a$, and hence $t_a=\tau(\mu_a)$ is also unique for any $\kappa$.
Thus, for any fixed $\kappa$, the two dates are unique at an interior solution.
If $x_b=1$, the binding participation constraint and the strict decrease of $v_L$ uniquely determine $t_b$.

It remains to consider the boundary $x_a=1$.
Under this policy, the agent stops at $t_a=\tau(\mu_b)$ unless he has already stopped at $t_b$.
Step~8 gives $t_a>t^*$, and thus $\mu_b\in(\mu^*,1)$.
For a fixed $\kappa$, $t_b$ uniquely maximizes $w(t_b)+\kappa v_L(t_b)$ by the monotonicity established above.
After choosing $t_b$, maximizing the Lagrangian with respect to $\mu_b$ is equivalent to maximizing $[\mathcal U_{\,\kappa}(\mu_b)-w(t_b)-\kappa v_L(t_b)]/\mu_b$.
The numerator of its derivative is $\mu_b(\mathcal U_{\,\kappa})'(\mu_b)-\mathcal U_{\,\kappa}(\mu_b)+w(t_b)+\kappa v_L(t_b)$, whose derivative with respect to $\mu_b$ is $\mu_b(\mathcal U_{\,\kappa})''(\mu_b)$.
Step~7 shows that $(\mathcal U_{\,\kappa})''$ is either positive throughout or changes sign once from negative to positive.
Hence the derivative numerator is either increasing throughout or first decreases and then increases, and therefore the ratio has at most one interior maximizer.
Thus $\mu_b$ and $t_a=\tau(\mu_b)$ are unique for every $\kappa$.

It remains to determine the probabilities.
When both dates carry positive stopping probability, the binding stopping and participation constraints form two linear equations in $x_a$ and $1-x_b$.
This linear system has a unique solution because
\begin{align*}
v_H'(t_a)\big[v_L(t_a)-v_L(t_b)\big]
-v_L'(t_a)\big[v_H(t_a)-v_H(\tau(1))\big]<0.
\end{align*}
The strict inequality follows because $v_H'(t_a)>0$, $v_L'(t_a)<0$, $v_L(t_a)<v_L(t_b)$, and $v_H(t_a)<v_H(\tau(1))$.
At a boundary, the remaining probability is pinned down by the binding participation constraint or Bayes' rule.
All optimal policies identified by such four-tuples therefore induce the same $F_H$ and $F_L$.

By the proof of Lemma~\ref{lem.static.reduction} and Steps~7 and 8, every optimal policy induces the same $F_H$ and $F_L$ as an optimal policy identified by such a four-tuple.
Hence every optimal implementation induces the same pair $\langle F_H^*,F_L^*\rangle$.

\subsection{Proof of Proposition \ref{prop.comparative}}

\subsubsection*{Step 1: Comparative statics when $\mu_0\geq \mu^*$}
If $\mu_0\geq\mu^*$, then $\tau(\mu_0)\geq t^*$, and Steps~2 and 3 of the proof of Theorem~\ref{thetheorem} imply $x_b=0$.
The remaining optimization is the static persuasion problem characterized in Proposition~\ref{proposition.static}, which establishes the result when $\mu_0\geq\mu^*$.

\subsubsection*{Step 2: Comparative statics in the interior regime}

We now consider $\mu_0<\mu^*$. 
Whenever $x_b>0$, we have $t_b>\tau(\mu_0)$. 
To see this, note that if equality held, receiving no stopping recommendation at $t_b=\tau(\mu_0)$ would give the agent a strictly positive continuation surplus, contradicting the binding participation constraint.

We first analyze the interior regime, in which none of the boundary constraints on $x_a$, $x_b$, $t_a$, and $t_b$ bind.
Both constraints in Theorem \ref{thetheorem} are then binding and give two linear equations in $x_a$ and $x_b$.
Solving these two equations, we obtain
\begin{align*}
  x_b=\frac{(1-\mu_0)\big(v_H(\tau(1))-v_H(t_a)\big) v_L'(t_a)+v_H'(t_a) \big(\mu_0  v_H(\tau(1))+(1-\mu_0)v_L(t_a)-V_{\text{NI}}(\mu_0)\big)}{(1-\mu_0) \big((v_H(\tau(1))-v_H(t_a)) v_L'(t_a)-(v_L(t_b)-v_L(t_a)) v_H'(t_a)\big)},
\end{align*}
and
\begin{align*}
  x_a=\frac{v_L'(t_a)\big(\mu_0v_H(\tau(1))+(1-\mu_0) v_L(t_b)-V_{\text{NI}}(\mu_0)\big)}{\mu_0  \big((v_H(\tau(1))-v_H(t_a)) v_L'(t_a)-(v_L(t_b)-v_L(t_a)) v_H'(t_a)\big)}.
\end{align*}

Substituting these two expressions into the principal's payoff and taking the first-order condition with respect to $t_b$, 
 we obtain
\begin{align*}
 &\bigg[v_L'(t_b)\big((w(\tau(1))-w(t_a))v_L'(t_a)+(w(t_a)-w(t_b))v_H'(t_a)\big)\\
 &-\bigg(w'(t_b)\big((v_H(\tau(1))-v_H(t_a))v_L'(t_a)+(v_L(t_a)-v_L(t_b))v_H'(t_a)\big)\bigg)\bigg]\cdot\\
 &\underbrace{\frac{(1-\mu_0)\big(v_H(\tau(1))-v_H(t_a)\big) v_L'(t_a)+v_H'(t_a) \big(\mu_0  v_H(\tau(1))+(1-\mu_0)v_L(t_a)-V_{\text{NI}}(\mu_0)\big)}{\left(\big(v_H(t_a)-v_H(\tau(1))\big)v_L'(t_a)+\big(v_L(t_b)-v_L(t_a)\big)v_H'(t_a)\right)^2}}_{\text{is equal to }0\text{ if and only if }x_b=0}=0,
\end{align*} 
Because $x_b>0$, the underbraced fraction is nonzero, and thus the bracketed expression in the first-order condition for $t_b$ must equal zero.
This equation depends only on $t_a$ and $t_b$.
Substituting the same expressions for $x_a$ and $x_b$ and differentiating with respect to $t_a$ gives
\begin{align*}
&\underbrace{\big[\mu_0v_H(\tau(1))+(1-\mu_0)v_L(t_b)-V_{\text{NI}}(\mu_0)\big]}_{\text{is equal to }0\text{ if and only if }x_a=0}\\
&\qquad\times\frac{d}{dt_a}\left[
\frac{\big(w(t_b)-w(t_a)\big)v_H'(t_a)+\big(w(t_a)-w(\tau(1))\big)v_L'(t_a)}
{\big(v_H(\tau(1))-v_H(t_a)\big)v_L'(t_a)+\big(v_L(t_a)-v_L(t_b)\big)v_H'(t_a)}
\right]=0.
\end{align*}
Because $x_a>0$, the underbraced factor is nonzero, and hence the derivative in this expression must equal zero. 
This gives a second equation involving only $t_a$ and $t_b$. 
For every $\kappa$, the strict decrease of $-w'/v_L'$ determines a unique $t_b$, while Step~7 establishes a unique interior contact point and hence a unique $t_a$. 
After substituting these dates, the remaining first-order condition is strictly decreasing in $\kappa$, and therefore the two equations have a unique interior solution. 
Since neither equation depends on $\mu_0$, the resulting pair $(t_a,t_b)$ is independent of $\mu_0$ throughout the interior regime.

Then taking the derivative of $x_b$ with respect to $\mu_0$, we obtain
\begin{align*}
 \frac{dx_b}{d\mu_0}=-\frac{(v_H(\tau(1))-v_H(\tau(\mu_0))) v_H'(t_a)}{(1-\mu_0 )^2 \left(v_L'(t_a)(v_H(t_a)-v_H(\tau(1))) +(v_L(t_b)-v_L(t_a)) v_H'(t_a)\right)},
\end{align*}
which is negative because $v_H(\tau(\mu_0))<v_H(\tau(1))$, $v_H'(t_a)>0$, $v_L'(t_a)<0$, $v_H(t_a)<v_H(\tau(1))$, and $v_L(t_b)>v_L(t_a)$.
Similarly, taking the derivative of $x_a$ with respect to $\mu_0$, we obtain
\begin{align*}
 \frac{dx_a}{d\mu_0}=\frac{(v_L(\tau(\mu_0))-v_L(t_b)) v_L'(t_a)}{\mu_0 ^2 \left((v_H(\tau(1))-v_H(t_a))v_L'(t_a)-(v_L(t_b)-v_L(t_a)) v_H'(t_a)\right)},
\end{align*}
which is positive because $v_L(\tau(\mu_0))>v_L(t_b)$, $v_L'(t_a)<0$, and the denominator is negative.

\subsubsection*{Step 3: Establishing the existence of thresholds $\mu_l$ and $\mu_h$}
The interior regime is optimal if and only if the combination $(t_a,t_b,x_a,x_b)$ satisfies the four border constraints strictly.
When they are not, one (and only one) of $x_b=0$, $x_a=1$ and $x_b=1$, which implies $x_a=0$ by (\ref{eq.constraint}), holds. 
Now we exclude the possibility that $x_b=0$. 
Indeed, by (\ref{eq.constraint}), it suffices to show that 
\begin{align*}
  -\frac{1-\mu_0}{\mu_0}\cdot\frac{v'_L(t_a)}{v'_H(t_a)}=-\frac{1-\mu_0}{\mu_0}\cdot\frac{(1-p_0)rz e^{\lambda t_a}+p_0(\lambda+r)z-p_0\lambda y_L}{(1-p_0) r ze^{\lambda  t_a}+p_0 (r+\lambda) z-p_0\lambda y_H}
  >1.
\end{align*}
This inequality holds if and only if $t_a\geq\tau(\mu_0)$, which is guaranteed by $\mu_0<\mu^*$.
Consequently, by the interiority of $t_a$ and $t_b$ and the monotonicity of $x_b$ and $x_a$, it fails by either $x_a=1$, which takes place when $\mu_0$ is sufficiently large, or $x_b=1$, which takes place when $\mu_0$ is sufficiently small.
Since $x_b$ is strictly decreasing and $x_a$ is strictly increasing in $\mu_0$, each of these constraints can bind at most once.
The priors for which the interior solution is feasible therefore form an interval, and continuity establishes its endpoints $\mu_l$ and $\mu_h$, with $x_b=1$ at the lower endpoint and $x_a=1$ at the upper endpoint.

\subsubsection*{Step 4: Comparative statics when  $x_a=1$}
When $x_a=1$, the optimization problem can be written as
\begin{align*}
  \max_{x_b,t_b}:&\mu_0 w(\tau(\mu_b(x_b)))+(1-\mu_0)\bigg(x_bw(t_b)+(1-x_b)w(\tau(\mu_b(x_b)))\bigg)\\
  &\text{s.t.: }\mu_0 v_H(\tau(\mu_b(x_b)))+(1-\mu_0)\bigg(x_bv_L(t_b)+(1-x_b)v_L(\tau(\mu_b(x_b)))\bigg)\geq V_{\text{NI}}(\mu_0),
\end{align*}
where $\mu_b(x_b)=\mu_0/(\mu_0+(1-\mu_0)(1-x_b))$.
Using $\mu_b$ as the choice variable in place of $x_b$, we can rewrite the problem as
\begin{align*}
  \max_{\mu_b\in[\mu_0,1],t_b\in(0,t^*)}:&\frac{\mu_0}{\mu_b} w(\tau(\mu_b))+\left(1-\frac{\mu_0}{\mu_b}\right)w(t_b)\\
  &\text{s.t.: }\frac{\mu_0}{\mu_b} V_{\text{NI}}(\mu_b)+\left(1-\frac{\mu_0}{\mu_b}\right)v_L(t_b)\geq V_{\text{NI}}(\mu_0).
\end{align*}
Let $\kappa \ge 0$ denote the Lagrange multiplier associated with this constraint. 
The Lagrangian $\mathcal{L}$ for the principal's problem is given by:
\begin{align*}
    \mathcal{L} = \frac{\mu_0}{\mu_b} \bigg[ w(\tau(\mu_b)) + \kappa V_{\text{NI}}(\mu_b) \bigg] + \left(1 - \frac{\mu_0}{\mu_b}\right) \bigg[ w(t_b) + \kappa v_L(t_b) \bigg] - \kappa V_{\text{NI}}(\mu_0).
\end{align*}

Now we show that in this regime, $t_b$ increases with $\mu_0$. 
Differentiating the first-order condition
$w'(t_b)+\kappa v_L'(t_b)=0$ with respect to $\kappa$ yields
\begin{align*}
  \frac{dt_b}{d\kappa}=-\frac{v'_L(t_b)}{w''(t_b)+\kappa v''_L(t_b)}.
\end{align*}
By the optimality of $t_b$, the denominator is negative, and thus $dt_b/d\kappa<0$.
Then it suffices to show that $d\kappa/d\mu_0<0$.
Let $t_b(\mu_0,\kappa)$ and $\mu_b(\mu_0,\kappa)$ be the maximizers of the Lagrangian for a given $\kappa$, and since the participation constraint binds at the optimum, we can define
\begin{align*}
  S(\kappa,\mu_0)\equiv \frac{\mu_0}{\mu_b(\mu_0,\kappa)} V_{\text{NI}}(\mu_b(\mu_0,\kappa))+\left(1-\frac{\mu_0}{\mu_b(\mu_0,\kappa)}\right)v_L(t_b(\mu_0,\kappa))-V_{\text{NI}}(\mu_0)
\end{align*}
as the net surplus function.
Since the incentive constraint binds, we have
\begin{align*}
  v_L(t_b)=\frac{\mu_b V_{\text{NI}}(\mu_0)-\mu_0V_{\text{NI}}(\mu_b)}{\mu_b-\mu_0}.
\end{align*}
By the envelope theorem, 
\begin{align*}
  \frac{\partial S(\kappa,\mu_0)}{\partial\mu_0}=\frac{V_{\text{NI}}(\mu_b)-v_L(t_b)}{\mu_b}-V_{\text{NI}}'(\mu_0)=\frac{V_{\text{NI}}(\mu_b)-V_{\text{NI}}(\mu_0)}{\mu_b-\mu_0}-V_{\text{NI}}'(\mu_0),
\end{align*}
which is positive by the convexity of $V_{\text{NI}}(\cdot)$.
Also, we can show that $\partial S(\kappa,\mu_0)/\partial\kappa>0$.
For any pair of multipliers $\kappa_1>\kappa_2\geq 0$, let $(t_b^1,\mu_b^1)$ and $(t_b^2,\mu_b^2)$ be the corresponding maximizers of the Lagrangian.
By the definition of the maximizers, we have
\begin{align*}
  &\frac{\mu_0}{\mu_b^1} \bigg[ w(\tau(\mu_b^1)) + \kappa_1 V_{\text{NI}}(\mu_b^1) \bigg] + \left(1 - \frac{\mu_0}{\mu_b^1}\right) \bigg[ w(t_b^1) + \kappa_1 v_L(t_b^1) \bigg] - \kappa_1 V_{\text{NI}}(\mu_0)\\
  &\geq \frac{\mu_0}{\mu_b^2} \bigg[ w(\tau(\mu_b^2)) + \kappa_1 V_{\text{NI}}(\mu_b^2) \bigg] + \left(1 - \frac{\mu_0}{\mu_b^2}\right) \bigg[ w(t_b^2) + \kappa_1 v_L(t_b^2) \bigg] - \kappa_1 V_{\text{NI}}(\mu_0),
\end{align*}
and
\begin{align*}
  &\frac{\mu_0}{\mu_b^2} \bigg[ w(\tau(\mu_b^2)) + \kappa_2 V_{\text{NI}}(\mu_b^2) \bigg] + \left(1 - \frac{\mu_0}{\mu_b^2}\right) \bigg[ w(t_b^2) + \kappa_2 v_L(t_b^2) \bigg] - \kappa_2 V_{\text{NI}}(\mu_0)\\
  &\geq \frac{\mu_0}{\mu_b^1} \bigg[ w(\tau(\mu_b^1)) + \kappa_2 V_{\text{NI}}(\mu_b^1) \bigg] + \left(1 - \frac{\mu_0}{\mu_b^1}\right) \bigg[ w(t_b^1) + \kappa_2 v_L(t_b^1) \bigg] - \kappa_2 V_{\text{NI}}(\mu_0).
\end{align*}
Adding these two inequalities yields
\begin{align*}
 (\kappa_1-\kappa_2)\bigg[\left[\frac{\mu_0}{\mu_b^2}V_{\text{NI}}(\mu_b^2)+\left(1 - \frac{\mu_0}{\mu_b^2}\right)v_L(t_b^2)\right]-\left[\frac{\mu_0}{\mu_b^1}V_{\text{NI}}(\mu_b^1)+\left(1 - \frac{\mu_0}{\mu_b^1}\right)v_L(t_b^1)\right]\bigg]\leq 0.
\end{align*}
Since $\kappa_1>\kappa_2$, we have $S(\kappa_2,\mu_0)\leq S(\kappa_1,\mu_0)$.
The inequality is strict because $\kappa_1\neq\kappa_2$ implies distinct maximizers via the first-order conditions.
Thus $\partial S(\kappa,\mu_0)/\partial\kappa>0$.
Thus, by the implicit function theorem,
\begin{align*}
  \frac{d\kappa}{d\mu_0}=-\frac{\partial S(\kappa,\mu_0)/\partial\mu_0}{\partial S(\kappa,\mu_0)/\partial\kappa}<0.
\end{align*}
Combining the above results that $dt_b/d\kappa<0$ and $d\kappa/d\mu_0<0$, we have $dt_b/d\mu_0>0$.

Next, we show that $x_b$ decreases with $\mu_0$.
Note that the optimality condition of $t_b$ is simply $w'(t_b)+\kappa v'_L(t_b)=0$, which is independent of $\mu_b$ and $\mu_0$.
Thus, we write $\Phi(\kappa)\equiv \max_{t_b} \left[ w(t_b) + \kappa v_L(t_b) \right]$, and then if $\mu_b$ maximizes the Lagrangian, it must maximize
\begin{align*}
  &\frac{\mu_0}{\mu_b} \bigg[ w(\tau(\mu_b)) + \kappa V_{\text{NI}}(\mu_b) \bigg] + \left(1 - \frac{\mu_0}{\mu_b}\right)\Phi(\kappa) - \kappa V_{\text{NI}}(\mu_0)\\
  &=\mu_0\cdot\frac{w(\tau(\mu_b)) + \kappa V_{\text{NI}}(\mu_b)-\Phi(\kappa)}{\mu_b}+\Phi(\kappa)-\kappa V_{\text{NI}}(\mu_0),
\end{align*}
which is equivalent to maximizing $(w(\tau(\mu_b)) + \kappa V_{\text{NI}}(\mu_b)-\Phi(\kappa))/\mu_b$.
Thus, we write $\mu_b(\mu_0,\kappa,t_b)$ simply as $\mu_b(\kappa)$.
That is, by the Bayes plausibility constraint, at the optimum:
\begin{equation*}
    x_b(\mu_0, \mu_b(\kappa)) = 1 - \frac{\mu_0}{1-\mu_0} \frac{1-\mu_b(\kappa)}{\mu_b(\kappa)}.
\end{equation*}
Taking the total derivative with respect to $\mu_0$:
\begin{equation*}
    \frac{d x_b}{d \mu_0} = \frac{\partial x_b}{\partial \mu_0} + \frac{\partial x_b}{\partial \mu_b} \frac{d \mu_b}{d \kappa} \frac{d \kappa}{d \mu_0}.
\end{equation*}
Obviously, $\partial x_b/\partial\mu_0<0$ and $\partial x_b/\partial\mu_b>0$.
Also, the former discussion shows that $d\kappa/d\mu_0<0$.
Thus, to establish that $dx_b/d\mu_0<0$, it suffices to show that $d\mu_b/d\kappa>0$.

Here, the first-order condition of $\mu_b$ yields
\begin{align*}
  &\frac{d}{d\mu_b}\left(\frac{w(\tau(\mu_b)) + \kappa V_{\text{NI}}(\mu_b)-\Phi(\kappa)}{\mu_b}\right)\\&\propto G(\mu_b,\kappa)\equiv \mu_b\bigg(w'(\tau(\mu_b))\tau'(\mu_b) + \kappa V'_{\text{NI}}(\mu_b)\bigg)-\left(w(\tau(\mu_b)) + \kappa V_{\text{NI}}(\mu_b)-\Phi(\kappa)\right)=0.
\end{align*}
By the Implicit Function Theorem, $d\mu_b/d\kappa = - (\partial G/\partial \kappa) / (\partial G/\partial \mu_b)$. 
The optimality of $\mu_b$ indicates that the denominator $\partial G/\partial \mu_b$ is negative by the second-order condition.
Let $\hat t_b$ be the maximizer of $w(t_b) + \kappa v_L(t_b)$, and then the numerator is given by
\begin{align*}
  \frac{\partial G(\mu_b,\kappa)}{\partial\kappa}&=\mu_b V'_{\text{NI}}(\mu_b)-V_{\text{NI}}(\mu_b)+v_L(\hat t_b)\\
  &=\mu_b V'_{\text{NI}}(\mu_b)-V_{\text{NI}}(\mu_b)+\frac{\mu_b V_{\text{NI}}(\mu_0)-\mu_0V_{\text{NI}}(\mu_b)}{\mu_b-\mu_0}\\
  &=\mu_b\left[V'_{\text{NI}}(\mu_b)-\frac{V_{\text{NI}}(\mu_b)-V_{\text{NI}}(\mu_0)}{\mu_b-\mu_0}\right]>0
\end{align*} 
by the convexity of $V_{\text{NI}}(\cdot)$.
Combining the calculations above, we have $d\mu_b/d\kappa>0$, which completes the proof that $dx_b/d\mu_0<0$.

\subsubsection*{Step 5: Comparative statics when  $x_b=1$}
Finally, when $x_b=1$, the optimization problem can be written as
\begin{align*}
  \max_{t_b}:&w(t_b)\\
  &\text{s.t.: }v_L(t_b)\geq \frac{V_{\text{NI}}(\mu_0)-\mu_0 v_H(\tau(1))}{1-\mu_0}.
\end{align*}
Since $v_L(\cdot)$ is decreasing, it suffices to show that the right-hand side decreases with $\mu_0$. 
Note that $v_H(\tau(1))=V_{\text{NI}}(1)$, and thus
\begin{align*}
  \frac{d}{d\mu_0}\left(\frac{V_{\text{NI}}(\mu_0)-\mu_0 V_{\text{NI}}(1)}{1-\mu_0}\right)=\frac{V'_{\text{NI}}(\mu_0)-\frac{V_{\text{NI}}(1)-V_{\text{NI}}(\mu_0)}{1-\mu_0}}{1-\mu_0}<0,
\end{align*}
by the convexity of $V_{\text{NI}}(\cdot)$, which completes the proof.


\subsection{Proof of Theorem \ref{proposition.gradual}}


We complete the proof in two steps.

\subsubsection*{Step 1: Disclosure before the principal's ideal stopping time}

We can calculate
\begin{align*}
 R(w,t)=\frac{p_0\lambda^2(Y-Z)-\left(p_0+(1-p_0)e^{\lambda t}\right)r_P^2 Z+p_0\lambda r_P(Y-2 Z)}{p_0\lambda(Y-Z)-\left(p_0+(1-p_0)e^{\lambda t}\right)r_P Z}
\end{align*}
and
\begin{align*}
 R(v_L,t)=\frac{p_0\lambda^2(y_L-z)-\left(p_0+(1-p_0)e^{\lambda t}\right)r_A^2 z+p_0\lambda r_A(y_L-2z)}{p_0\lambda(y_L-z)-\left(p_0+(1-p_0)e^{\lambda t}\right)r_A z}.
\end{align*}
Function $R(v_L,t)$ is increasing in $r_A$ if and only if
\begin{align*}
 r_A\geq\frac{\lambda\left(\sqrt{p_0(1-p_0)e^{\lambda t} (z-y_L)z}+p_0(y_L-z)\right)}{z\left[p_0+(1-p_0)e^{\lambda t}\right]}
\end{align*}
if $y_L<z$, and is always increasing in $r_A$ if $y_L\geq z$.
Also, $R(v_L,t)$ goes to infinity when $r_A$ goes to infinity.
Thus, for any time $t\in[\tau(\mu_0),t^*)$, $R(v_L,t)>R(w,t)$ when $r_A$ is sufficiently large.

Next, we show that if we fix the discount rate $r_A$, the set of dates at which $R(v_L,t)\geq R(w,t)$ is an interval.
When $\tau(0)>0$, it suffices to show that $R(w,t)$ is convex but $R(v_L,t)$ is concave. 
We calculate
\begin{align*}
 &\frac{d^2R(w,t)}{dt^2}=\\&\quad\frac{\lambda^3(1-p_0)p_0r_P e^{\lambda t}(\lambda Y-(\lambda+r_P)Z)Z\left(p_0(\lambda Y-(\lambda+r_P)Z)+(1-p_0)r_Pe^{\lambda t}Z\right)}{\left(p_0(\lambda Y-(\lambda+r_P)Z)-(1-p_0)r_Pe^{\lambda t}Z\right)^3},
\end{align*}
and since $t^*>0$, this expression is positive when $t<t^*$.
Meanwhile, 
\begin{align*}
 &\frac{d^2R(v_L,t)}{dt^2}=\\&\quad\frac{\lambda^3(1-p_0)p_0r_Ae^{\lambda t}(\lambda y_L-(\lambda+r_A)z)z\left(p_0(\lambda y_L-(\lambda+r_A)z)+(1-p_0)r_Ae^{\lambda t}z\right)}{\left(p_0(\lambda y_L-(\lambda+r_A)z)-(1-p_0)r_Ae^{\lambda t}z\right)^3},
\end{align*}
which is strictly negative since $t\geq \tau(\mu_0)>\tau(0)$.
When $\tau(0)=0$, take $r_A$ sufficiently large that $r_A>\lambda(y_L/z-1)$, or equivalently, $(\lambda+r_A)z-\lambda y_L>0$, and it then suffices to show that $R(w,t)$ is increasing in $t$ while $R(v_L,t)$ is decreasing. 
This follows from
\begin{align*}
    \frac{dR(w,t)}{dt}=\frac{\lambda^2 (1-p_0) p_0 r_P e^{\lambda t} (\lambda Y-(\lambda+r_P)Z)}{\left(\lambda p_0(Y-Z)-r_PZ \left((1-p_0) e^{\lambda t}+p_0\right)\right)^2}>0,
\end{align*}
and
\begin{align*}
    \frac{dR(v_L,t)}{dt}=-\frac{\lambda^2 (1-p_0) p_0 r_A e^{\lambda t} ((\lambda+r_A)z-\lambda y_L)}{\left(\lambda p_0(y_L-z)-r_A z \left((1-p_0) e^{\lambda t}+p_0\right)\right)^2}<0.
\end{align*}

Before applying the curvature results, note that $F_H$ is fixed before $t^*$ and the principal chooses only $F_L$.
Lemma \ref{lemma.useful} gives
\[
 \bar v_L(t,s)=\frac{v_L(s)-v_L(t)}{\gamma(0,t)},
\]
where $\gamma(0,t)>0$ is evaluated using the agent's discount rate $r_A$.
Thus Assumptions \ref{ass:statewise-separable} and \ref{ass:multi-separable} hold for the reduced single-state problem.

We have established the existence of $[\underline{t}_g,\bar{t}_g]$, and thus by Lemma \ref{lem:no-rent}, if $dF_L(t)>0$ for $t\in[\underline{t}_g,\bar{t}_g]$, the incentive constraint at time $t$ must be binding.
Now we show that under the optimal information policy $\mathcal{P}=\langle F_H,F_L\rangle$, if $\mathcal{S}(\mathcal{P})\cap [\underline{t}_g,\bar{t}_g]$ is not a singleton, it must be an interval.
Suppose not.
Then there exist time instants $t_1,t_2\in[\underline{t}_g,\bar{t}_g]$ ($t_1<t_2$), such that $dF_L(t_1)>0$ and $dF_L(t_2)>0$ but $dF_L(t)=0$ for all $t\in(t_1,t_2)$.
Then the incentive constraint at $t_1$ and $t_2$ must be binding, but it is not necessarily so for $t\in(t_1,t_2)$.
Notice that the incentive constraint at time $t$ holds if and only if
\begin{equation*}
\begin{aligned}
  \underbrace{\mu_0\int_{t_2}^{\infty}v_H(s)dF_H(s)+(1-\mu_0)\int_{t_2}^{\infty}v_L(s)dF_L(s)}_{\text{irrelevant to }t}&\\
 \geq\widehat{V}(t)\equiv\mu_0v_H(t)+(1-\mu_0)&(1-F_L(t_1))v_L(t).   
\end{aligned}
\end{equation*}
Here, the left-hand side is irrelevant to $t$, and the derivative of $\widehat{V}(t)$ with respect to $t$ is given by
\begin{align*}
    \widehat{V}'(t)=e^{-(r_A+\lambda)t}\bigg(&p_0\lambda\big( y(\mu_0)-(1-\mu_0) F_L(t_1)y_L\big)\\&-\big(p_0(\lambda+r_A)+(1-p_0)r_Ae^{\lambda t}\big)\big(1-(1-\mu_0)F_L(t_1)\big)\cdot z\bigg),
\end{align*}
whose term in parentheses is strictly decreasing in $t$ and determines the sign of $\widehat V'(t)$.
Thus, $\widehat{V}(t)$ is either a monotonic or a hump-shaped function. 
However, because the continuation constraints bind at $t_1$ and $t_2$, feasibility requires $\widehat V(t_1)=\widehat V(t_2)\geq\widehat V(t)$ for every $t\in(t_1,t_2)$.
The monotonic case is incompatible with equal endpoint values.
In the hump-shaped case, $\widehat V(t)>\widehat V(t_1)=\widehat V(t_2)$ for every $t\in(t_1,t_2)$, again a contradiction.

Thus, by Lemma \ref{lem:no-rent}, for all $t\in\mathcal{S}(\mathcal{P})\cap[\underline{t}_g,\bar{t}_g]$, we have
\begin{align*}
 \mu_0\int_t^\infty\bar{v}_H(t,s)dF_H(s)+(1-\mu_0)\int_t^\infty\bar{v}_L(t,s)dF_L(s)\equiv 0.
\end{align*}
Lemma \ref{lemma.useful}, together with $F_H(t)\equiv 0$ for all $t<t^*$, indicates that
\begin{align*}
 \mu_0\int_{t^*}^\infty\big(v_H(s)-v_H(t)\big)dF_H(s)+(1-\mu_0)\int_t^\infty\big(v_L(s)-v_L(t)\big)dF_L(s)\equiv 0.
\end{align*}
Taking derivatives with respect to $t$ of both sides, we get
\begin{align*}
 -\mu_0v'_H(t)-(1-\mu_0)\int_t^\infty v'_L(t)dF_L(s)=-\mu_0v'_H(t)-(1-\mu_0)(1-F_L(t))v'_L(t)\equiv 0,
\end{align*}
which produces the desired form.

We now turn to part (iv).
Assumption \ref{ass:multi-separable} already holds for the reduced problem, and Step~5 of the proof of Theorem~\ref{thetheorem} verifies Assumption \ref{ass:no-disclosure}.
To match the reduced problem with the general formulation in \eqref{eq:multistate-problem}, take $\psi\equiv0$.
By \eqref{eq:psi-perspective}, its homogeneous extension is $\Psi(q_L)=q_L\psi(1)=0$ for every $q_L\geq0$, and thus $\Psi$ is differentiable with $\partial_{q_L}\Psi(q_L)=0$.
Since $\hat v_L(t,t)=\bar v_L(t,t)=0$ for every $t$, \eqref{eq:continuation-pointwise} holds with equality whenever $q_L>0$, verifying Assumption \ref{ass:continuation-order}.
Since Assumptions \ref{ass:multi-separable}, \ref{ass:no-disclosure}, and \ref{ass:continuation-order} hold and part (i) gives $R(v_L,t)<R(w,t)$ on $[\tau(\mu_0),\underline t_g)$ and $(\bar t_g,t^*)$, Lemma \ref{lem:bounded-interval} implies that each of $\mathcal{S}(\mathcal{P})\cap[\tau(\mu_0),\underline t_g)$ and $\mathcal{S}(\mathcal{P})\cap(\bar t_g,t^*)$ contains at most one date, establishing part (iv).

\subsubsection*{Step 2: Disclosure after the principal's ideal stopping time}

Finally, consider the disclosure chosen in the dissuasion subproblem.
As established in Step~6 of the proof of Theorem~\ref{thetheorem}, this continuation problem can be implemented by a static signal at $t^*$.
For a fixed interim belief $\mu_{t^*}$ and promised continuation payoff $u$, the principal's problem is given by \eqref{eq.kg}.
Let $\kappa \geq 0$ be the multiplier on the promised-payoff constraint.
Because the state space is binary and the problem has one constraint in addition to Bayes plausibility, \citet[Corollary~3.1]{doval2024constrained} guarantees an optimal signal with at most three posterior beliefs.

The constrained static problem may require three posterior beliefs.
We first show that any such solution must be supported on $\{0,\mu_a,1\}$, and then use the policy chosen before the principal's ideal stopping time to exclude this configuration from the complete dynamic optimum.
If $\kappa=0$, the dissuasion problem reduces to an ordinary Bayesian persuasion problem with fixed mean $\mu_{t^*}$.
By Carath\'eodory's theorem, it admits an optimal signal with at most two posterior beliefs.
Suppose $\kappa>0$.
Both $W_{\mathrm{NI}}(t^*,\mu)$ and $V_{\mathrm{NI}}(t^*,\mu)$ are constant on $[0,\mu^*]$.
On $(\mu^*,1)$, the sign of the second derivative of
\[
W_{\mathrm{NI}}(t^*,\mu)+\kappa V_{\mathrm{NI}}(t^*,\mu)
\]
is the sign of
\[
Z\frac{r_P}{\lambda}\left(1+\frac{r_P}{\lambda}\right)
\left[
e^{\lambda(\tau(\mu)-t^*)}
-\frac{2\lambda+r_P}{\lambda+r_P}
\right]
+\kappa z\frac{r_A}{\lambda}
e^{(\lambda+r_P-r_A)(\tau(\mu)-t^*)}.
\]
As a function of $e^{\lambda(\tau(\mu)-t^*)}$, this expression is either increasing or has a single minimum.
Its negative region is therefore connected.
An affine upper support can touch the Lagrangian payoff at a smooth interior posterior only within this region.
Moreover, it cannot touch at two distinct posteriors in the region, since strict concavity would place the Lagrangian payoff strictly above the supporting line between the two contact points.
Thus, any affine upper support has at most one contact $\mu_a$ in $(\mu^*,1)$.

All contacts in $[0,\mu^*]$ can be combined into a single contact.
A three-contact upper support can arise only when the Lagrangian payoff is convex, then concave, and then convex on $(\mu^*,1)$.
Since the payoff is constant on $[0,\mu^*]$ and initially increases to the right of $\mu^*$, such a supporting line has a positive slope and can touch the constant region only at $0$.
Its strictly concave region supplies at most one interior contact, and therefore the remaining contact must be $1$.
Therefore, every signal that genuinely requires three posterior beliefs is supported on $\{0,\mu_a,1\}$ for some $\mu_a\in(\mu^*,1)$.

The following comparison will be used below to evaluate the immediate-stopping deviations.
For any $s\neq t^*$, the principal's normalized loss from moving the stopping date from $t^*$ to $s$, divided by the agent's normalized gain, is
\begin{equation}
\frac{-\left[e^{-r_P(s-t^*)}-\dfrac{r_P}{\lambda+r_P}e^{-(\lambda+r_P)(s-t^*)}-\dfrac{\lambda}{\lambda+r_P}\right]}{\dfrac{r_A}{\lambda+r_A}e^{\lambda(s-t^*)}+\dfrac{\lambda}{\lambda+r_A}e^{-r_A(s-t^*)}-1}.
\label{eq:normalized-loss-gain-ratio}
\end{equation}
The ratio in \eqref{eq:normalized-loss-gain-ratio} converges to $r_P/r_A$ as $s$ approaches $t^*$.
A strict three-contact upper support requires the Lagrangian payoff to be convex, then concave, and then convex.
Evaluating the unique possible interior minimum of its second derivative shows that this curvature pattern can arise only if $\frac{2\lambda+r_P}{\lambda+r_P}\frac{r_A-r_P-\lambda}{r_A-r_P}>1$, or equivalently $r_A>2(r_P+\lambda)$.

We now compare the ratio in \eqref{eq:normalized-loss-gain-ratio} at any $s<t^*$ with its value at $s=\tau(\mu_a)$.
We show that the former is no greater than $r_P/r_A$, whereas the latter is no smaller than $r_P/r_A$.

For $s<t^*$, the denominator in \eqref{eq:normalized-loss-gain-ratio} is positive.
Hence, the desired inequality is equivalent to showing that $r_P/r_A$ times the denominator in \eqref{eq:normalized-loss-gain-ratio}, minus its numerator, is nonnegative.
At $t^*$, this quantity and its first derivative are both zero.
Its second derivative has the same sign as
\begin{equation}
e^{(r_P-r_A+\lambda)(s-t^*)}
-\left(\frac{r_P}{\lambda}+2\right)
+\left(\frac{r_P}{\lambda}+1\right)e^{\lambda(s-t^*)}>0
\quad\text{for }s<t^*.
\label{eq:tail-left-comparison}
\end{equation}
Integrating \eqref{eq:tail-left-comparison} from $s$ to $t^*$ shows that the derivative of $r_P/r_A$ times the denominator in \eqref{eq:normalized-loss-gain-ratio}, minus its numerator, is negative at $s$.
Integrating once more shows that $r_P/r_A$ times the denominator, minus the numerator, is positive at $s$.
Hence, the ratio in \eqref{eq:normalized-loss-gain-ratio} is no greater than $r_P/r_A$ for every $s<t^*$.

Since $\mu_a>\mu^*$ implies $\tau(\mu_a)>t^*$, it remains to show that the ratio in \eqref{eq:normalized-loss-gain-ratio} is still increasing at $s=\tau(\mu_a)$.
For $s>t^*$, differentiating the numerator and denominator in \eqref{eq:normalized-loss-gain-ratio} gives
\begin{equation}
\frac{r_P(\lambda+r_A)}{r_A\lambda}
\frac{e^{-r_P(s-t^*)}-e^{-(\lambda+r_P)(s-t^*)}}
{e^{\lambda(s-t^*)}-e^{-r_A(s-t^*)}}.
\label{eq:tail-derivative-ratio}
\end{equation}
The logarithmic derivative of \eqref{eq:tail-derivative-ratio} is
\begin{equation}
 r_A-\lambda-r_P
 +\frac{\lambda e^{\lambda(s-t^*)}}{e^{\lambda(s-t^*)}-1}
 -\frac{(r_A+\lambda)e^{(r_A+\lambda)(s-t^*)}}
 {e^{(r_A+\lambda)(s-t^*)}-1}.
\label{eq:tail-log-derivative}
\end{equation}
Its derivative is
\begin{equation}
-\frac{\lambda^2e^{\lambda(s-t^*)}}
{\left(e^{\lambda(s-t^*)}-1\right)^2}
+\frac{(r_A+\lambda)^2e^{(r_A+\lambda)(s-t^*)}}
{\left(e^{(r_A+\lambda)(s-t^*)}-1\right)^2}<0
\quad\text{for }s>t^*.
\label{eq:tail-right-curvature}
\end{equation}
The inequality follows because $z/\sinh(z(s-t^*)/2)$ is strictly decreasing in $z>0$.
The expression in \eqref{eq:tail-log-derivative} converges to $r_A/2-\lambda-r_P>0$ as $s$ approaches $t^*$ and to $-\lambda-r_P<0$ as $s$ tends to infinity.
Equations~\eqref{eq:tail-log-derivative}--\eqref{eq:tail-right-curvature} therefore show that \eqref{eq:tail-derivative-ratio} first increases and then decreases.
Because the numerator and denominator in \eqref{eq:normalized-loss-gain-ratio} both vanish at $t^*$ and the derivative of the denominator is positive for $s>t^*$, the ratio in \eqref{eq:normalized-loss-gain-ratio} is the average of \eqref{eq:tail-derivative-ratio} over $(t^*,s)$, weighted by the denominator derivative.
As $s$ increases, this average first increases and then decreases as well.

To determine whether the ratio is still increasing at $s=\tau(\mu_a)$, substitute $s=\tau(\mu)$ into \eqref{eq:normalized-loss-gain-ratio}.
The value obtained differs only by a positive constant from $[Z-W_{\mathrm{NI}}(t^*,\mu)]/V_{\mathrm{NI}}(t^*,\mu)$.
The contact conditions at $0$ and $\mu_a$ give
\[
\left.\frac{d}{d\mu}\left[\frac{Z-W_{\mathrm{NI}}(t^*,\mu)}{V_{\mathrm{NI}}(t^*,\mu)}\right]\right|_{\mu=\mu_a}
=\frac{[\partial_\mu W_{\mathrm{NI}}+\kappa\partial_\mu V_{\mathrm{NI}}][\mu_a\partial_\mu V_{\mathrm{NI}}-V_{\mathrm{NI}}]}{V_{\mathrm{NI}}^2}>0,
\]
where every function and derivative on the right-hand side is evaluated at $(t^*,\mu_a)$.
The factor $\partial_\mu W_{\mathrm{NI}}+\kappa\partial_\mu V_{\mathrm{NI}}$ is positive because it is the slope of the supporting line.
The factor $\mu_a\partial_\mu V_{\mathrm{NI}}-V_{\mathrm{NI}}$ is positive because $V_{\mathrm{NI}}(t^*,0)=0$ and $V_{\mathrm{NI}}(t^*,\mu)$ is strictly convex on $(\mu^*,1)$.
Thus, the ratio in \eqref{eq:normalized-loss-gain-ratio} is still increasing at $s=\tau(\mu_a)$.
Combining the two comparisons gives
\begin{equation}
\left.\text{the ratio in \eqref{eq:normalized-loss-gain-ratio}}\right|_{s<t^*}
\leq \frac{r_P}{r_A}
\leq
\left.\text{the ratio in \eqref{eq:normalized-loss-gain-ratio}}\right|_{s=\tau(\mu_a)}.
\label{eq:tail-ratio-comparison}
\end{equation}

We use \eqref{eq:tail-ratio-comparison} to rule out the three-contact signal across all possible histories leading to $t^*$:

\medskip
\noindent\textit{Case 1: A continuation constraint binds at some $t<t^*$, and $t$ is the last such date. Suppose no stopping date in $(t,t^*)$ is used with positive probability.}\par
\smallskip
Replace the continuation policy after $t$ by immediate stopping at $t$, with the stopping recommendation sent independently of the state.
The posterior following this recommendation equals the interim belief $\mu_{t^*}$.
The no-disclosure property verified in Step~1 and the absence of stopping in $(t,t^*)$ imply that this belief and the state-specific survival probabilities remain constant on that interval.
Because the continuation constraint binds at $t$, feasibility at every $s\in[t,t^*)$ requires the no-information continuation payoff at $s$ to be no greater than its value at $t$.
This payoff is single-peaked and maximized at $\tau(\mu_{t^*})$, and thus $\tau(\mu_{t^*})\leq t$.
The recommendation to stop at $t$ is therefore obedient.
The binding continuation constraint also implies that immediate stopping and the original continuation give the agent the same payoff at $t$.
Lemma~\ref{lemma.useful} then implies that every earlier continuation constraint is unchanged, and hence the replacement is feasible.

It remains to compare the principal's payoffs.
First set the probability of posterior $1$ to zero, and thus the signal at $t^*$ is supported on $0$ and $\mu_a$.
Relative to stopping at $t^*$, immediate stopping at $t$ and continuation until $\tau(\mu_a)$ change the agent's payoff in opposite directions.
Because the continuation constraint at $t$ binds, the two changes have the same absolute value.
The principal's payoff under the continuation policy minus her payoff from immediate stopping at $t$ therefore has the sign of
\[
\left.\text{the ratio in \eqref{eq:normalized-loss-gain-ratio}}\right|_{s=t}
-
\left.\text{the ratio in \eqref{eq:normalized-loss-gain-ratio}}\right|_{s=\tau(\mu_a)}
\leq 0,
\]
where the inequality follows from \eqref{eq:tail-ratio-comparison}.
Thus, immediate stopping at $t$ gives the principal at least as much as the boundary signal.

Now assign probability $\varepsilon>0$ to posterior $1$ while preserving Bayes plausibility.
Relative to the boundary signal, the probability assigned to $\mu_a$ falls by $\varepsilon/\mu_a$.
The agent's promised payoff therefore increases by
\[
\varepsilon\left[V_{\mathrm{NI}}(t^*,1)-\frac{V_{\mathrm{NI}}(t^*,\mu_a)}{\mu_a}\right]>0.
\]
The inequality follows from $V_{\mathrm{NI}}(t^*,0)=0$ and the strict convexity of $V_{\mathrm{NI}}(t^*,\mu)$ on $(\mu^*,1)$.
All three posteriors lie on the same supporting line for $W_{\mathrm{NI}}+\kappa V_{\mathrm{NI}}$.
Bayes plausibility keeps the expected value of this supporting line fixed, and hence the principal's payoff falls by $\kappa$ times the displayed increase in the agent's payoff.
Since $\kappa>0$, a signal assigning positive probability to all three posteriors gives the principal strictly less than immediate stopping at $t$, a contradiction.

\medskip
\noindent\textit{Case 2: A continuation constraint binds at some $t<t^*$, and $t$ is the last such date. Suppose some stopping date $s\in(t,t^*)$ is used with positive probability.}\par
\smallskip
The stopping recommendation at $s$ reveals the low state, and posterior $0$ in the signal at $t^*$ also contains only the low state.
We alter these two positive stopping probabilities together.
Move a small amount of the low-state probability stopping at $s$ to $s+\varepsilon$, and move a small amount of the low-state probability stopping at $t^*$ to $t^*-\delta$, where $s+\varepsilon<t^*-\delta$.
Choose the two amounts so that the loss in the agent's date-zero payoff from the first change equals the gain from the second change.
Equivalently, the two amounts are chosen in inverse proportion to $v_L(s)-v_L(s+\varepsilon)$ and $v_L(t^*-\delta)-v_L(t^*)$.
Both differences are positive because stopping obedience gives $s\geq\tau(0)$ and $v_L$ decreases after $\tau(0)$.

We first verify that this change is feasible.
At every date weakly before $s$, Lemma~\ref{lemma.useful} implies that the two changes in the continuation payoff are the corresponding date-zero changes multiplied by the same positive factor.
They therefore cancel.
The continuation constraint at $s$ is slack because $t$ is the last date at which a continuation constraint binds.
Hence the moved probabilities can be chosen small enough to preserve the constraints between $s$ and $s+\varepsilon$.
At every date from $s+\varepsilon$ to $t^*-\delta$, the first moved probability has already stopped, while moving the other low-state probability from $t^*$ to $t^*-\delta$ raises the agent's continuation payoff.
At every date between $t^*-\delta$ and $t^*$, that low-state probability no longer continues to $t^*$, which removes a negative continuation payoff.
All later continuation constraints are unchanged.
Both new stopping recommendations reveal the low state and occur weakly after $\tau(0)$, and thus the stopping constraints also remain satisfied.

It remains to compare the principal's payoff.
Per unit of the agent's payoff lost by delaying stopping at $s$, the principal gains
\[
\frac{w(s+\varepsilon)-w(s)}{v_L(s)-v_L(s+\varepsilon)}>0.
\]
Per unit of the agent's payoff restored by moving stopping from $t^*$ to $t^*-\delta$, the principal loses
\[
\frac{w(t^*)-w(t^*-\delta)}{v_L(t^*-\delta)-v_L(t^*)}.
\]
The second ratio converges to zero as $\delta\downarrow0$, since $w'(t^*)=0$ and $v_L'(t^*)<0$.
Fixing a sufficiently small $\varepsilon>0$ and then choosing $\delta>0$ sufficiently small makes the principal's gain from the first change strictly larger than her loss from the second.
The two moved probabilities can then be scaled to preserve the agent's payoff exactly.
This feasible change strictly raises the principal's payoff, contradicting optimality.

\medskip
\noindent\textit{Case 3: No continuation constraint binds at any $t<t^*$.}\par
\smallskip
Suppose first that some low-state probability stops at a date $s<t^*$.
Apply the same two-date change used in Case~2.
Move a small amount of this probability from $s$ to $s+\varepsilon$, move a small amount of the posterior-$0$ probability from $t^*$ to $t^*-\delta$, and choose the two amounts so that their effects on the agent's date-zero payoff cancel.
Lemma~\ref{lemma.useful} then leaves every continuation constraint weakly before $s$ unchanged.
All constraints after $s$ are slack in Case~3, and hence the moved probabilities can be chosen small enough to preserve them.
The two new stopping recommendations reveal the low state and occur after $\tau(0)$, and thus they are obedient.
The payoff comparison in Case~2 applies without change and gives the principal a strict gain.
Hence the original policy cannot stop with positive probability before $t^*$.

It remains to consider a policy under which both states reach $t^*$.
In particular, both states reach $\tau(\mu_0)$.
Lemma~\ref{lemma.useful} implies that the agent's continuation surplus at that date equals
\[
\frac{\mu_0\int v_H(s)dF_H(s)+(1-\mu_0)\int v_L(s)dF_L(s)-V_{\mathrm{NI}}(\mu_0)}{\gamma(0,\tau(\mu_0))}.
\]
The continuation constraint at $\tau(\mu_0)$ is slack in Case~3, and hence the numerator is strictly positive.
The original policy therefore gives the agent strictly more than his no-information outside option, and initial participation is slack.

We now change only the probabilities assigned to the three posteriors at $t^*$.
Increase the probability of $\mu_a$ by a small amount and reduce the probabilities of $0$ and $1$ in the proportions $1-\mu_a$ and $\mu_a$.
Total probability and the posterior mean are unchanged.
Strict convexity gives $V_{\mathrm{NI}}(t^*,\mu_a)<(1-\mu_a)V_{\mathrm{NI}}(t^*,0)+\mu_aV_{\mathrm{NI}}(t^*,1)$, and thus this change strictly reduces the promised continuation payoff.
The three posterior beliefs and their induced stopping dates remain unchanged, and therefore all stopping and continuation decisions following the signal remain obedient.

Finally, at each of the three posteriors, $W_{\mathrm{NI}}(t^*,\mu)+\kappa V_{\mathrm{NI}}(t^*,\mu)$ equals the same linear function of $\mu$.
The change preserves both total probability and the posterior mean, and thus the expectation of this linear function is unchanged.
Since the change lowers the agent's payoff and $\kappa>0$, it strictly raises the principal's payoff.
All continuation constraints before $t^*$ and initial participation are strictly slack, and hence a sufficiently small change preserves them.
The resulting policy is feasible and strictly improves the principal's payoff, contradicting optimality.

\medskip
To summarize, the three cases rule out any optimal signal that assigns positive probability to $0$, $\mu_a$, and $1$ simultaneously.
An optimal signal at $t^*$ can therefore be chosen with at most two posterior beliefs.

The same deviations rule out a binary signal supported by $0$ and an interior posterior whenever stopping occurs before $t^*$.
In the only remaining subcase of Case~3, both states reach $t^*$ and all earlier continuation constraints and initial participation are slack.
Replacing a small part of the binary signal by immediate stopping at $t^*$ is then feasible and strictly profitable because $w$ is maximized at $t^*$.
A degenerate signal is included below by setting $x_a=1$.
Hence every remaining nondegenerate signal is supported by an interior posterior $\mu_a$ and posterior $1$.

Let $t_a\in[t^*,\tau(1)]$ be the stopping date induced by $\mu_a$.
Because posterior $1$ contains only state $H$, Bayes plausibility assigns all remaining probability of state $L$ to $\mu_a$.
The probability of state $H$ is split between the two posteriors.
Let $x_a\in[0,1]$ be the probability assigned to $\mu_a$.
The agent then stops at $t_a$ with probability one in state $L$ and with probability $x_a$ in state $H$.
The remaining high-state probability stops at $\tau(1)$, and no further disclosure is required.

\subsection{Proof of Proposition \ref{proposition.nocommitment}}

With a slight abuse of notation, we define
 \begin{align*}
  w(t,s)=p_t\left(1-e^{-(\lambda+r)(s-t)}\right)\frac{\lambda Y}{\lambda+r}+\left(1-p_t+p_te^{-\lambda(s-t)}\right)e^{-r(s-t)}Z
 \end{align*}
 as the principal's payoff at date $t$ when the agent stops at date $s\geq t$.
 Also, given belief $\mu$, when there is no subsequent information, the agent stops at date $\tau(\mu)$. Therefore, at any date $t$, if the agent holds belief $\mu$, he is willing to continue at this instant if and only if
 \begin{align*}
  \tau(\mu) \geq t\Rightarrow\mu\geq\bar\mu_t \equiv\frac{1}{y_H-y_L}\left(\left(1+\frac{r(p_0+(1-p_0)e^{\lambda t})}{p_0\lambda}\right)z-y_L\right).
 \end{align*}
 It is straightforward that $\bar\mu_t$ is increasing with $t$.
 Given $t\leq\tau(\mu)$, the principal's payoff at date $t$ from the agent with belief $\mu$ can be expressed as $w(t,\tau(\mu))$, which is equal to
 \begin{equation}
 \begin{aligned}
  W_{\text{NI}}(t,\mu)=&\left(1-e^{-(\lambda+r)(\tau(\mu)-{t})}\right)\frac{p_{t}\lambda Y}{\lambda+r}\\&+\left(1-p_{t}+p_{t}\cdot e^{-\lambda(\tau(\mu)-{t})}\right)\cdot e^{-r(\tau(\mu)-{t})}\cdot Z.
 \end{aligned}
 \end{equation}

\subsubsection*{Step 1: The disclosure when $\mu_0>\mu^*$}
 We first show that the concavification result of $w(t,\tau(\cdot))$ is time-invariant.
 Replicating the proof of Lemma \ref{lemma.useful} step by step, we know that $w(0,s)=w(0,t)+\gamma(0,t)\cdot [w(t,s)-Z]$, and therefore
 \begin{align*}
  \frac{\partial w(t,s)}{\partial s}=\frac{w'(0,s)}{\gamma(0,t)}.
 \end{align*}
 Thus, $w(t,\cdot)$ is decreasing when $t>t^*$, and similarly, $w(t,\tau(\cdot))$ decreases with $\tau(\cdot)$ given that $\tau(\cdot)$ is increasing with $\mu$.
 Also, we have identified a belief $\mu_H$ in Proposition \ref{proposition.static} such that the concavification of $w\circ\tau$ is either non-disclosure ($\mu\le\mu_H$), or splitting the belief into $\mu_H$ and $1$ ($\mu>\mu_H$).
 At any time $t$, since the sign of $w''(t,\cdot)$ is identical to $w''(0,\cdot)$, the concavification result can be derived in a similar vein, such that the threshold of non-disclosure is determined by choosing $\mu$ to minimize
 \begin{align*}
     \frac{w(t,\tau(1))-w(t,\tau(\mu))}{1-\mu}=\frac{w(\tau(1))-w(\tau(\mu))}{\gamma(0,t)(1-\mu)},
 \end{align*}
 the solution to which is irrelevant to $t$.

 At any time $t$, we derive the equilibrium strategy when $\mu_t>\mu^*$ as follows. 
 Obviously, if $\mu_t\le\bar{\mu}_t$, the agent would have stopped, and therefore there is no information disclosure. 
 If $\mu_t>\bar{\mu}_t$, suppose there is no subsequent information thereafter, and then the optimal disclosure at time $t$ can be derived by concavifying $\max\{Z,w(t,\tau(\mu))\}$, which is continuous for all $\mu\in[\mu^*,1]$.
 Therefore, when $\bar\mu_t\in[\mu_H,1)$, the optimal disclosure at $\mu\in(\bar{\mu}_t,1)$ must split $\mu$ into  $\bar\mu_t$ and $1$.
 As $\bar{\mu}_t$ is increasing, after time $\tau(\mu_H)$ with $\bar\mu_{\tau(\mu_H)}=\mu_H$, there will be no information thereafter on the equilibrium path.
 When $t\in[0,\tau(\mu_H)]$, since the concavification of $w(t,\tau(\cdot))$ is invariant in $t$ and $\max\{Z,w(t,\tau(\mu))\}$ is non-decreasing before $\mu^*$, the optimal disclosure for $\mu\geq\mu^*$ always splits $\mu$ into $\mu_H$ and $1$ when $\mu>\mu_H$ and is non-disclosure otherwise.

\subsubsection*{Step 2: When $\mu_0<\mu^*$, the optimal policy in Theorem \ref{thetheorem} cannot be implemented}

Suppose, for the sake of contradiction, that there exists a belief martingale $\{\mu_t\}_t$ that implements the optimal policy in Theorem \ref{thetheorem} without dynamic commitment.

At every continuation history reached with positive probability, sequential optimality requires the agent to be indifferent between continuing and stopping. Otherwise, the principal can profitably revise the continuation policy to extract the remaining rent.

Fix any such history at date $t$. Conditional on no breakthrough over $[t,t+dt)$, every successor history either stops, yielding $z$, or continues with zero continuation rent, also yielding $z$. Therefore, regardless of how the belief tree branches after this history, the agent's indifference condition is
\begin{align*}
 p_t\lambda dt\,y(\mu_t)
 +(1-p_t\lambda dt)(1-rdt)z
 +o(dt)
 =z.
\end{align*}

Letting $dt\to0$ gives
\begin{align*}
 y(\mu_t)
 =
 \left(1+\frac{r}{p_t\lambda}\right)z.
\end{align*}

Since $y(\cdot)$ is strictly increasing, this condition uniquely determines the belief at every continuation history at date $t$:
\begin{align*}
 \mu_t
 =
 \hat{\mu}_t
 \equiv
 y^{-1}\left(
 \left(1+\frac{r}{p_t\lambda}\right)z
 \right).
\end{align*}

Moreover, $\hat{\mu}_t$ is strictly increasing because $p_t$ is strictly decreasing.

The optimal policy in Theorem \ref{thetheorem} has a one-shot accelerator at date $t_b$. A behaviorally equivalent implementation may send interim messages before $t_b$, but it cannot induce the agent to stop before that date. Consider any $t,t+dt\in[\tau(\mu_0),t_b)$. Conditional on no breakthrough, every successor history at date $t+dt$ is therefore a continuation history. By the uniqueness result above, every such history must induce the same belief $\hat{\mu}_{t+dt}$, regardless of how the information structure branches.

Because breakthrough arrival is independent of quality, conditioning on no breakthrough does not change the belief-martingale condition. It therefore requires
\begin{align*}
 \hat{\mu}_t
 =
 E_t\mu_{t+dt}
 =
 \hat{\mu}_{t+dt},
\end{align*}
which contradicts the strict increase of $\hat{\mu}_t$. Hence, no belief martingale can implement the one-shot accelerator without dynamic commitment.

\subsubsection*{Step 3: The disclosure when $\mu_0<\mu^*$}

First, by concavification, the principal has no incentive to generate a belief higher than $\mu^*$, and therefore, the agent stops no later than $t^*$ with probability $1$.

Second, since we can consider the strategy of dynamic persuasion as constructed by a series of simple recommendations at each date, the principal's decision at date $t$ can be understood as allocating the unit probability mass for each state, $\theta=\{H,L\}$, between stopping and continuing.
For the same reason as in Step 2 of the proof of Theorem~\ref{thetheorem}, the principal never stops the agent before $t^*$ when $\theta=H$ in equilibrium. 
That is, $F_H(t)$, the cumulative probability that the agent stops no later than $t$ when $\theta=H$, is equal to $0$ when $t<t^*$ and jumps to $1$ when $t\geq t^*$.

Third, it is also straightforward that no information is disclosed when $t<\tau(\mu_0)$, where the agent is willing to experiment with probability $1$ even without information disclosure.

Then it remains to specify $F_L$.
In this no-commitment setting, the principal's inability to credibly promise future informational rewards forces her to optimize her disclosure strategy at each instant. 
This sequential optimization leads to a complete extraction of the agent's informational rents. 
Consequently, for the persuasion to be effective for any $t\in[\tau(\mu_0),t^*)$, the agent must be kept indifferent between continuing and stopping. 
This indifference condition is mathematically expressed by the agent's expected continuation payoff being zero:
\begin{align*}
 \mu_0\int_t^\infty \bar{v}_H(t,s)dF_H(s)+(1-\mu_0)\int_t^\infty\bar{v}_L(t,s)dF_L(s)=0,\quad\forall t\in[\tau(\mu_0),t^*).
\end{align*}
Replicating the proof in Theorem \ref{proposition.gradual}, we obtain the desired form (\ref{eq.closeform}).
Consider an on-path history at date $t$ with posterior belief $\mu_t$. 
Future disclosure decisions are taken as given at this history, and thus the principal chooses only the stopping probability assigned at date $t$ in each state. 
Since both $w$ and $v_H$ are increasing before $t^*$, stopping in state $H$ is strictly Pareto dominated, and hence $dF_H(t)=0$. 
Thus, the remaining choice is $dF_L(t)$, the probability to confirm the bad state at time $t$. 
Given the future continuation policy, increasing the low-state stopping probability at $t$ replaces a strictly negative low-state continuation payoff with the payoff from stopping immediately, which is zero. 
The principal, by contrast, strictly prefers this mass to continue because $t<t^*$, and therefore the continuation constraint must bind. 
The binding continuation constraint gives the desired form (\ref{eq.closeform}) for $dF_L(t)$.


\section{Applications of the Arrow--Pratt Approach}
\label{sec:applications}

We apply the three lemmas in Appendix A to eight settings drawn from the models of \citet{liu2023motivating}, \citet*{saeedi2025getting}, and \citet{koh2024attention}.  
Table C.1 summarizes the eight settings considered below and records which of the three lemmas applies in each case.

\begin{table}[H]
\centering
\small
\renewcommand{\arraystretch}{1.15}
\caption*{\textbf{Table C.1.} Applicability of Lemmas A.1--A.3}
\begin{tabular}{@{}>{\raggedright\arraybackslash}p{0.25\textwidth}>{\raggedright\arraybackslash}p{0.39\textwidth}>{\raggedright\arraybackslash}p{0.29\textwidth}@{}}
\toprule
\textbf{Application} & \textbf{Setting} & \textbf{Applicable lemmas} \\
\midrule
Motivating Effort with Information about Future Rewards
& 
& A.1--A.3 \\
\midrule
Getting the Agent to Wait
& Exponential discounting under common priors
& A.1 if $r_P<r$; A.2 if $r_P>r$ \\
\addlinespace
 
& General discounting under common priors
& A.1 where $R(w,t)<R(D,t)$; A.2 on an interval that exhausts all remaining stopping probability where $R(w,t)>R(D,t)$ \\
\addlinespace
 
& $D\circ w^{-1}$ first convex and then concave
& A.1 where \((D\circ w^{-1})''>0\); A.2 on the final interval on an interval that exhausts all remaining stopping probability and satisfies \((D\circ w^{-1})''<0\) \\
\addlinespace
 
& Heterogeneous priors
& A.1 if $r_P<r$ \\
\midrule
Attention Capture
& The principal is more sensitive to stopping-time risk
& A.2 on a final disclosure interval \\
\addlinespace
 
& The agent is more sensitive to stopping-time risk
& A.1 \\
\addlinespace
 
& \(R(w,t)-R(v,t)\) changes sign
& A.1 where \(R(w,t)<R(v,t)\); A.2 on the final disclosure interval \\
\bottomrule
\end{tabular}
\end{table}

Among the applications, Assumption \ref{ass:no-disclosure} is satisfied whenever Assumption \ref{ass:multi-separable} makes a common-date contraction available.
Assumption \ref{ass:statewise-separable} allows stopping probability in a given state to be shifted to earlier and later dates while keeping the relevant average of $v$ unchanged. 
Meanwhile, Assumption \ref{ass:multi-separable} provides the common payoff scale needed to move stopping probability from all states to the same date without changing earlier continuation payoffs.
Assumption \ref{ass:continuation-order} is additionally needed when stopping probability remains after the contracted interval.
It holds with equality in the reduced problems of the main model and \citet{liu2023motivating} because $\psi\equiv0$.
It fails in \textit{Getting the Agent to Wait} and \textit{Attention Capture} at beliefs where full information has positive value, although their final-interval contractions do not require it.
Under heterogeneous priors, Assumption \ref{ass:multi-separable} already fails and a joint contraction is unavailable.

The results in \citet{koh2024attention} and \citet*{saeedi2025getting} may require the Arrow--Pratt ranking to be applied interval by interval.
In their results, however, every contraction applies to the final disclosure interval, and hence Lemma~\ref{lem:final-interval} applies without Assumption~\ref{ass:continuation-order}.
Both models violate Assumption~\ref{ass:continuation-order} at beliefs where full information has positive value, and the counterexamples below show that moving stopping probability within a bounded interval to its certainty-equivalent date can violate a later continuation constraint.

\subsection{\textit{Motivating Effort with Information about Future Rewards}}

In \citet{liu2023motivating}, the agent's payoff from working from date $t$ to date $s$ in state $\theta\in\{H,L\}$ is
\begin{equation*}
 v_\theta(t,s)
 =\frac{\bigl(1-e^{-(\lambda_\theta+r)(s-t)}\bigr)
 (\lambda_\theta y_\theta-z)}{\lambda_\theta+r}.
\end{equation*}
Thus $v_\theta(t,s)$ is the function denoted by $\hat v_\theta(t,s)$ in
Appendix~\ref{appendix.general}.
The task-completion rate $\lambda_\theta$ captures task feasibility and is perfectly correlated with task quality.
We consider the case
\[
 \lambda_Hy_H>z>\lambda_Ly_L,
\]
so the agent prefers to work longer in state $H$ and to stop sooner in state $L$.
The principal initially chooses the pair $\langle F_H,F_L\rangle$.
In state $H$, both players prefer continued effort, and thus the high-state schedule is pinned down at its efficient value.
The remaining policy choice is $F_L$.
The principal's payoff from stopping at $t$ in the reduced low-state problem is
\begin{equation*}
 w(t)
 =\frac{\lambda_L
 \bigl(1-e^{-(\lambda_L+r_P)t}\bigr)b_L}
 {\lambda_L+r_P},
\end{equation*}
which is increasing in $t$.
Here, $b_L>0$ is the principal's payoff from task completion in state $L$.
The high-state contribution is constant, and the positive low-state prior weight does not affect the optimizer.
The principal's payoff in the reduced problem is therefore state-irrelevant, as required by \eqref{eq:multistate-problem}.
Because $b_L$ is positive and constant, it does not affect the curvature comparison.
Bayes' rule gives
\begin{equation}
\label{eq:liu-belief}
 \mu_t
 =\frac{\mu_0e^{-\lambda_Ht}}
 {\mu_0e^{-\lambda_Ht}
 +(1-\mu_0)e^{-\lambda_Lt}\bigl(1-F_L(t)\bigr)}.
\end{equation}
Substituting \eqref{eq:liu-belief} into the continuation constraint reduces the two-state problem to
\begin{equation*}
\begin{aligned}
 \max_{F_L}\quad &\int_0^\infty w(t)\,dF_L(t)\\
 \text{subject to}\quad
 &\int_{[t,\infty)} v_L(t,s)\,dF_L(s)\geq C(t),
 \qquad t\geq0,
\end{aligned}
\end{equation*}
where
\begin{equation*}
C(t)
=-\frac{\mu_0e^{-(\lambda_H-\lambda_L)t}}{1-\mu_0}
 \frac{\lambda_Hy_H-z}{\lambda_H+r}.
\end{equation*}

All three assumptions hold in the reduced problem.
For Assumption \ref{ass:multi-separable}, let $v(s)=v_L(0,s)$.
The agent's payoff satisfies
\[
 v_L(t,s)=e^{(\lambda_L+r)t}v(s)
 +\frac{\lambda_Ly_L-z}{\lambda_L+r}
 \bigl(1-e^{(\lambda_L+r)t}\bigr),
\]
so the payoff depends on $s$ through $v(s)$, its date-$t$ factor is positive, and $v$ is strictly decreasing because $z>\lambda_Ly_L$.

For Assumption \ref{ass:no-disclosure}, fix any interval $[\underline t,\bar t)$ to which the policy assigns no stopping probability.
For every $t$ in this interval,
\[
 \int_{[t,\infty)}v(s)\,dF_L(s)
 =
 \int_{[\underline t,\infty)}v(s)\,dF_L(s).
\]
Substituting the separable form of $v_L(t,s)$ into the continuation constraint and dividing by $e^{(\lambda_L+r)t}$ gives the equivalent constraint
\[
 \int_{[t,\infty)}v(s)\,dF_L(s)
 \geq
 -\frac{\mu_0}{1-\mu_0}
 \frac{\lambda_Hy_H-z}{\lambda_H+r}
 e^{-(\lambda_H+r)t}
 +\bigl(1-F_L(\underline t)\bigr)
 \frac{\lambda_Ly_L-z}{\lambda_L+r}
 \bigl(1-e^{-(\lambda_L+r)t}\bigr).
\]
Using \eqref{eq:liu-belief} and the absence of stopping probability within the interval, the derivative of the right-hand side equals
\[
 \bigl(1-F_L(\underline t)\bigr)e^{-(\lambda_L+r)t}
 \left[
 \frac{\mu_t}{1-\mu_t}(\lambda_Hy_H-z)
 +\lambda_Ly_L-z
 \right].
\]
Starting from any belief $\mu$, the agent's interior no-information stopping time, measured from that history, is
\begin{equation}
 \tau(\mu)
 =\frac{1}{\lambda_H-\lambda_L}
 \ln\left[
 \frac{\mu}{1-\mu}
 \frac{\lambda_Hy_H-z}{z-\lambda_Ly_L}
 \right].
\end{equation}
At this stopping time, the bracketed term equals zero.
Because $\lambda_H>\lambda_L$, the posterior $\mu_t$ falls while no disclosure occurs, and hence the bracket is nonpositive at every later date.
Thus, on every no-disclosure interval beginning at or after the no-information stopping date, its first date is the tightest one.
This verifies Assumption \ref{ass:no-disclosure} for the reduced problem.

For Assumption \ref{ass:continuation-order}, the reduced problem has $\psi\equiv0$, and thus $\Psi\equiv0$.
Moreover, $\hat v_L(t,t)=v_L(t,t)=0$.
Hence \eqref{eq:continuation-pointwise} holds with equality.
The inequality $z>\lambda_Ly_L$ also implies $v_L(t,s)\leq v_L(t,t)=0$ for every $s\geq t$, which is the payoff comparison used in Lemma \ref{lem:bounded-interval}.

Assumption \ref{ass:multi-separable} implies Assumption \ref{ass:statewise-separable}, and the reduced problem also satisfies Assumptions \ref{ass:no-disclosure} and \ref{ass:continuation-order}.
The three lemmas therefore apply under their respective curvature and interval conditions. 
The relevant Arrow--Pratt coefficients are
\begin{equation*}
 R(w,t)=\lambda_L+r_P,
 \qquad
 R(v,t)=\lambda_L+r.
\end{equation*}
Maintain $\lambda_H>\lambda_L$.
If $r_P>r$, the relevant interval exhausts the remaining low-state stopping probability, and hence Lemma \ref{lem:final-interval} gives one-shot disclosure.
If $r_P<r$, Lemma \ref{lem:no-rent} makes the continuation constraint bind wherever the gradual policy has positive density.
Differentiating this identity gives the Poisson disclosure rate characterized in Proposition~5 of \citet{liu2023motivating}.

\subsection{\textit{Getting the Agent to Wait}}

\subsubsection{Exponential discounting under common priors}
\citet*{saeedi2025getting} call the two parties the principal and the agent.
The principal's payoff from stopping at date $t$ is $w(t)=1-e^{-r_Pt}$.
After stopping, the agent chooses a static action with indirect payoff $V(\mu)$, where $V$ is strictly convex, differentiable, and symmetric around $1/2$.
A stopping recommendation inducing an interior posterior $\mu_s\in(0,1)$ can be split, at the same date, into posteriors $0$ and $1$ while preserving its mean.
This raises the agent's stopping payoff because $\mu_sV(1)+(1-\mu_s)V(0)>V(\mu_s)$ by strict convexity, without changing the principal's payoff.
We therefore restrict attention to stopping recommendations that fully reveal the state; that is, a recommendation sent in state $H$ induces posterior $1$, while one sent in state $L$ induces posterior $0$.
Symmetry gives $V(0)=V(1)$, and thus the discounted value of full revelation at date $s$, evaluated at date $t$, is $e^{-r(s-t)}V(1)$, independent of the state revealed at $s$.
Hence stopping at $s$ enters both states' payoffs through
$v(s)=e^{-rs}V(1)$,
up to the positive date-$t$ factor and the current stopping payoff $V(\mu_t)$.

Let $F_H$ and $F_L$ be the state-contingent stopping distributions.
The no-message belief is
\begin{equation*}
 \mu_t
 =\frac{\mu_0\bigl(1-F_H(t)\bigr)}
 {(1-\mu_0)\bigl(1-F_L(t)\bigr)
 +\mu_0\bigl(1-F_H(t)\bigr)}.
\end{equation*}
The continuation constraint is
\begin{equation}
 \int_{[t,\infty)} e^{-r(s-t)}
 \left[\mu_tV(1)\,dF_H(s\mid s\geq t)
 +(1-\mu_t)V(0)\,dF_L(s\mid s\geq t)\right]
 \geq V(\mu_t).
\label{eq:sss-ic}
\end{equation}
Symmetry permits the aggregate stopping distribution $F(t)=\mu_0F_H(t)+(1-\mu_0)F_L(t)$.
Multiplying \eqref{eq:sss-ic} by the aggregate probability $1-F(t)$ of reaching $t$ gives
\begin{equation*}
 \int_{[t,\infty)}
 \bigl(e^{-r(s-t)}V(1)-V(\mu_t)\bigr)\,dF(s)
 \geq0.
\end{equation*}
For Assumption \ref{ass:multi-separable}, take $A(t)=e^{rt}$, $v(s)=e^{-rs}V(1)$, and $B_\theta(t)=0$.
On a no-disclosure interval with a fixed belief, $e^{-r(s-t)}V(1)$ increases with the current date $t$ for every fixed disclosure date $s$, while the immediate stopping payoff remains $V(\mu)$.
The interval's first date is therefore the tightest one, which verifies Assumption \ref{ass:no-disclosure}.

Assumption \ref{ass:continuation-order} is not satisfied.
To see why, fix $q_H,q_L>0$ and let $\mu=q_H/(q_H+q_L)$.
Then $\Psi(q)=(q_H+q_L)V(\mu)$.
Consider first an increase in $q_H$ while holding $q_L$ fixed.
The quotient rule gives
\[
 \frac{\partial\mu}{\partial q_H}
 =\frac{(q_H+q_L)-q_H}{(q_H+q_L)^2}
 =\frac{1-\mu}{q_H+q_L}.
\]
Strict convexity implies $V'(\mu)<[V(1)-V(\mu)]/(1-\mu)$.
Substituting the derivative above into the product rule gives
\[
\begin{aligned}
 \partial_{q_H}\Psi(q)
 &=V(\mu)+(q_H+q_L)V'(\mu)
 \frac{\partial\mu}{\partial q_H}=V(\mu)+(1-\mu)V'(\mu)\\
 &<V(\mu)+(1-\mu)
 \frac{V(1)-V(\mu)}{1-\mu}=V(1)=\hat v_H(t,t).
\end{aligned}
\]
For the same reason, $\partial_{q_L}\Psi(q)=V(\mu)-\mu V'(\mu)<V(0)=\hat v_L(t,t)$.  
Thus \eqref{eq:continuation-pointwise} is reversed in both states.  
The reversed inequalities reflect the strict convexity generated by the positive value of information.

The stopping-time risk coefficients are $R(w,t)=r_P$ and $R(v,t)=r$.
When $r_P>r$, Lemma \ref{lem:final-interval} gives a global one-shot disclosure at a common full-revelation date $\hat t$.
Assumption \ref{ass:no-disclosure} makes the date-zero continuation constraint the tightest.
Since the principal prefers delay, this constraint binds at $e^{-r\hat t}V(1)=V(\mu_0)$.
Taking logs gives $\hat{t}=\frac{1}{r}\ln\frac{V(1)}{V(\mu_0)}$.

If $r_P<r$, Lemma \ref{lem:no-rent} makes the continuation constraint bind at every interior disclosure date.
It remains to show that the disclosure support cannot contain a gap.
If $(t_1,t_2)$ separated two disclosure dates, the constraint would bind at $t_2$ but fail at every earlier date in the gap by Assumption \ref{ass:no-disclosure}.
The support is therefore connected, and hence disclosure is gradual.

This gradual policy has two phases.
Suppose $\mu_0>1/2$.
Strict convexity and symmetry of $V$ imply that $V(\mu)$ is minimized at $1/2$, and thus the gain from full information, $V(1)-V(\mu)$, is greatest there.
Full revelation gives the agent the same payoff $V(1)=V(0)$ in either state.
An $H$-state disclosure, however, makes the absence of disclosure bad news, moving the posterior toward $1/2$ and lowering the payoff $V(\mu)$ from stopping immediately.
An $L$-state disclosure would move the posterior farther above $1/2$ and raise that payoff.
The first phase therefore discloses only state $H$.
To obtain a closed-form solution for $F$, note that the continuation constraint, by Lemma \ref{lem:no-rent}, binds at every interior date of this phase.
\begin{equation}
 V(1)\int_{[t,\infty)}e^{-rs}\,dF(s)
 =e^{-rt}(1-F(t))V(\mu_t).
\label{eq:sss-first-phase-binding}
\end{equation}
Differentiating both sides of \eqref{eq:sss-first-phase-binding} gives
\[
 -e^{-rt}V(1)F'(t)
 =-re^{-rt}(1-F(t))V(\mu_t)-e^{-rt}F'(t)V(\mu_t)
 +e^{-rt}(1-F(t))V'(\mu_t)\dot\mu_t.
\]
Because only state $H$ is disclosed, the undisclosed probability of state $L$ remains constant.
Bayes' rule therefore gives
\begin{equation}
 (1-\mu_t)(1-F(t))=1-\mu_0
 \quad\Longrightarrow\quad
 \frac{F'(t)}{1-F(t)}=-\frac{\dot\mu_t}{1-\mu_t}.
\label{eq:sss-first-phase-bayes}
\end{equation}
Using \eqref{eq:sss-first-phase-bayes} to eliminate $F'(t)$ and rearranging yields
\[
 -r
 =\frac{\dot\mu_t}{V(\mu_t)}
 \left[
  \frac{V(1)-V(\mu_t)}{1-\mu_t}-V'(\mu_t)
 \right].
\]
This equation determines the no-message belief, and Bayes' rule gives the
state-$H$ disclosure rate $-\dot\mu_t/[\mu_t(1-\mu_t)]$.
The absence of
disclosure lowers the posterior until it reaches $1/2$.
If $\mu_0<1/2$,
the argument is symmetric and only state $L$ is disclosed in the first phase.

Once the posterior reaches $1/2$, continuing to disclose only one state would move it away from the point at which full information is most valuable.
The second phase therefore discloses both states at the same rate, which keeps the no-message belief at $1/2$.
The binding continuation constraint becomes
\begin{equation*}
 \int_{[t,\infty)} e^{-rs}\,dF(s)
 =e^{-rt}\bigl(1-F(t)\bigr)\frac{V(1/2)}{V(1)}.
\end{equation*}
Differentiating both sides gives
\[
 -e^{-rt}F'(t)
 =-e^{-rt}\bigl[r(1-F(t))+F'(t)\bigr]
 \frac{V(1/2)}{V(1)}.
\]
Cancelling $e^{-rt}$ and rearranging gives the differential equation
\[
 \frac{F'(t)}{1-F(t)}
 =r\frac{V(1/2)}{V(1)-V(1/2)}.
\]
Thus disclosure arrives at a constant Poisson rate.

\paragraph{Why an internal contraction can fail.}
The common-prior model gives an explicit example.
Set $r=1$, $r_P=1/2$, $\mu_0=1/2$, and $V(\mu)=\mu^2+(1-\mu)^2$.
The optimal gradual policy has constant hazard one, and hence $\tau\sim\operatorname{Exp}(1)$ and its continuation constraint binds at every date.
Contract the stopping probability in $[0,1]$ to its $e^{-t}$-certainty-equivalent date $\hat t$, where
\[
 e^{-\hat t}
 =\mathbb E[e^{-\tau}\mid\tau\in[0,1]]
 =\frac{1+e^{-1}}{2},
 \qquad \hat t\simeq0.380.
\]
At date $t=0.7$, the contracted policy retains only the original tail after date $1$, and
$\mathbb E[e^{-(\tau-t)}\mid\tau>1]=e^{-0.3}/2<1/2=V(1/2)/V(1)$.
Thus the contraction violates the continuation constraint even though it moves both states proportionally and leaves the posterior unchanged.

\subsubsection{General discounting under common priors}
Let $w(t)$ be the principal's increasing payoff from stopping at date $t$, and let $D(t)$ be
the agent's decreasing discount factor.
Taking the principal's payoff as
the time index, the transformed discount factor $D\circ w^{-1}$ satisfies
\[
 (D\circ w^{-1})''(w(t))
 =\frac{D'(t)}{[w'(t)]^2}
 \bigl[R(w,t)-R(D,t)\bigr].
\]
Because $D'(t)<0$, $D\circ w^{-1}$ is convex when $R(w,t)<R(D,t)$ and
concave when $R(w,t)>R(D,t)$.  

\subsubsection{Convex-concave discounting}
Proposition 8 of \citet{saeedi2025getting} studies the case in which $D\circ w^{-1}$ is first convex and then concave. 
The two gradual phases occur in the initial convex region, where the first phase moves the belief gradually toward $1/2$ by disclosing only the state that is currently more likely, while the second phase discloses both states at the same rate to keep the belief at $1/2$.
In this region, the equation governing the no-message belief before it reaches \(1/2\) replaces $-r$ by $D'(t)/D(t)$.  
After the posterior reaches $1/2$, both states are disclosed at the common rate
\[
 -\frac{D'(t)}{D(t)}
 \frac{V(1/2)}{V(1)-V(1/2)}.
\]

The second phase ends before the inflection date, after which no information is disclosed until a final full-revelation date. 
The interval contracted in their optimal policy is therefore the final disclosure interval and exhausts all remaining stopping probability. 
Lemma \ref{lem:final-interval} applies precisely because the contraction leaves no later continuation constraint to check.

This terminal result cannot be extended directly to an arbitrary interval.  
If stopping probability remains after the contracted interval, the later continuation constraints must still be satisfied.  
The present model does not satisfy Assumption \ref{ass:continuation-order}, and the exponential-discounting counterexample shows that moving such an interval to its certainty-equivalent date can violate one of those later constraints.

\subsubsection{Heterogeneous priors}
We now allow the principal and the agent to hold different prior beliefs about the state.
Let $\mu_P$ and $\mu_A$ be the principal's and the agent's initial
probabilities of state $H$, and define
$\ell:=\frac{\mu_A}{1-\mu_A}\frac{1-\mu_P}{\mu_P}$.
Taking the principal's prior as $\mu_0$, use the surviving masses from
Appendix~\ref{appendix.general}, and thus
$q_H(t)=\mu_P[1-F_H(t)]$ and
$q_L(t)=(1-\mu_P)[1-F_L(t)]$.
Up to a common positive scale, the agent's
surviving mass is $\ell q_H(t)+q_L(t)$, and his no-message belief is
\[
 \mu_t^A=\frac{\ell q_H(t)}{\ell q_H(t)+q_L(t)}.
\]
The principal chooses $q_H$ and $q_L$ to maximize
\[
 \int_0^\infty e^{-r_Pt}
 \bigl[q_H(t)+q_L(t)\bigr]dt
\]
subject to
\[
 -V(1)\int_{[t,\infty)}e^{-r(s-t)}
 d\bigl[\ell q_H(s)+q_L(s)\bigr]
 \geq \bigl[\ell q_H(t)+q_L(t)\bigr]V(\mu_t^A)
 \qquad\text{for every }t\geq0.
\]

This problem has the form of \eqref{eq:multistate-problem}.
It has the
state-irrelevant payoff $w(t)=1-e^{-r_Pt}$ and
\[
 \hat v_H(t,s)=\ell e^{-r(s-t)}V(1),
 \qquad
 \hat v_L(t,s)=e^{-r(s-t)}V(1),
\]
while
\[
 \psi(\mu)
 =\bigl[\ell\mu+1-\mu\bigr]
 V\left(\frac{\ell\mu}{\ell\mu+1-\mu}\right),
 \qquad C(t)=0.
\]
It therefore satisfies Assumption \ref{ass:statewise-separable} with
$v(s)=e^{-rs}V(1)$, $A_H(t)=\ell e^{rt}$, and $A_L(t)=e^{rt}$.
When
$\ell\neq1$, however, $A_H$ and $A_L$ are different, and hence Assumption
\ref{ass:multi-separable} fails.

This distinction matters even under a uniform Arrow--Pratt ranking.
Direct calculation gives
\[
 R(w,t)=r_P,
 \qquad
 R(v,t)=r.
\]
Assumption \ref{ass:statewise-separable} is sufficient for Lemma
\ref{lem:no-rent}.
Hence, if $r_P<r$, Lemma \ref{lem:no-rent} remains
applicable over the entire horizon.
When $r_P>r$, moving the two
state-contingent stopping distributions to a common date requires Assumption
\ref{ass:multi-separable}, which is not satisfied.  
Proposition 3 of \citet{saeedi2025getting} illustrates this distinction.  
It assumes $r_P>r$ and $\ell<1$, and thus the principal is more sensitive to stopping-time risk at every date.  
Yet its optimal policy first places a positive atom of state-$L$ disclosure at an internal date $t_1^*$, then discloses state $L$ gradually on $[t_1^*,t_2^*]$, and finally reveals state $H$ at $t_2^*$.

\subsection{\textit{Attention Capture}}

\subsubsection{The principal is more sensitive to stopping-time risk}
\citet{koh2024attention} call the two parties the designer and the decision maker.
We refer to them as the principal and the agent, respectively.
The state is $\theta\in\{0,1\}$.
At each date, the agent either continues to observe the information chosen by the principal or stops and takes an action.
Let $u(a,\theta)$ be his payoff from action $a$ in state $\theta$.
If he stops with posterior $\mu$, his optimal payoff is
\[
 \psi(\mu):=\max_a\mathbb E_\mu[u(a,\theta)].
\]
If he learns the state before acting, his payoff in state $\theta$ is $u_\theta^*:=\max_a u(a,\theta)$.
The value of full information at belief $\mu$ is therefore
\[
 \varphi(\mu)
 :=\sum_\theta\mu(\theta)u_\theta^*-\psi(\mu).
\]
The agent also bears an attention cost while waiting.
We write the accumulated cost at date $t$ as $-v(t)$, where $v$ is smooth and strictly decreasing, and hence waiting from date $t$ to date $s$ costs $v(t)-v(s)$.
The principal receives an increasing payoff $w(t)$ when the agent stops at date $t$.

Represent a policy by the state-contingent stopping distributions $F_1$ and $F_0$.
The principal's objective and the agent's continuation constraints then take the form of \eqref{eq:multistate-problem}, as verified below.

If full information arrives at date $s\geq t$, the agent's date-$t$ payoff in state $\theta$ is
$\hat v_\theta(t,s)=u_\theta^*+v(s)-v(t)$.
Hence Assumption \ref{ass:multi-separable} holds with $A(t)=1$, $B_\theta(t)=u_\theta^*-v(t)$, and $C(t)=0$.

On a no-disclosure interval, the posterior is fixed.
Waiting at date $t$ for full information at a fixed date $s$ rather than acting immediately gives the agent $\varphi(\mu)+v(s)-v(t)$.
This gain increases with $t$, and thus the interval's first date is its tightest continuation constraint.
This verifies Assumption \ref{ass:no-disclosure}.

Assumption \ref{ass:continuation-order} is not satisfied.
At the belief $\mu=q/\sum_\theta q_\theta$, homogeneity of $\Psi$ and Euler's identity imply $\sum_\theta\mu(\theta)\partial_{q_\theta}\Psi(q)=\psi(\mu)$.
At $s=t$, the expected payoff from full revelation is $\sum_\theta\mu(\theta)\hat v_\theta(t,t)=\sum_\theta\mu(\theta)u_\theta^*$.
Hence, at any belief with $\varphi(\mu)>0$,
\[
 \sum_\theta\mu(\theta)
 \bigl[\hat v_\theta(t,t)-\partial_{q_\theta}\Psi(q)\bigr]
 =\varphi(\mu)>0.
\]
At least one bracketed difference is therefore positive, whereas Assumption \ref{ass:continuation-order} requires every difference to be nonpositive.

Using accumulated cost as the time index transforms the principal's payoff into $w\circ(-v)^{-1}$.
Its curvature satisfies
\[
 (w\circ(-v)^{-1})''(-v(t))
 =\frac{w'(t)}{[v'(t)]^2}
 \bigl[R(v,t)-R(w,t)\bigr].
\]
Thus the curvature of $w\circ(-v)^{-1}$ gives exactly the Arrow--Pratt comparison used in the two lemmas.

When $R(w,t)>R(v,t)$ on an interval $I$ that exhausts all remaining stopping probability, Lemma \ref{lem:final-interval} moves that probability to the date $\hat t$ satisfying
\[
 v(\hat t)=\mathbb E[v(\tau)\mid\tau\in I]
\]
and raises the principal's payoff.
If the stopping-time distribution is concentrated at a single date $\hat t$, its binding date-zero constraint is $v(\hat t)=v(0)-\varphi(\mu_0)$.

\subsubsection{The agent is more sensitive to stopping-time risk}
When $R(w,t)<R(v,t)$, Lemma \ref{lem:no-rent} instead makes the continuation constraint bind at every interior date with positive stopping density.
To describe the resulting gradual policy, let $\varphi^*:=\max_\mu\varphi(\mu)$ and $\Phi^*:=\arg\max_\mu\varphi(\mu)$.
Conditional on receiving no stopping recommendation by date $t$, Bayes' rule gives
\begin{equation}
\label{eq:attention-fixed-posterior}
 \mu_t=
 \frac{\mu_0\bigl(1-F_1(t)\bigr)}
 {\mu_0\bigl(1-F_1(t)\bigr)
 +(1-\mu_0)\bigl(1-F_0(t)\bigr)}.
\end{equation}
In the gradual policy characterized by \citet{koh2024attention}, whenever this posterior moves, it does so in a direction that weakly raises $\varphi(\mu_t)$.
In the binary model, this movement is generated by recommending stopping in only one state.
Once the posterior reaches $\Phi^*$, stopping can be recommended in both states in proportions that keep the ratio in \eqref{eq:attention-fixed-posterior}, and hence the posterior, unchanged.
Let $F(t):=\mu_0F_1(t)+(1-\mu_0)F_0(t)$ denote the unconditional distribution of the stopping date $\tau$.
During this fixed-posterior phase, its binding continuation constraint is
\begin{equation}
\label{eq:attention-binding}
 -\int_t^\infty v'(s)\bigl(1-F(s)\bigr)\,ds
 =\varphi^*\bigl(1-F(t)\bigr).
\end{equation}
Differentiating both sides of \eqref{eq:attention-binding} gives
\[
 \frac{F'(t)}{1-F(t)}=-\frac{v'(t)}{\varphi^*}.
\]
Stopping therefore occurs at the constant rate $1/\varphi^*$ when
accumulated cost is used as the time index.
For the linear specification
$v(t)=-ct$, the calendar-time hazard is $c/\varphi^*$.
This is the
continuous-time version of the geometric stopping-time distribution in
\citet{koh2024attention}.

\subsubsection{S-shaped principal payoff}
Proposition 2(ii) of \citet{koh2024attention} combines the two comparisons when $w\circ(-v)^{-1}$ is S-shaped.
Once the posterior in that policy has entered $\Phi^*$, stopping occurs gradually until a date $t^*$.
Conditional on reaching $t^*$, no stopping occurs until all remaining stopping probability is concentrated at the date $\hat t$ satisfying $v(\hat t)=v(t^*)-\varphi^*$.
Under the linear specification $v(t)=-ct$, this fixed-posterior gradual phase has the constant hazard $c/\varphi^*$ and the final stopping date is $t^*+\varphi^*/c$.
Lemma \ref{lem:no-rent} explains why the continuation constraint binds during the gradual phase, while Lemma \ref{lem:final-interval} explains the concentration of the final stopping interval.
The transition date $t^*$ is instead determined by the conditional concavification of $w\circ(-v)^{-1}$ over future dates.

For an arbitrary increasing principal payoff, the general result in \citet{koh2024attention} restricts an optimum to block structures but does not determine the number or location of the blocks.
The Arrow--Pratt comparison likewise does not characterize the complete optimal policy for an arbitrary payoff function.
Because Assumption \ref{ass:continuation-order} fails here, it cannot in general justify contracting an internal interval; the counterexample below shows that such a contraction can violate a later continuation constraint.

\paragraph{Why internal contraction can fail.}
Concentrating a bounded interval at its $v$-certainty-equivalent date need not preserve later constraints.
The failure can already be seen under the linear specification $v(t)=-t$.
Fix any $\mu_0\in\Phi^*$, set $\varphi^*=10$, and place stopping probabilities $100/102$, $1/102$, and $1/102$ at dates $1$, $5$, and $15$.
Feasibility requires the expected remaining attention cost, conditional on reaching date $t$, to be at most $10$.
Before date $1$ this expectation is at most $120/102$; between dates $1$ and $5$ it is $10-t$; after date $5$ it is $15-t\leq10$.

Contract the stopping probability in $[1,5]$ to its mean $\hat t=(100+5)/101=105/101<2$.
At date $2$, the original conditional expected remaining time is $(5+15)/2-2=8$.
After the contraction, only date $15$ remains, and hence the expected remaining time is $15-2=13>10$.
Applying the same distribution in both states keeps the posterior at $\mu_0$, and thus the contraction violates the continuation constraint without changing beliefs.

\end{document}